\documentclass[useAMS,usenatbib,usegraphicx]{mn2e}
\usepackage{tikz}
\usepackage{color}

\title[Optical and NIR IFU spectroscopy of NGC 7582]
  {Optical and near-infrared IFU spectroscopy of the nuclear region of the AGN-starburst galaxy NGC 7582 }
\author[Ricci et al.]
  {T.V.~Ricci\thanks{tiago.ricci@uffs.edu.br}$^{1}$,
  J.E.~Steiner$^2$, D.~May$^3$, A.~Garcia-Rissmann$^3$, 
  R.B.~Menezes$^2$ \\
  $^1$Universidade Federal da Fronteira Sul, Campus Cerro Largo, RS 97900-000, Brasil \\
  $^2$Instituto de Astronomia, Geof\'isica e Ci\^encias Atmosf\'ericas, Universidade de S\~ao Paulo, 05508-900, S\~ao Paulo, Brasil\\
  $^3$Laborat\'orio Nacional de Astrof\'isica/MCTI, Rua dos Estados Unidos, 154, Bairro das Na\c{c}\~oes, 37500-000, Itajub\'a, MG, Brazil
  }
\date{Accepted 2017 October 18.}
\pagerange{\pageref{firstpage}--\pageref{lastpage}} \pubyear{2017}

\def\LaTeX{L\kern-.36em\raise.3ex\hbox{a}\kern-.15em
    T\kern-.1667em\lower.7ex\hbox{E}\kern-.125emX}

\begin{document}

\label{firstpage}

\maketitle

\begin{abstract}

NGC 7582 is an SB(s)ab galaxy which displays evidences of simultaneous nuclear activity and star formation in its centre. Previous optical observations revealed, besides the H II regions, an ionization cone and a gas disc in its central part. \textit{Hubble Space Telescope (HST)} images in both optical and infrared bands show the active galactic nuclei (AGNs) and a few compact structures that are possibly associated with young stellar clusters. In order to study in detail both the AGN and evidence for star formation, we analyse optical (Gemini Multi-Object Spectrograph) and near-infrared (Spectrograph for Integral Field Observations in the Near Infrared) archival data cubes. We detected five nebulae with strong He II$\lambda$4686 emission in the same region where an outflow is detected in the [O III]$\lambda$5007 kinematic map. We interpreted this result as clouds that are exposed to high-energy photons emerging from the AGN throughout the ionization cone. We also detected Wolf$-$Rayet features which are related to emission of one of the compact clusters seen in the \textit{HST} image. Broad H$\alpha$ and Br$\gamma$ components are detected at the position of the nucleus. [Fe II]$\lambda$1.644 $\mu$m, H$_2\lambda$2.122 $\mu$m and Br$\gamma$ flux maps show two blobs, one north and the other south from the nucleus, that seem to be associated with five previously detected mid-infrared sources. Two of the five He II nebulae are partially ionized by photons from starbursts. However, we conclude that the main source of excitation of these blobs is the AGN jet/disc. The jet orientation indicates that the accretion disc is nearly orthogonal to the dusty torus. 

\end{abstract}


\section{Introduction} \label{sec:intro}

\begin{figure*}
    \includegraphics[scale=0.4]{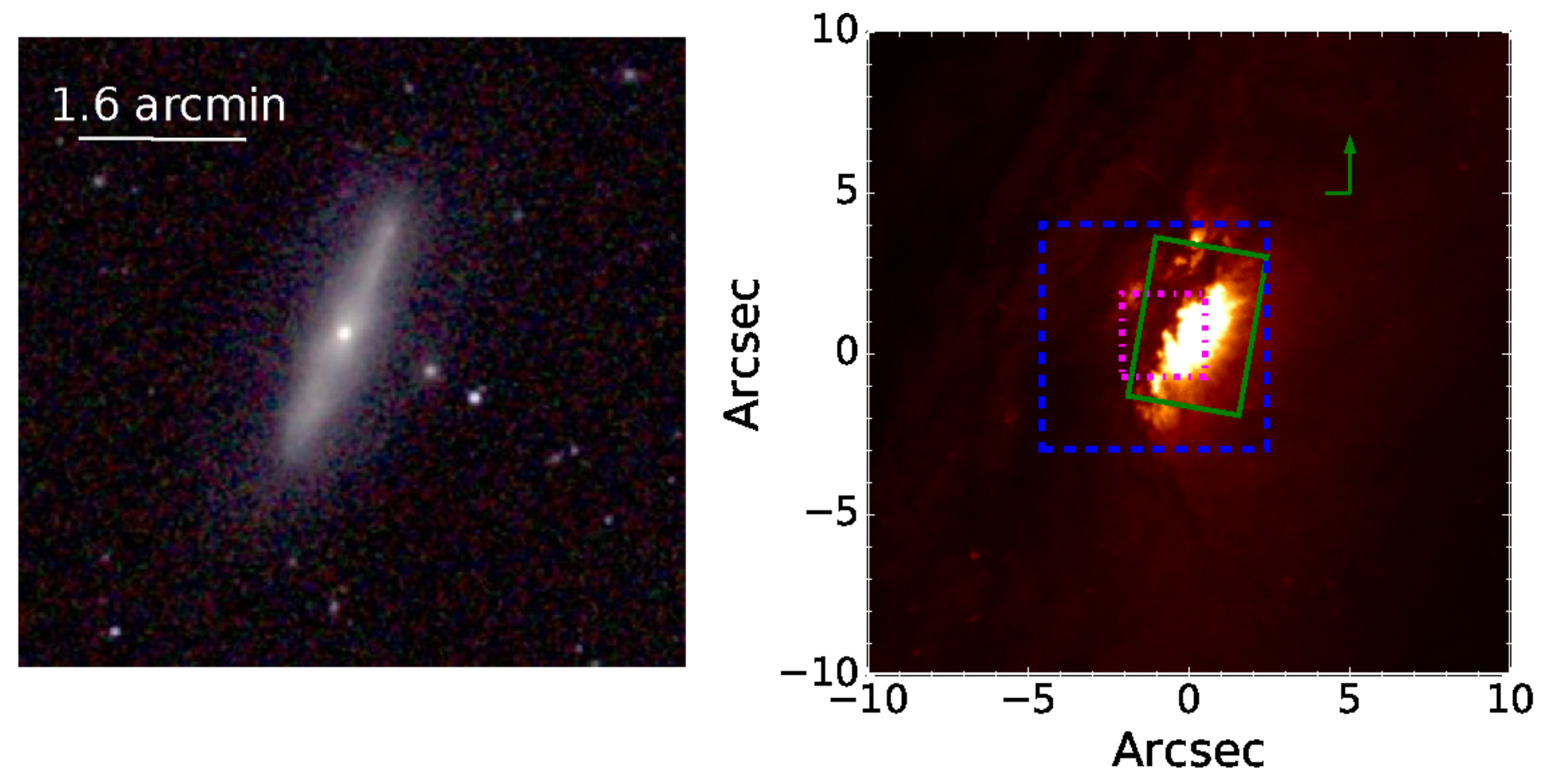}
\caption{Left: composed \textit{JHK} image of NGC 7582 from Two Micron All Sky Survey \citep{2006AJ....131.1163S}. Right: optical image from \textit{HST}. The solid green rectangle represents the GMOS FOV while the dashed blue and the dotted magenta ones correspond to both SINFONI FOVs used in this work. For this galaxy, 1 arcsec = 110 pc. \label{fig:ngc7582} }
\end{figure*}

Active galactic nuclei (AGNs) are usually strong hard X-ray sources, but only a few galaxies present this characteristic and have optical spectra that are dominated by H II regions. NGC 7582 [SB(s)ab, according to RC 3 \citep{1991trcb.book.....D}] is one of them. This galaxy (see Fig. \ref{fig:ngc7582}) is a member of the Grus quartet, the other components being NGC 7552, NGC 7590 and NGC 7599. NGC 7582 was identified by \citet{1978ApJ...223..788W} as a hard X-ray source observed by the \textit{Uhuru} and \textit{Ariel V} satellites. This identification was based on the coincidence of the source position in both bands and also in the presence of weak He II$\lambda$4686 and [Ne V]$\lambda$3425 emission in the optical spectrum, otherwise dominated by lines emitted by H II regions. 

Follow-up observations have also indicated the AGN nature of this source. Optical spectra showed strong deviation in velocity and profile of the [O III]$\lambda$5007 line, when compared to simple H II regions \citep{1981A&A....97...71V,1985MNRAS.216..193M}. Evidence of an ionization cone \citep{1985MNRAS.216..193M,1991MNRAS.250..138S,2016ApJ...824...50D} is clearly suggested when comparing the [O III] line to H$\alpha$. Soft X-ray observations also reveal the ionization cone \citep{2007MNRAS.374..697B,2017A&A...600A.135B}. Based on observations using an optical integral field unit (IFU) with a field of view (FOV) of 3.6 $\times$ 3.9 kpc$^2$ and a resolution of 130 pc, \citet{2016ApJ...824...50D} proposed that radiation pressure is dominant over the extended narrow-line region (NLR) of NGC 7582 across the ionization cone and that this may drive the outflows detected in this region.

The position of the AGN was defined from infrared observations \citep{2002ApJ...571L...7P,2006MNRAS.369L..47W}. \textit{V}-band observations made with the \textit{Hubble Space Telescope (HST)} revealed that the AGN is variable in the optical region \citep{2007MNRAS.374..697B}. \citet{2006A&A...460..449W} used diffraction-limited long-slit observations in the mid-infrared (MIR) region to model the kinematics of the [Ne II]$\lambda$12.8$\mu$m gas in the centre of NGC 7582. They derived a black hole mass of 5.5$\times$10$^7$ M$_\odot$. 

Although such observations point towards the presence of a hidden broad-line region (BLR) in the context of the unified model \citep{1985ApJ...297..621A,1993ARA&A..31..473A}, broad H$\alpha$ emission was not detected by spectropolarimetric measurements \citep{1997Natur.385..700H}, as one would expect in such cases. On the other hand, \citet{2007A&A...466..855P} and \citet{2015ApJ...815...55R}, interpreted their X-ray observations as a nucleus that is obscured by the Compton-thick material of the torus plus a reflection feature of the X-ray corona. A hidden BLR was proposed by \citet{2009ApJ...695..781B} and \citet{2015ApJ...815...55R} to explain the short-term spectral variation of NGC 7582.

The detection of a sudden broad emission in a few optical permitted lines that appeared in 1998 \citep{1999ApJ...519L.123A} was interpreted as a supernova (SN) II near the nucleus of this object. This is not surprising as star formation activity is clearly present \citep{1981A&A....97...71V,1985MNRAS.216..193M}. However, because these spectroscopic observations lacked high spatial resolution, the precise location of the 1998 flare remains unknown, and it is possible that it coincides with the AGN.

\citet{2001ApJS..136...61S} and \citet{2009MNRAS.393..783R} presented both ionized gas and molecular flux maps of the centre of NGC 7582 using near-infrared (NIR) IFU data. In particular, \citet{2009MNRAS.393..783R} proposed the existence of an inflow of gas through a nuclear bar and associated the kinematics of the ionized gas with a gaseous disc plus outflows (feedback) from the nucleus. Besides, their nuclear spectrum shows a broad component in the Br$\gamma$ line and a continuum emission from the hot dust that possibly comes from the inner walls of the torus (r $\leq$ 1 pc). \citet{2006MNRAS.369L..47W} detected, besides the AGN, other six compact structures in the central region of NGC 7582 using an MIR image of the [Ne II]$\lambda$12.8$\mu$m line. They interpreted these emissions as being produced by young massive stellar clusters deeply embedded in dust. 

A Seyfert-like jet radio structure is seen in 3 and 6 cm maps with a spatial resolution of $\sim$ 1.2 and 2.1 arcsec, respectively, as shown in \citet{1998MNRAS.300..757F}. However, based on joint radio, [Fe II] and MIR analysis, these authors suggested that the radio emission is caused by star forming regions in the central region of NGC 7582. \citet{1999A&AS..137..457M} analysed the 3.5 cm radio emission of NGC 7582 with a resolution of $\sim$ 1.0 arcsec. They argued that most of the galaxies (including NGC 7582) in their sample have radio structures that are similar to what is seen in Seyfert galaxies, although it would be hard to distinguish, with their spatial resolution, jets powered by AGNs from compact nuclear starbursts. This also seems to be the case also for the radio emission presented by \citet{1998MNRAS.300..757F}. 

Being one of the nearest galaxies ($d$ = 21 Mpc) in which a starburst galaxy has an obscured AGN, IFU observations both in the optical and in the NIR with high spatial resolution are important to better characterize the complex nature of such objects. A first goal is to clarify if this galaxy is indeed a Seyfert 1 by confirming a broad component in H$\alpha$ in its nucleus. High spatial resolution 2D spectra will also allow resolving the NLR in individual clouds and distinguishing them from H II regions. In addition, the NIR less obscured view may clarify whether there is a Seyfert jet or not. In order to achieve these goals we will analyse for the first time combined optical and NIR data cubes with unprecedented spatial resolution. The optical data were obtained with the IFU of the Gemini Multi-Object Spectrograph (GMOS) with an FOV of 3.5$\times$5.0 arcsec$^2$ (385 $\times$ 550 pc$^2$). The NIR archival data were obtained with the Spectrograph for Integral Field Observations in the Near Infrared (SINFONI) with two resolutions; an FOV of 8$\times$8 arcsec$^2$ (880 $\times$ 880 pc$^2$) and an FOV of 3$\times$3 arcsec$^2$ (330 $\times$ 330 pc$^2$). Our present analysis will focus on the very central region of NGC 7582, covering scales of a few hundred parsecs and a spatial resolution of ∼ 65 pc in the optical and ∼ 30 pc in the NIR.

The paper is structured as follows: Section \ref{sec:obs_data_reduction} shows details of the observations and data reduction; Section \ref{sec:EL_parameters} shows maps of parameters related to the emission lines that were detected in NGC 7582; Section \ref{sec:spectral_analysis} presents a spectral analysis of the ionized gas; Section \ref{sec:discussion} shows a discussion of the results and Section \ref{sec:conclusions} presents our main conclusions of this work.

\section{Observations and data reduction} \label{sec:obs_data_reduction}

A summary about all observations used in this work is shown in Table \ref{tab:summary_observations}. In the next sections, we present detailed information for each data set.

\begin{table*}
\centering
\caption{Summary of the observations. For the HST-WFPC2 instrument, the wavelength range is the effective range. The total bandwidth for the F606W filter is 1500 \AA\ for a central wavelength of 5860 \AA. The PSFs are in units of arcsec. For the SINFONI data, the PSFs have elliptical shapes with a PA = 0$^o$, obtained from measurements around 2.3 $\mu$m.\label{tab:summary_observations}}
\begin{tabular}{cccccc}
\hline
Instrument & Observation date & Filter or grating & Spectral resolution  & Wavelength range & PSF  \\
 &  &  & (FWHM) &  & (FWHM) \\
\hline
GMOS$-$IFU & 2004 July 17 & B600-G5323 &  1.8 \AA &  4240$-$6850 \AA & 0.61 \\
SINFONI 100 mas & 2007 July 30 & H+K & 125 km s$^{-1}$  &  1.45$-$2.45 $\mu$m & 0.24$\times$0.28  \\
SINFONI 250 mas & 2007 August 10 & H+K & 125 km s$^{-1}$ &  1.45$-$2.45 $\mu$m  & 0.29$\times$0.31\\
HST$-$WFPC2 & 2001 July 24 & F606W & $-$ &5100$-$6600 \AA$^*$& 0.09 \\
HST$-$NICMOS & 1997 September 16 & F160W & $-$ & 1.4$-$1.8 $\mu$m& 0.14 \\
\hline

\end{tabular}
\end{table*}

\subsection{GMOS reduction} \label{sec:gmos_reduction}

NGC 7582 was observed on 2004 July 17 with the Gemini-South telescope (programme GS-2004A-Q-35), using the GMOS-IFU instrument \citep{2002PASP..114..892A,2004PASP..116..425H}, in one-slit mode. With such a configuration, 750 micro lenses located at the focal plane of the telescope divide the image into slices of 0.2 arcsec, of which 500 are used for the observation of the object and 250 are destined to measure the sky emission. The micro lenses are coupled to a set of optical fibres that are arranged linearly at the nominal position of the spectrograph slit (pseudo-slit). Each spectrum (object and sky) is horizontally aligned in a mosaic of three CCDs. The final product is a data cube with an FOV of 3.5 $\times$ 5.0 arcsec$^2$ and one spectral dimension. Three exposures of 720 s were taken from the central part of NGC 7582. The B600$-$G5323 grating was positioned at the central wavelength of 4900 \AA, covering the spectral interval of 4230$-$7060 \AA\ with a resolution of 1.8\AA\ (full width at half-maximum, FWHM), as estimated from the CuAr lamp lines. The seeing of this observation was estimated to be 0.65 arcsec using the acquisition image of NGC 7582, obtained with the GMOS imager in the $r$ filter (SDSS system).

The basic reduction procedures (bias, flat-field, spectra extraction, wavelength calibration, dispersion correction and flux calibration) were performed using the standard Gemini packages under the {\sc iraf}\footnote{{\sc iraf} is distributed by the National Optical Astronomy Observatory, which is operated by the Association of Universities for Research in Astronomy (AURA) under cooperative agreement with the National Science Foundation.} environment. We removed cosmic rays using the LACOS algorithm \citep{2001PASP..113.1420V}. We built three data cubes, one for each exposure, that were corrected for the effects of the differential atmospheric refraction using an algorithm developed by us, based on the equations by \citet{1982PASP...94..715F} and \citet{1998Metro..35..133B}. After this, we averaged the three data cubes, since they were already flux calibrated.

In addition to the basic reduction procedures, complementary techniques were performed in order to improve the quality of the data cube of NGC 7582. First, we removed high-frequency spatial noise using a Butterworth filter \citep{2008gonzaleswoods}, with a cut-off frequency of 0.20 F$_{NY}$, where F$_{NY}$ is the Nyquist frequency, and a filter index $n$ = 2. A low-frequency instrumental fingerprint was identified both in the spectral and spatial dimensions and was removed using Principal Component Analysis (PCA) Tomography \citep{2009MNRAS.395...64S}. At this stage, telluric lines were removed and then all spectra of the data cube were corrected for reddening effects caused by the Milky Way, assuming $A_V = 0.038$ \citep{2011ApJ...737..103S} and using the extinction curves of \citet{1989ApJ...345..245C}. Finally, we deconvolved each image of the data cube using the Richardson-Lucy technique \citep{1972JOSA...62...55R,1974AJ.....79..745L}, assuming a Gaussian point spread function (PSF) with FWHM = 0.65 arcsec and 10 iterations. The final PSF of the data cube was estimated to be 0.61 arcsec. An average image of the stellar continuum of this data cube is shown in Fig. \ref{HST_image}. The spectrum that results from the sum of all spectra of the data cube is presented in Fig. \ref{fig:espectros_cubos}. More details on all these complementary treatments performed on the data cube are given in \citet{2014MNRAS.440.2419R} and \citet{2014MNRAS.438.2597M,2015MNRAS.450..369M}.

\subsection{\textit{HST} data} \label{sec:hst}

\begin{figure*}
\hspace{-1.5cm}
\includegraphics[scale=0.4]{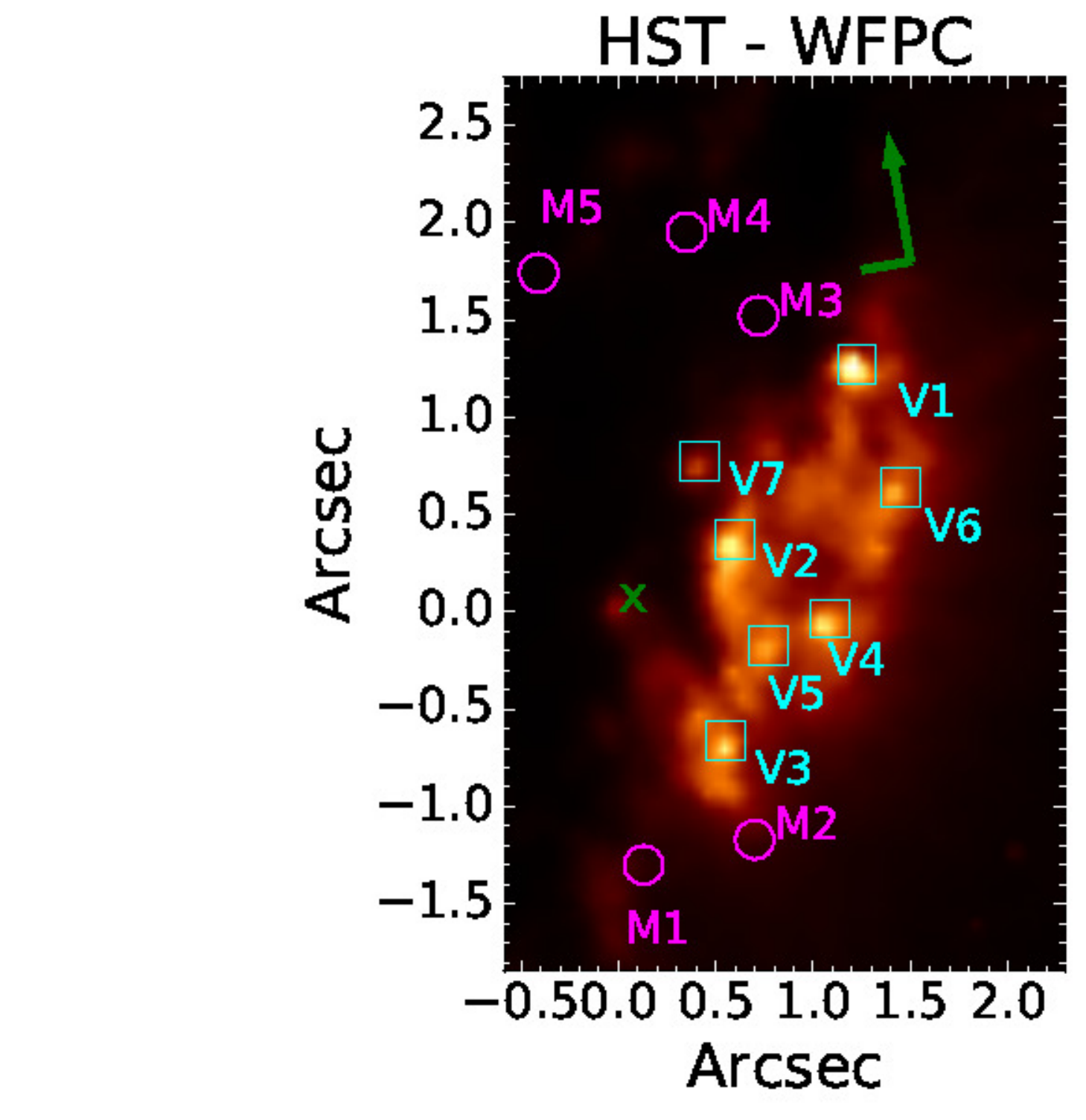}
\hspace{-0.3cm}
\includegraphics[scale=0.4]{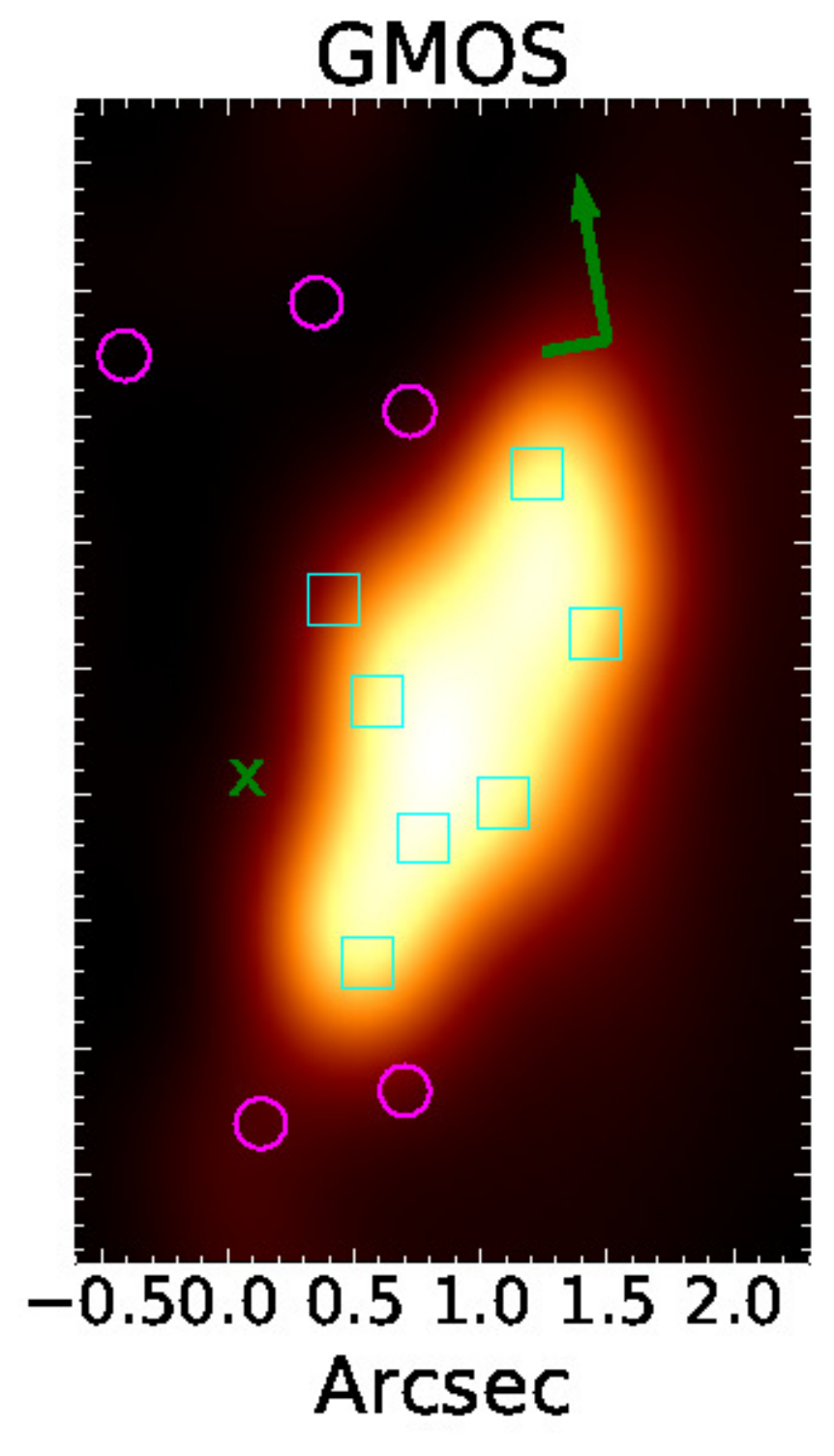}
\hspace{-0.3cm}
\includegraphics[scale=0.4]{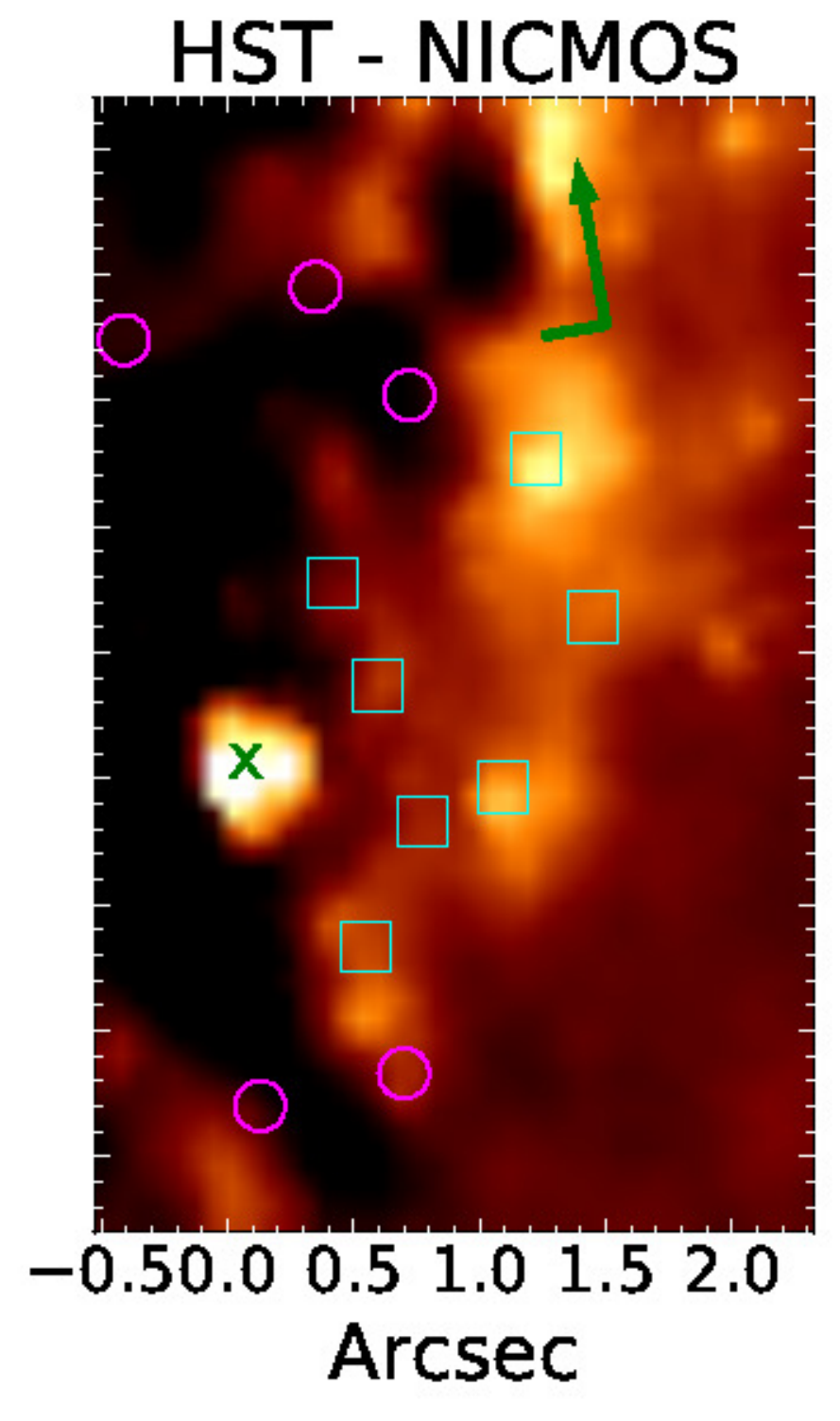}
\hspace{-0.3cm}
\includegraphics[scale=0.29]{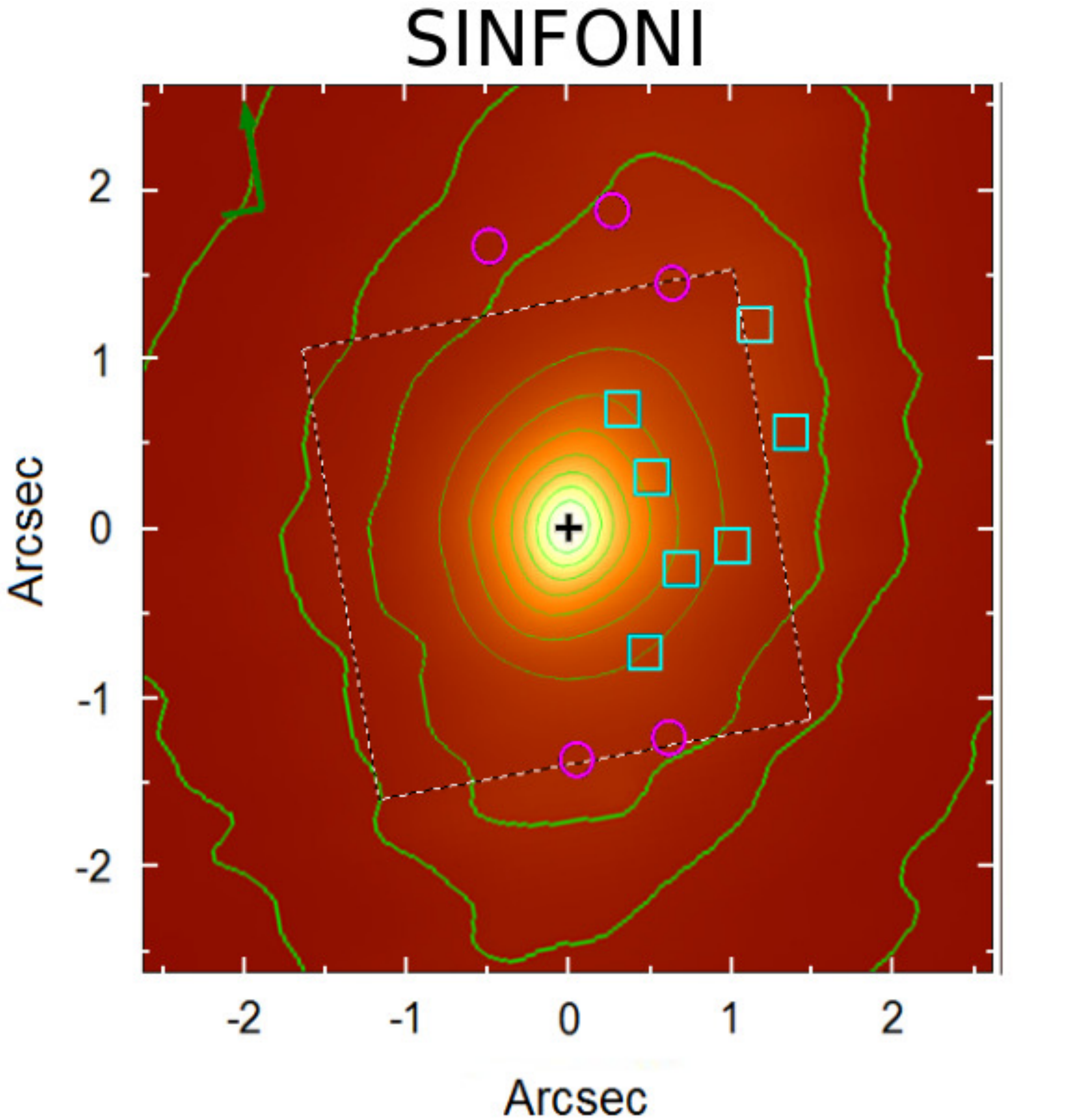}
\caption{Left: WFPC2 image of NGC 7582 in the GMOS FOV. This observation was made using the planetary camera in the wide \textit{V} band. Centre left: an average image of the GMOS data cube taken from 5160 to 6400 \AA. Centre right: Structure map built using a NICMOS image of NGC 7582 in the GMOS FOV. The F160W filter was used in this observation. Right: an average image of the SINFONI data. The green contours are isophotes of the image. All images correspond to the stellar continuum of NGC 7582. The green $\times$ indicates the position of the AGN, as identified by \citet{2007MNRAS.374..697B} in the WFPC2 image. We call a few compact structures that are seen in the WFPC2 image as V1, V2, V3, V4, V5, V6 and V7. We marked their positions with cyan squares. The magenta circles are related to the compact MIR sources detected in the image of the [Ne II]12.8$\mu$m by \citet{2006MNRAS.369L..47W}. The green arrow is the north direction, with east to the left.  \label{HST_image} }
\end{figure*}

\begin{figure*}
\hspace{-1.5cm}
\includegraphics[scale=0.53]{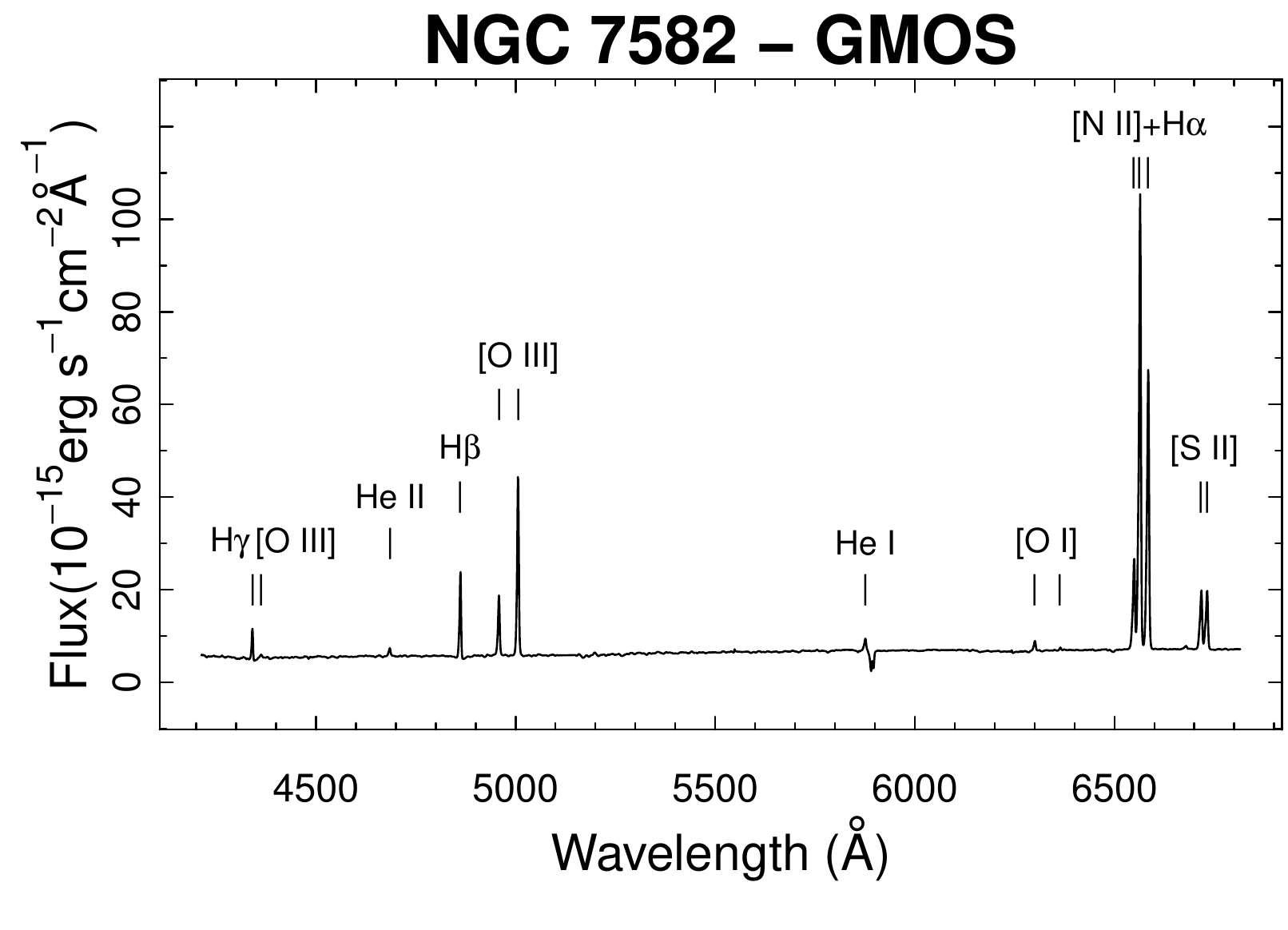}
\hspace{-0.1cm}
\includegraphics[scale=0.53]{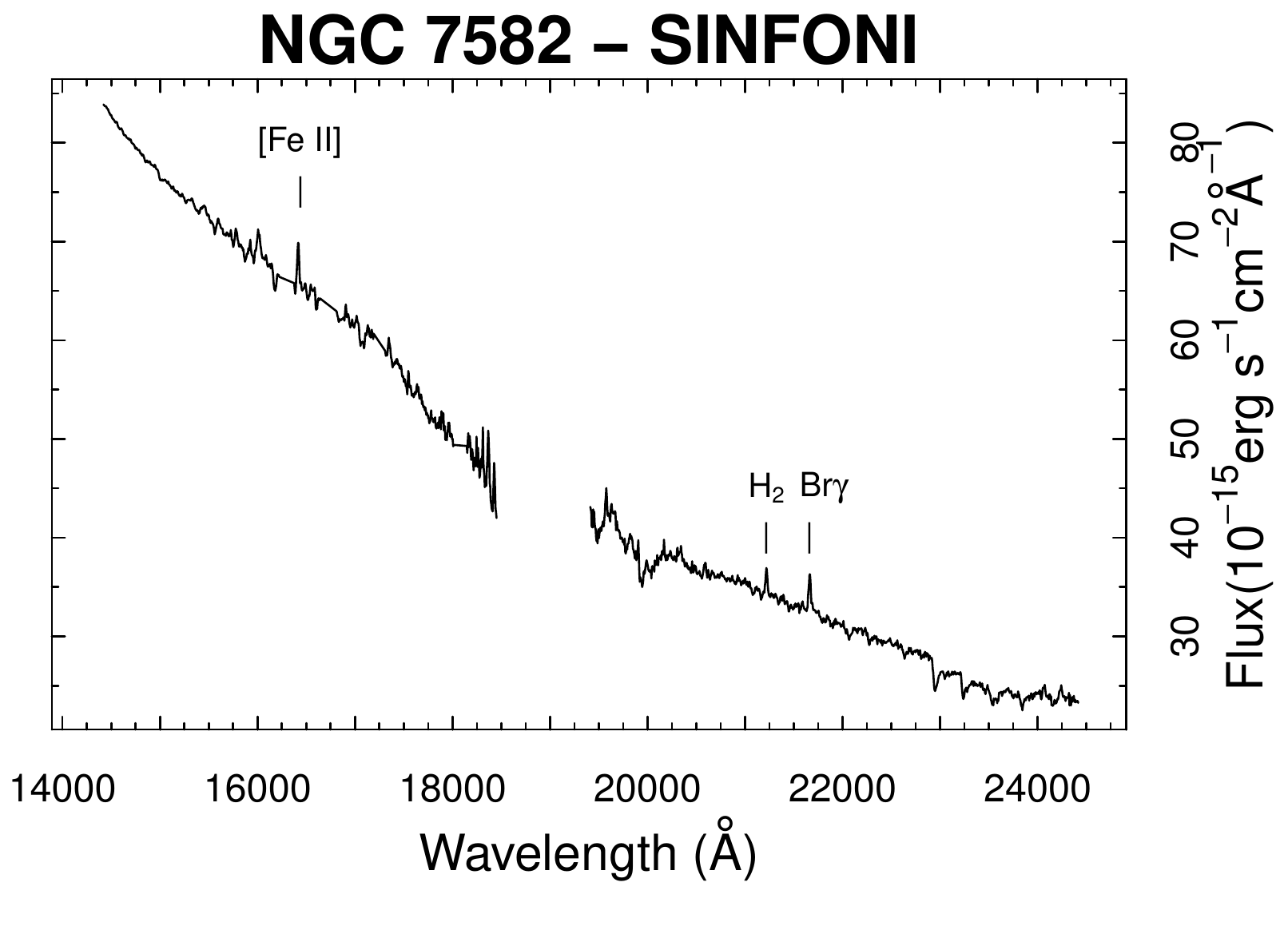}

\caption{Spectra of NGC 7582 in the optical region (left) and in the NIR region (right).  \label{fig:espectros_cubos} }
\end{figure*}

NGC 7582 was observed with the \textit{HST} using the Wide Field Planetary Camera 2 (WFPC2, programme SNAP 8597, PI: Michael W. Regan) and the Near Infrared Camera and Multi Object Spectrograph (NICMOS, programme SNAP 7330, PI: John Mulchaey). The F606W filter was used for the WFPC2 observation, which has a central wavelength of 5860 \AA, a bandwidth of 1500 \AA\ (wide V) and an effective wavelength of 5997 \AA. This spectral band is related to the stellar continuum of NGC 7582. The NICMOS observation was made with the F160W filter, which is centred in 1.55 $\mu$m, with a range of 1.4$-$1.8 $\mu$m and an effective wavelength of 1.77 $\mu$m. We present the WFPC2 image from the PC chip, together with an average image of the GMOS data cube taken from 5160 to 6400 \AA, in Fig. \ref{HST_image}. For the NICMOS observation, we show in Fig. \ref{HST_image} the structure map, which was made using the method suggested by \citet{2002ApJ...569..624P}, assuming a PSF related to the position of the nucleus of NGC 7582 calculated with the Tiny Tim code \citep{2011hst..psf}. All images are presented with the same FOV. The position of the AGN, as set by \citet{2007MNRAS.374..697B}, is indicated in the WFPC2 image. We also indicate the positions of the structures detected with an MIR image of the [Ne II]12.8$\mu$m line by \citet{2006MNRAS.369L..47W}. We used the same nomenclature as these authors for the MIR structures: M1, M2, M3, M4 and M5 (their M6 is not covered by our FOV).

We identified seven compact structures that are clearly seen in the WFPC2 image. They were named as V1, V2, V3, V4, V5, V6 and V7. According to \citet{2006MNRAS.369L..47W}, the AGN in the NICMOS observation is the brightest compact structure, therefore we matched both the WFPC 2 and NICMOS images using the position of the nucleus. We will show in the next sections that some of these structures that are seen in the \textit{HST}  are related to the emission of ionized gas detected with the GMOS data.

\subsection{SINFONI data} \label{sec:sinfoni_reduction}

NGC 7582 was observed in the NIR with the Very Large Telescope (VLT) using the SINFONI spectrograph \citep{2003SPIE.4841.1548E,2004Msngr.117...17B}. In this instrument, the FOV on the sky is sliced into 32 slitlets that are recorded into 64 detector pixels. This results in a grid of 32 $\times$ 64 rectangular spaxels. We retrieved two data sets from the European Southern Observatory (ESO) science archive facility. In both cases, the $H+K$ band was used, which produced a spectral range of 1.45$-$2.45 $\mu$m, with a resolution $R \sim$ 2400, corresponding to FWHM$\sim$125 km $s^{-1}$. One data set refers to an observation made in 2007 August 10 with a fore optics that produced an FOV of 8 $\times$ 8 arcsec$^2$ and spaxels of 0.125 $\times$ 0.250 arcsec$^2$ [programme 079.C-0328(B), PI: Krabbe]. Given the bad weather conditions, only one exposure of 60 s was usable with this setup with a seeing of 1.9 arcsec, although the observation has been assisted by the adaptive optics system of the VLT. From now on, we will refer to this data set as SINFONI 250. The observations related to the second data set were made in 2007 July 30 with a fore-optics that resulted in a FOV of 3$\times$3 arcsec$^2$ and spaxels of 0.05 $\times$ 0.100 arcsec$^2$ [programme 079.C-0328(A), PI: Krabbe]. Four exposures of 200 s also assisted by the adaptive optics system, were used with this setup in this work, with an average seeing of 2.1 arcsec. Hereinafter, we will refer to this data set as SINFONI 100.

Both data sets were reduced using the {\sc esoreflex} software, which includes standard procedures such as bad pixel removal, correction for the linearity of the detector, flat-field correction, spatial rectification, wavelength calibration, sky subtraction and data cube reconstruction. The spectrophotometric standard A0V star HD\,216009 was used for flux calibration. At this point, we have one data cube with the SINFONI 250 setup and four data cubes with the SINFONI 100 setup. For this last case, the observations were made with a maximum dithering of 4 pixels (0.2 arcsec) from each other, both in the $x$ and $y$ directions. This strategy allows one to obtain the median of the four data cubes with the purpose of avoiding both CCD defects and cosmic rays without losing a big area of the FOV. The final dimensions of the FOV here were $2.6 \times 2.6$ arcsec$^2$. For the data cube of the SINFONI 250 setup, all cosmic rays were successfully removed through individual selections, given that the time exposure in this case was short. Moreover, we verified that additional CCD defects did not compromise the regions of interest of NGC 7582. The final dimensions for the FOV in this case are $7 \times 7$ arcsec$^2$. 

Also in case of the SINFONI observations, additional techniques were used in order to improve the quality of the data cubes \citep{2014MNRAS.438.2597M,2015MNRAS.450..369M}. The first step was the spatial resampling of the data, followed by a quadratic interpolation. This procedure preserves the flux density of the images and improves the visualization of the contours of the structures. The new spatial sampling is 0.025 $\times$0.025 arcsec$^2$ for the SINFONI 100 data set and 0.0625$\times$0.0625 arcsec$^2$ for the SINFONI 250 data set. After this, we removed the high-frequency noise from the spatial dimension using a Butterworth filter. For the SINFONI 100 data set, we used a cut-off frequency of 0.32 F$_{NY}$ for the $x$-axis and of 0.28 F$_{NY}$ for the $y$-axis and a filter index $n$ = 5; for the SINFONI 250 data set, a cut-off frequency of 0.40 F$_{NY}$ for the $x$-axis and of 0.45 F$_{NY}$ for the $y$-axis was used with a filter index $n$ = 5. A low-frequency instrumental fingerprint in both spatial and spectral dimensions was also identified in both data sets and removed using PCA Tomography. 

Finally, we deconvolved both data cubes using the Richardson$-$Lucy technique. However, the situation here is not as simple as it is for the GMOS data cubes, since the PSF of the SINFONI data cubes is more difficult to model. PSFs extracted from the science data cubes themselves are recommended for SINFONI observations \citep{2015MNRAS.450..369M}. In NGC 7582, we have two options: the image of the broad component of the Br$\gamma$ line or the continuum emission from the hot dust of the nuclear torus. However, the image of the broad component of the Br$\gamma$ line is very noisy. Thus, we were left to use the emission from the hot dust. To obtain a reliable image of the hot dust and, consequently, a reliable PSF around 2.3 $\mu$m, we used the PCA Tomography technique. We refer the reader to \citet{2014MNRAS.438.2597M,2015MNRAS.450..369M} for more details on this issue. The PSFs of the SINFONI data cubes were elliptical before the deconvolution, with FWHM($x$) = 0.26 arcsec and FWHM($y$) = 0.32 arcsec for the SINFONI 100 data set and FWHM($x$) = 0.34 arcsec and with FWHM($y$) = 0.40 arcsec for the SINFONI 250 data set. The deconvolution procedure improved the Strehl ratio from 0.041 to 0.048 for the SINFONI 100 data set and from 0.014 to 0.017 for the SINFONI 250 data set. The new PSFs remained elliptical, with FWHM($x$) = 0.24 arcsec and FWHM($y$) = 0.28 arcsec for the SINFONI 100 data set and with FWHM($x$) = 0.29 arcsec and FWHM($y$) = 0.31 arcsec for the SINFONI 250 data set. An image of the stellar continuum combining both SINFONI 100 and SINFONI 250 is shown in Fig. \ref{HST_image}. The spectrum that results from the sum of all spectra of the data cube from the SINFONI 100 data set is presented in Fig. \ref{fig:espectros_cubos}.  

\section{Maps of the emission-line parameters} \label{sec:EL_parameters}

\subsection{Spectral synthesis} \label{sec:spectral_synthesis}

In order to study the optical emission-line properties of the central region of NGC 7582, we subtracted the stellar component from each spaxel of the GMOS data cube by means of the spectral synthesis using the software {\sc starlight} \citep{2005MNRAS.358..363C}. This code fits the stellar spectrum of a galaxy with a combination of different simple stellar populations (SSPs) taken from an established base. We used an updated version of the \citet{2003MNRAS.344.1000B} base (CB2007), which used stellar spectra from MILES and GRANADA libraries \citep{2006MNRAS.371..703S,2005MNRAS.358...49M}. A total of 150 SSPs with a spectral resolution of 2.51 \AA\ \citep{2011A&A...531A.109B,2011A&A...532A..95F} were used, with metallicities $Z$ = 0.0001, 0.0004, 0.004, 0.008, 0.02 and 0.05, and ages between 1 Myr and 14 Gyr. We used the \citet{1989ApJ...345..245C} curve to fit the extinction. All emission lines were masked before the fitting procedures, as well as the Na I$\lambda\lambda$5889, 5895 region and also the spectral regions that were affected by the gaps between the three CCDs of the GMOS instrument. Subtracting the model spectrum from the observed one leaves only the gaseous emission component of NGC 7582. From now on, we will call this particular data cube as the gas cube.

\begin{figure*}

\includegraphics[scale=0.4]{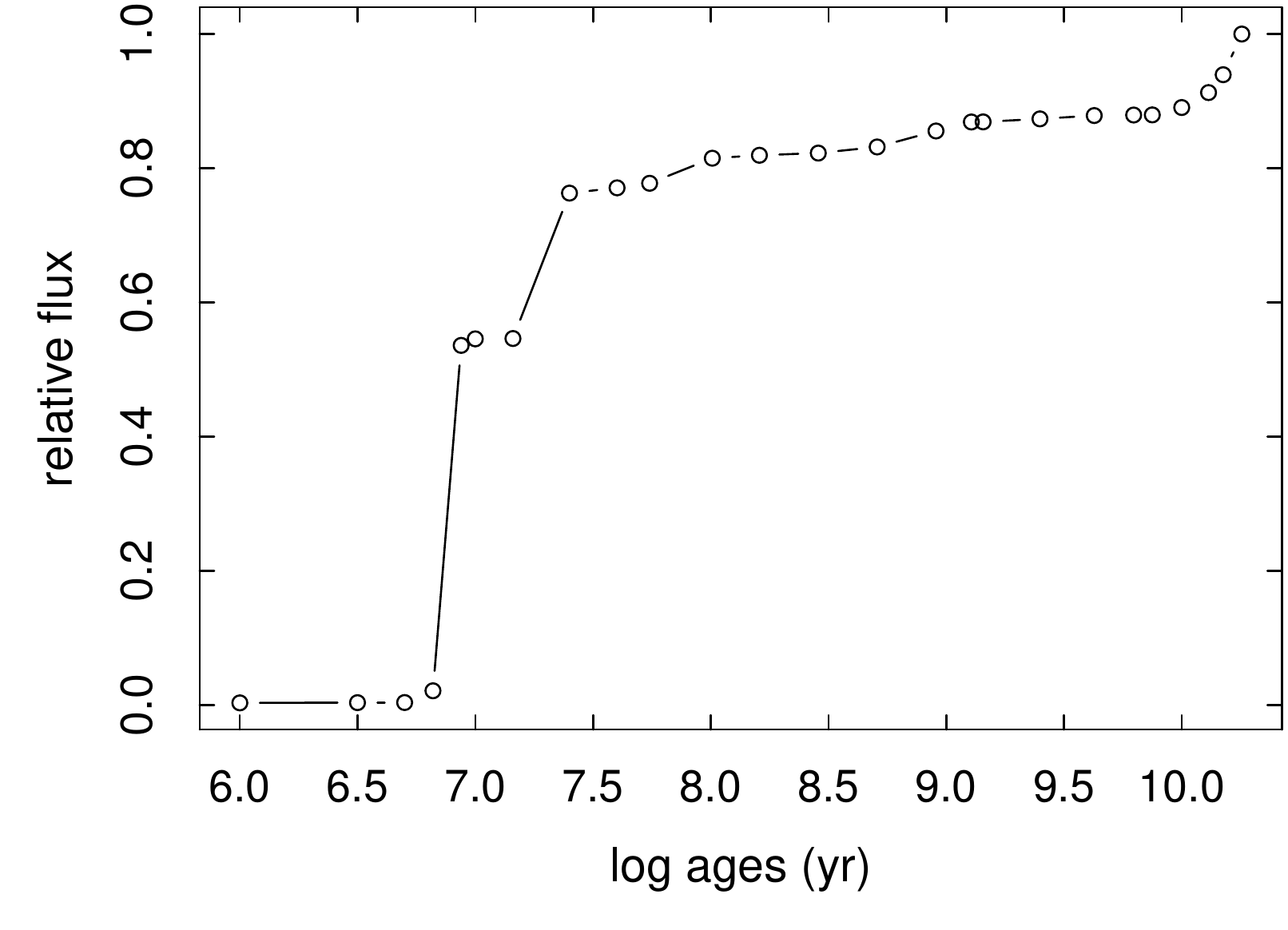}
\includegraphics[scale=0.4]{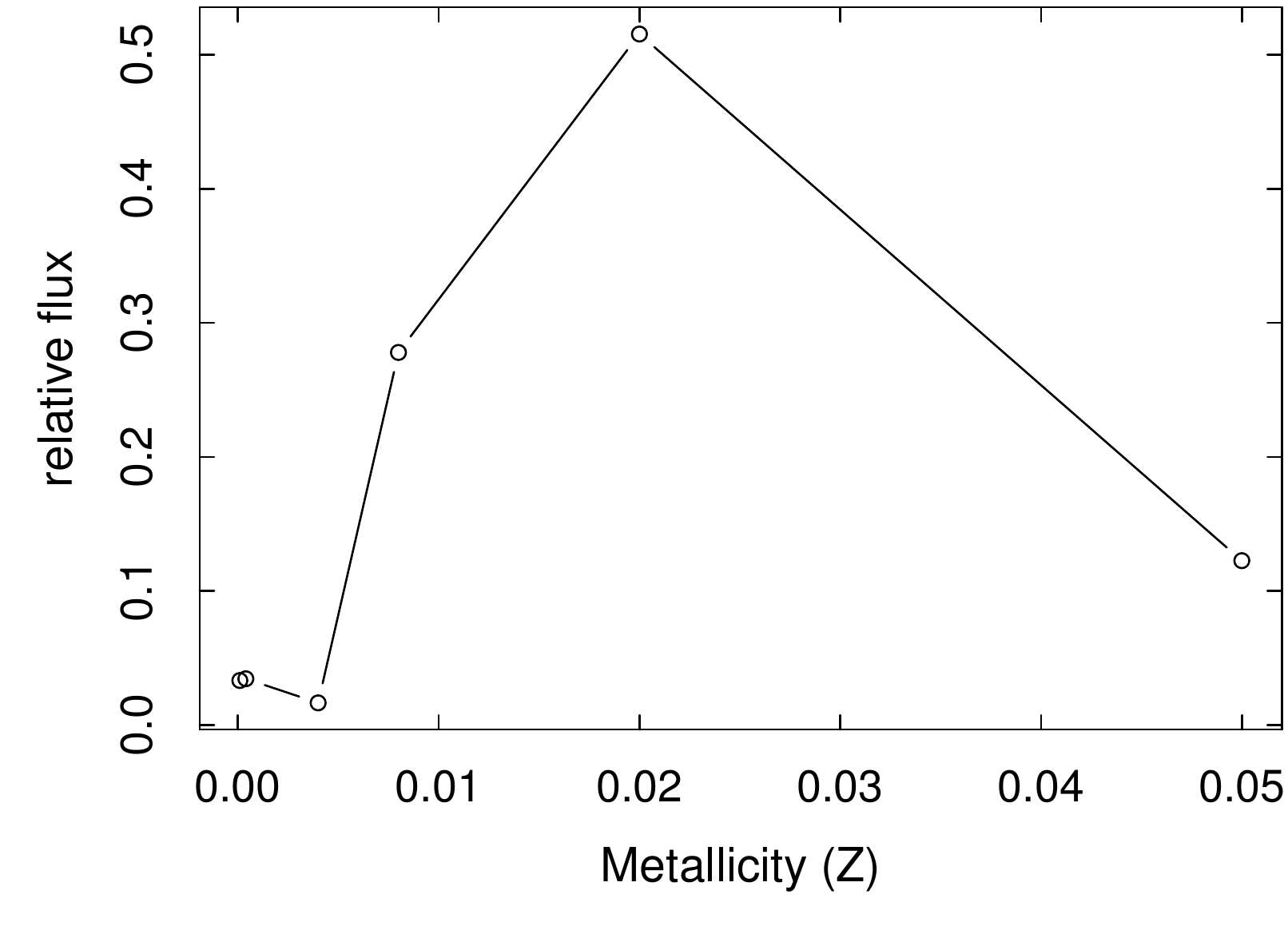}

\caption{Left: distribution of the stellar flux contribution at $\lambda$6420 \AA\ as a function of age. This graph shows the star formation history of the central part of NGC 7582. Much of the stellar light comes from a young stellar population. This is expected since a great fraction of the ionized gas emission is produced by starbursts. Right: flux contribution of the stellar populations of all ages and as a function of metallicity. \label{fig:stellar_pop_results}}
\end{figure*}

We briefly discuss the results that were obtained with this spectral synthesis. In particular, we show in Fig. \ref{fig:stellar_pop_results} the stellar flux contribution at $\lambda$6420 \AA\ of the whole FOV as a function of the population ages and metallicities. Half of the stellar flux comes from populations with ages $<$ 1.0$\times$10$^7$ yr. This is expected since a great fraction of the ionized gas emission is produced by young stellar populations. It is worth mentioning that \citet{1999MNRAS.303..173S} found that stellar populations with 1.0$\times$10$^7$ yr contribute with $\sim$ 42 \% of the stellar flux at $\lambda$5870 \AA. Considering the second half of the stellar flux, $\sim$ 30 \% is related to populations between 1.0$\times$10$^7$ and 2.5$\times$10$^7$ yr and 20 \% is associated with populations between 2.5$\times$10$^7$ and 1.4$\times$10$^{10}$ yr. When it comes to metallicity, $\sim$ 80 \% has $Z$ = 0.008 and 0.02, while the other 20 \% has $Z$ = 0.05 and $Z$ $<$ 0.004. It is beyond the scope of this paper to verify how much of these results are affected by the age$-$metallicity extinction degeneracy and other uncertainties that may arise from spectral synthesis (see e.g. \citealt{2014A&A...561A.130C}), thus Fig. \ref{fig:stellar_pop_results} should be analysed with caution.

\begin{figure*}

\includegraphics[scale=0.4]{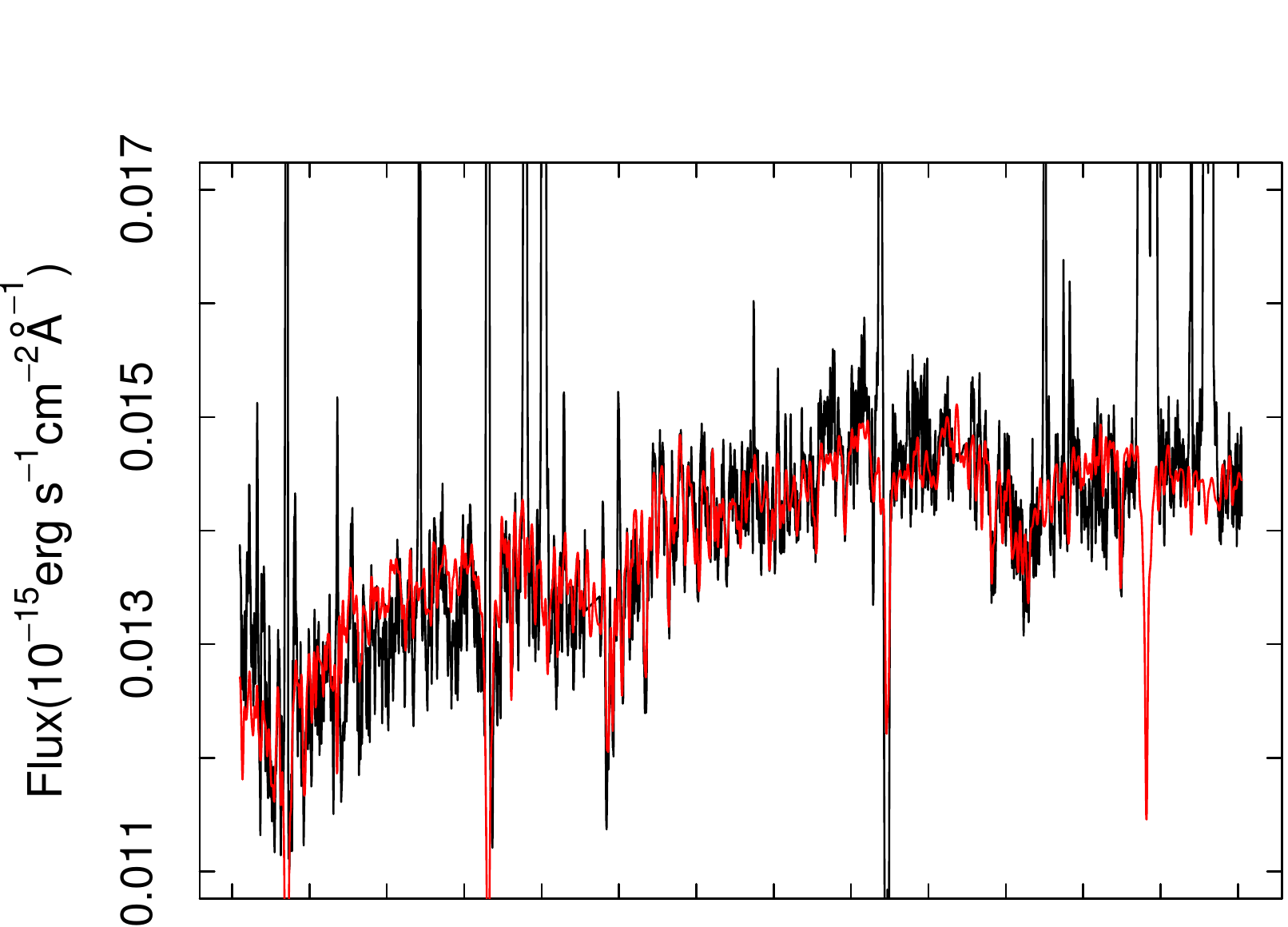}
\includegraphics[scale=0.4]{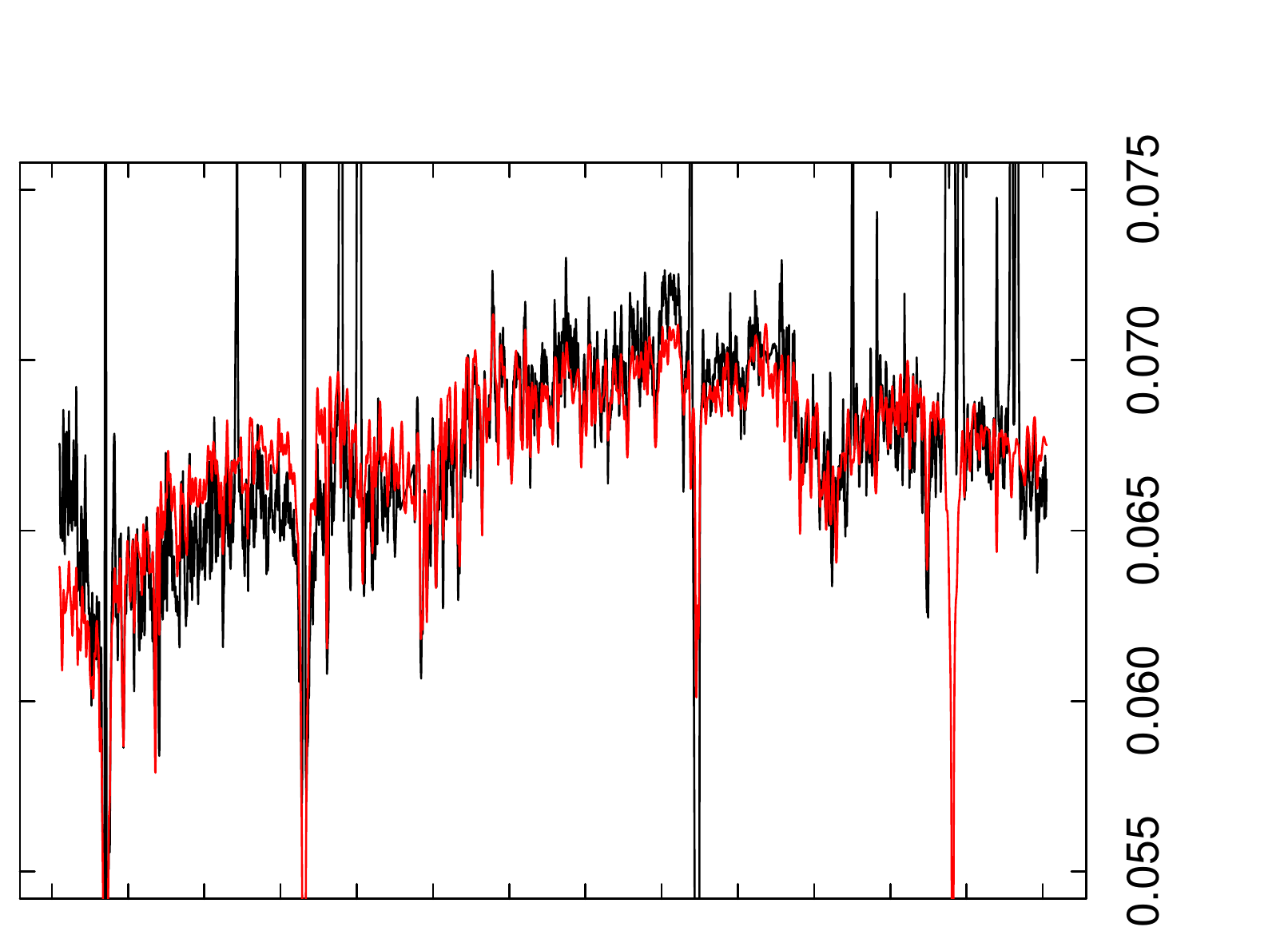}

\includegraphics[scale=0.4]{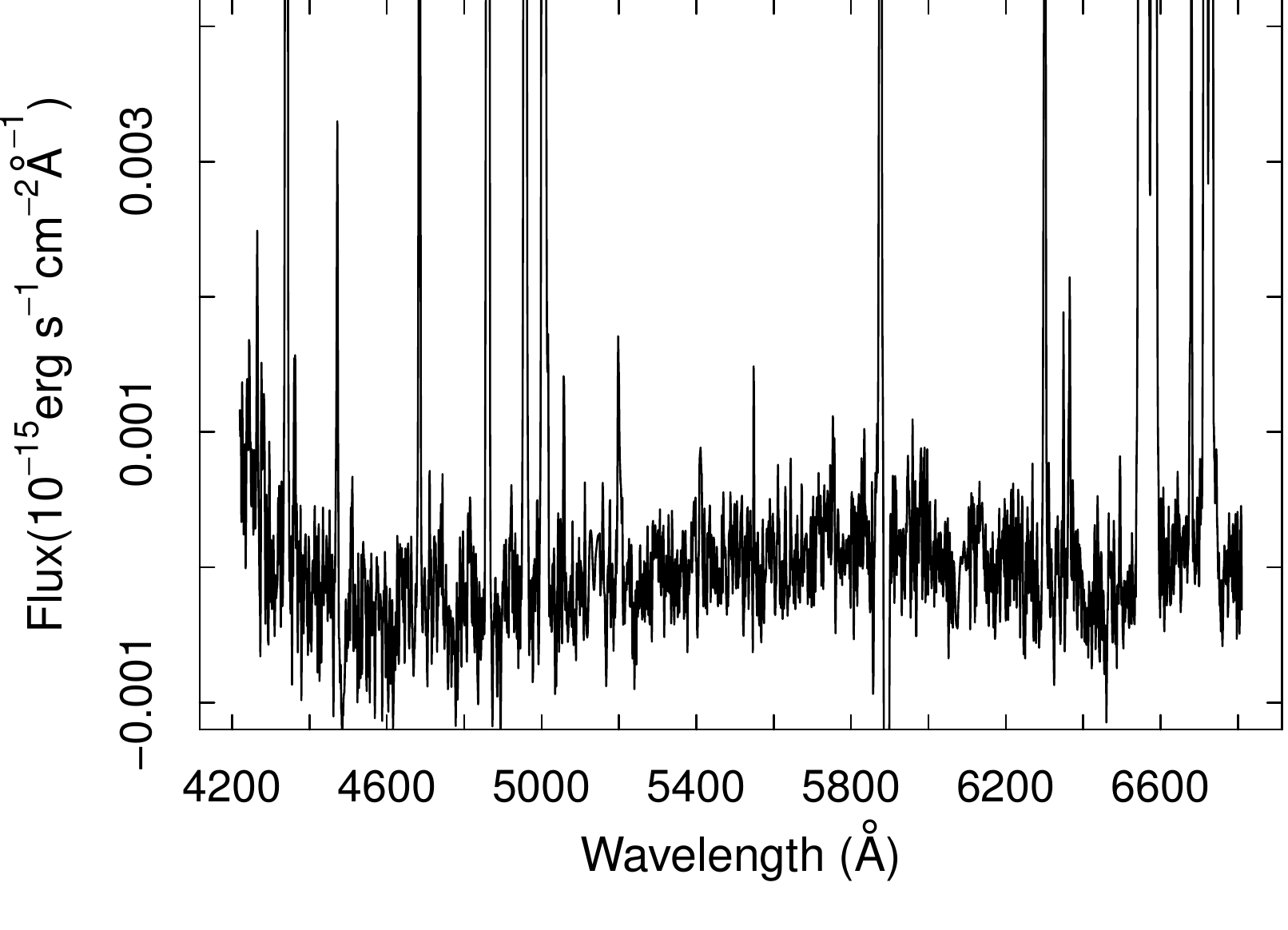}
\includegraphics[scale=0.4]{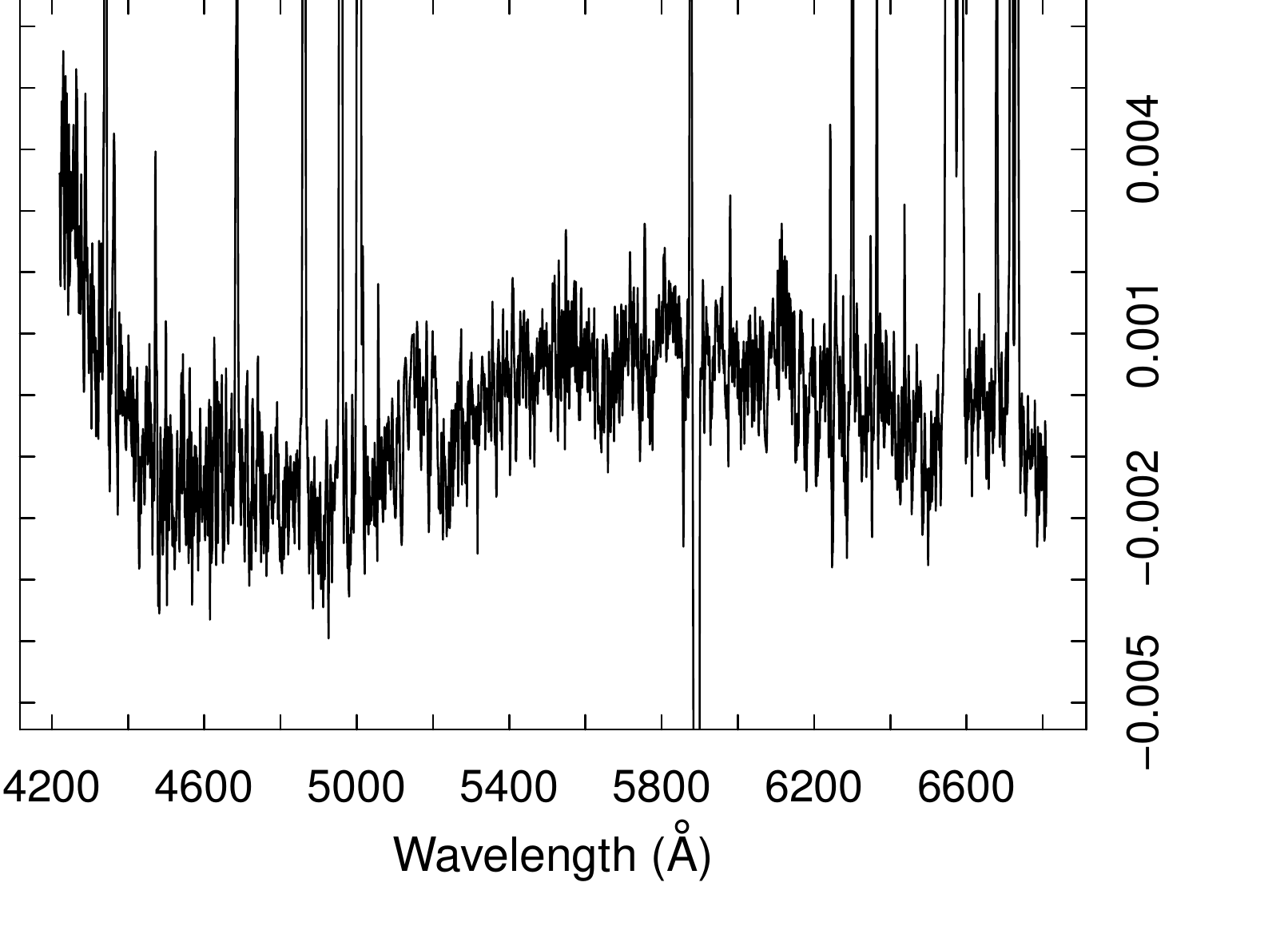}

\caption{Top: spectra (black) and spectral synthesis results (red) of two representative spectra of NGC 7582. The spectrum in the left was taken from the centre of the FOV and has a higher signal-to-noise ratio than the spectrum in the right, taken from the upper right region of the FOV. Bottom: residual spectra (observed$-$synthetic spectrum). The highest discrepancy is $\sim$ 6 \% in the 4200 \AA\ region. For the rest of the spectra, the difference is between 1 and 2 \%. \label{fig:stellar_pop_spectra}}
\end{figure*}

We show the fitting results for two representative spectra of the data cube of NGC 7582 in Fig. \ref{fig:stellar_pop_spectra}. One spectrum was taken from the centre of the FOV and represents information with higher signal-to-noise ratio, and the other spectrum has lower signal-to-noise ratio and is related to a region at the upper right position of the FOV. One may see, in blue, the regions that were given zero weight to the fitting procedure (emission lines and CCD gaps). The residuals are also shown in Fig. \ref{fig:stellar_pop_spectra}. The highest discrepancy is $\sim$ 6 \% in the 4200 \AA\ region. Along the other spectral regions, the difference is between 1 and 2 \% for both low and high signal-to-noise ratio spectra.

For the NIR data cubes SINFONI 100 and SINFONI 250, we subtracted their continuum emission in two different ways: (1) in the $K$ band we performed the penalized pixel fitting (\textsc{ppxf}), implemented by \citet{2004PASP..116..138C}, using the Near-Infrared Integral-Field Spectrograph stellar template V2.0 described in \citet{2009ApJS..185..186W}, ranging from 2.02 to 2.43 $\mu$m; and (2) for the remainder blue part of the $K$ band spectrum and the $H$ band each spaxel had a spline function fitted to its spectrum that was latter subtracted from the original one, producing an NIR gas cube.

\subsection{Line fitting of the optical data} \label{sec:line_fittings}

We fitted Gaussian functions to the H$\alpha$, [N II]$\lambda\lambda$6548, 6583, [S II]$\lambda\lambda$6713, 6730, [O I]$\lambda\lambda$6300, 6363, H$\beta$, [O III]$\lambda\lambda$4959, 5007, He II$\lambda$4686 and [N I]$\lambda\lambda$5198, 5200 lines of all spectra of the optical gas cube. We used the Levenberg$-$Marquardt algorithm to perform the fitting procedures. The kinematic parameters of the gas [peak position and dispersion ($\sigma$) of the Gaussian functions] were set as free parameters for the H$\alpha$, [N II]$\lambda\lambda$6548, 6583, and [O III]$\lambda\lambda$4959, 5007 lines. The profile fits of the [S II]$\lambda\lambda$6713, 6730, [O I]$\lambda\lambda$6300, 6363, He I$\lambda$5875 and [N I]$\lambda\lambda$5198, 5200 lines were bounded to the results found for the [N II]$\lambda\lambda$6548, 6583 doublet, while the He II$\lambda$4686 line was assumed to have the same profile as the [O III]$\lambda\lambda$4959, 5007 doublet. The profile fit of H$\beta$ was linked to the results found for H$\alpha$. We also assumed that the theoretical ratios [N II]$\lambda$6583/[N II]$\lambda$6548 = 3.06, [O I]$\lambda$6300/[O I]$\lambda$6363 = 3.05 and [O III]$\lambda$5007/[O I]$\lambda$4959 = 2.92 \citep{2006agna.book.....O} are fixed for the fitting procedures. Thus, only one amplitude is a free parameter for each of these doublets. With the final results, we built maps of the gas kinematic parameters and of the integrated flux of the optical emission lines. Relevant emission line ratios and the electron density were also mapped. These results correspond only to the narrow components of the emission lines. Since we detected a broad-line component in H$\alpha$ in the nuclear spectrum (see Section \ref{sec:nuclear_spectrum}), we subtracted this feature before performing the fitting procedure all over the gas cube. 

\subsection{Line flux maps} \label{sec:EL_flux}

\subsubsection{Optical data} \label{sec:el_flux_optical}

\begin{figure*}
\hspace{-1.0cm}
\includegraphics[scale=0.37]{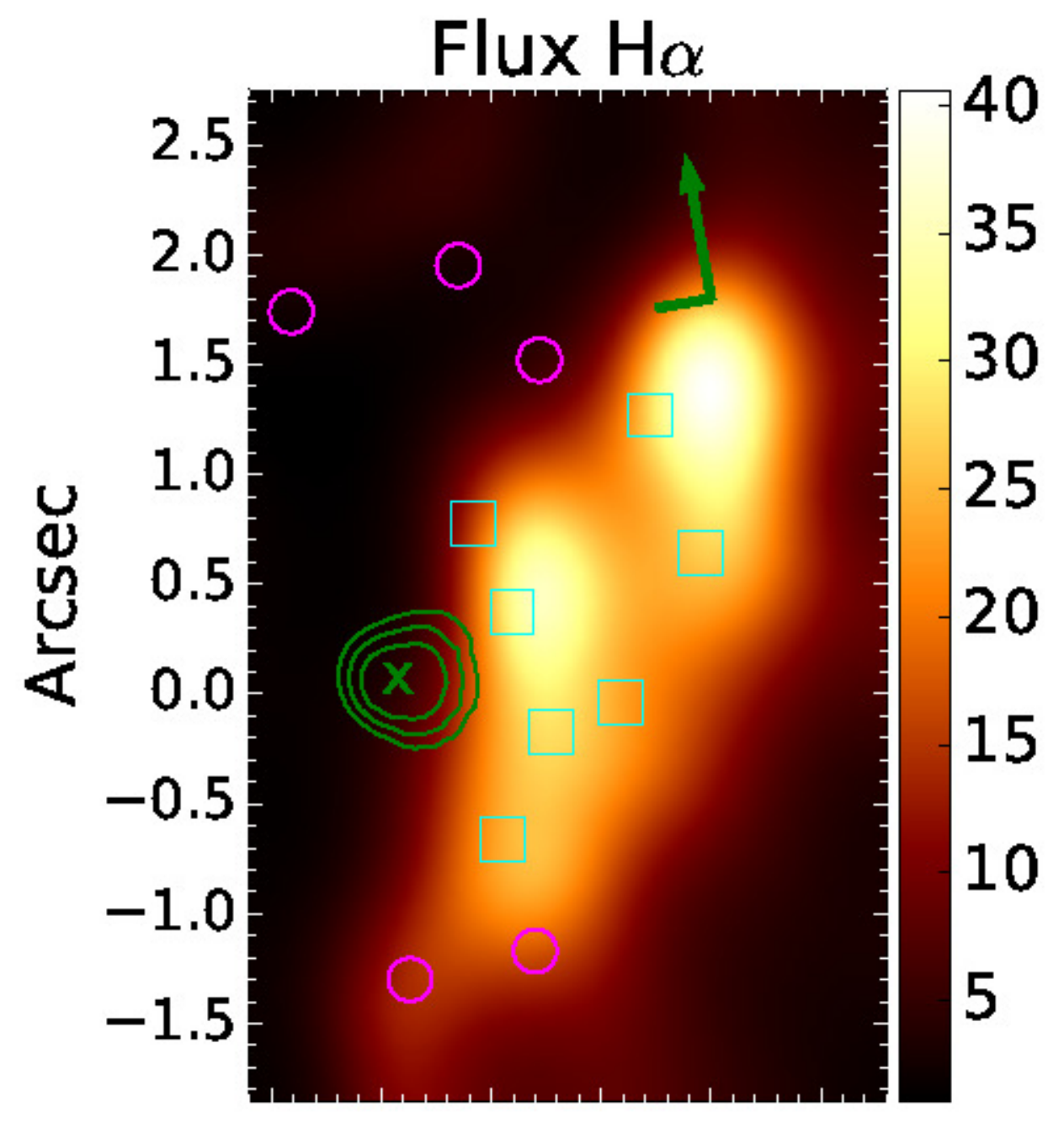}
\hspace{0.2cm}
\includegraphics[scale=0.37]{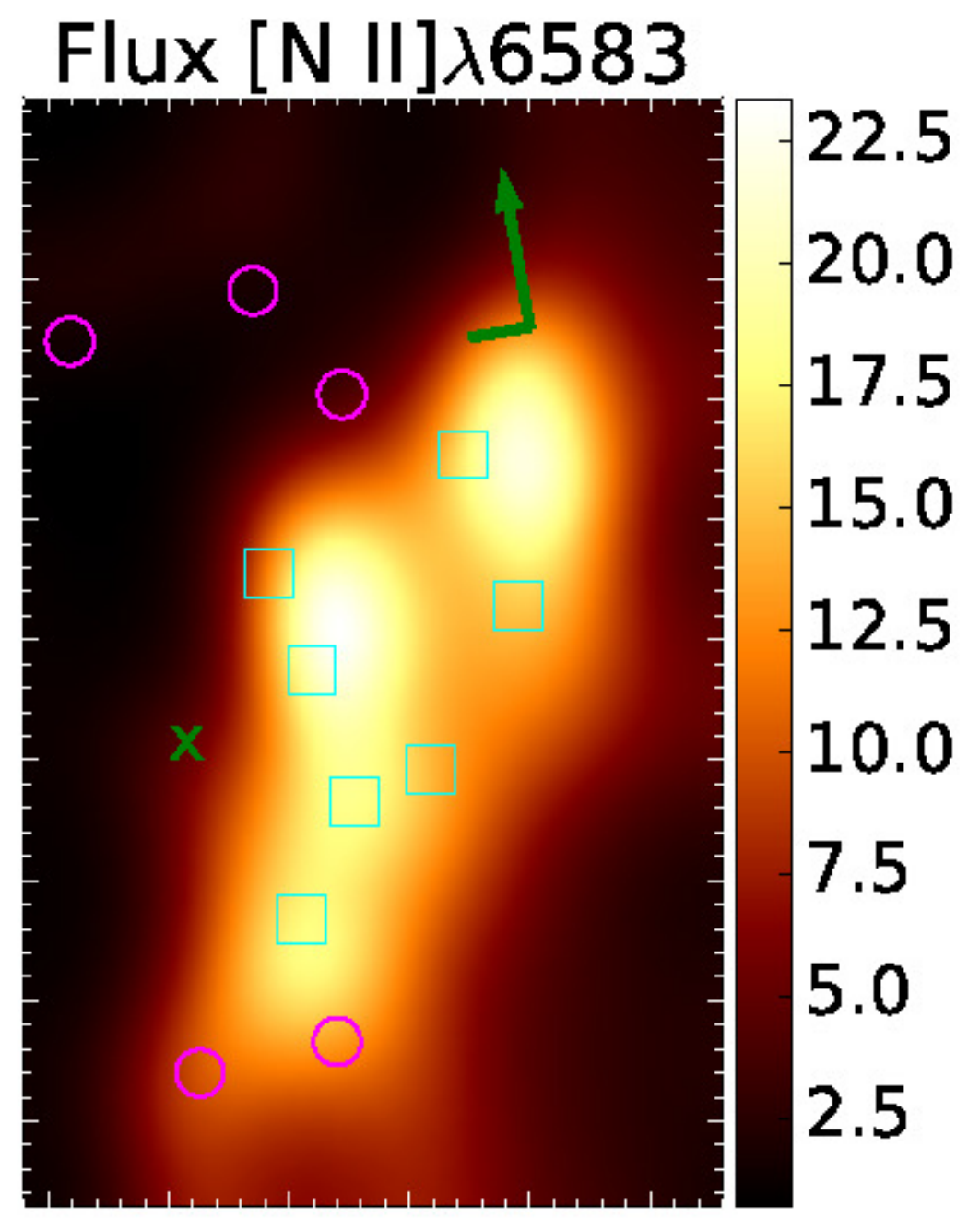}
\hspace{0.1cm}
\includegraphics[scale=0.37]{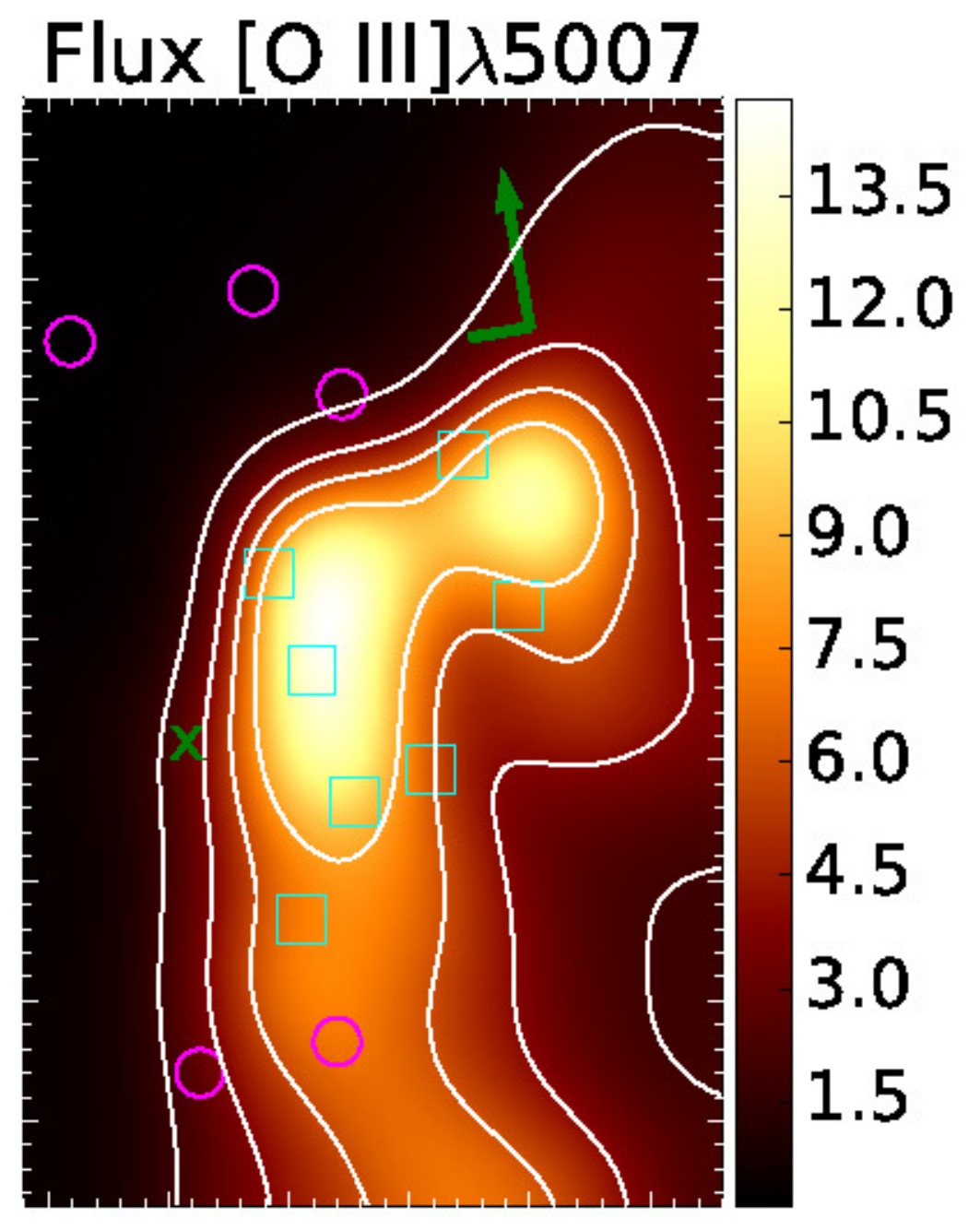}
\hspace{0.1cm}
\includegraphics[scale=0.37]{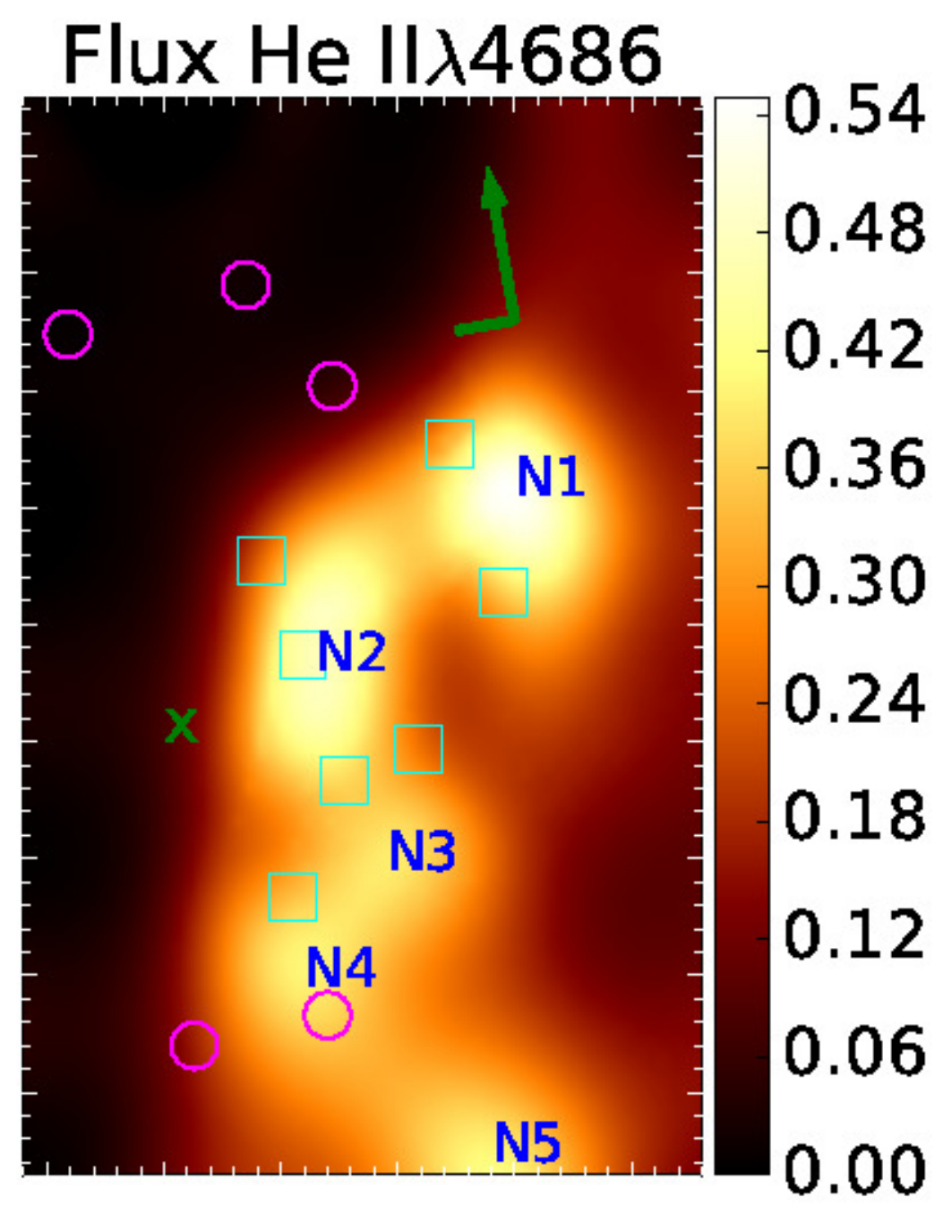}

\hspace{-0.8cm}
\includegraphics[scale=0.37]{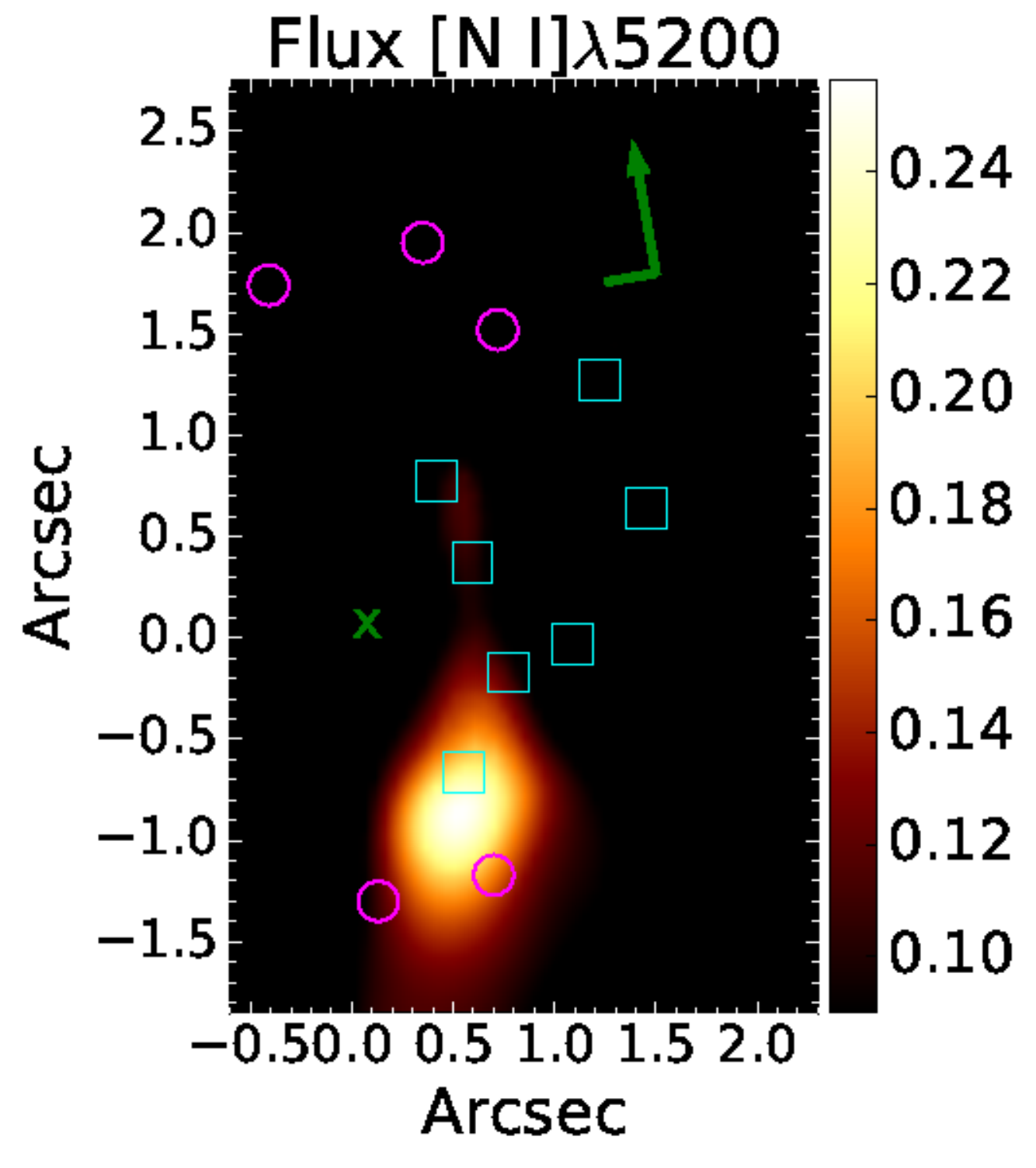}
\includegraphics[scale=0.37]{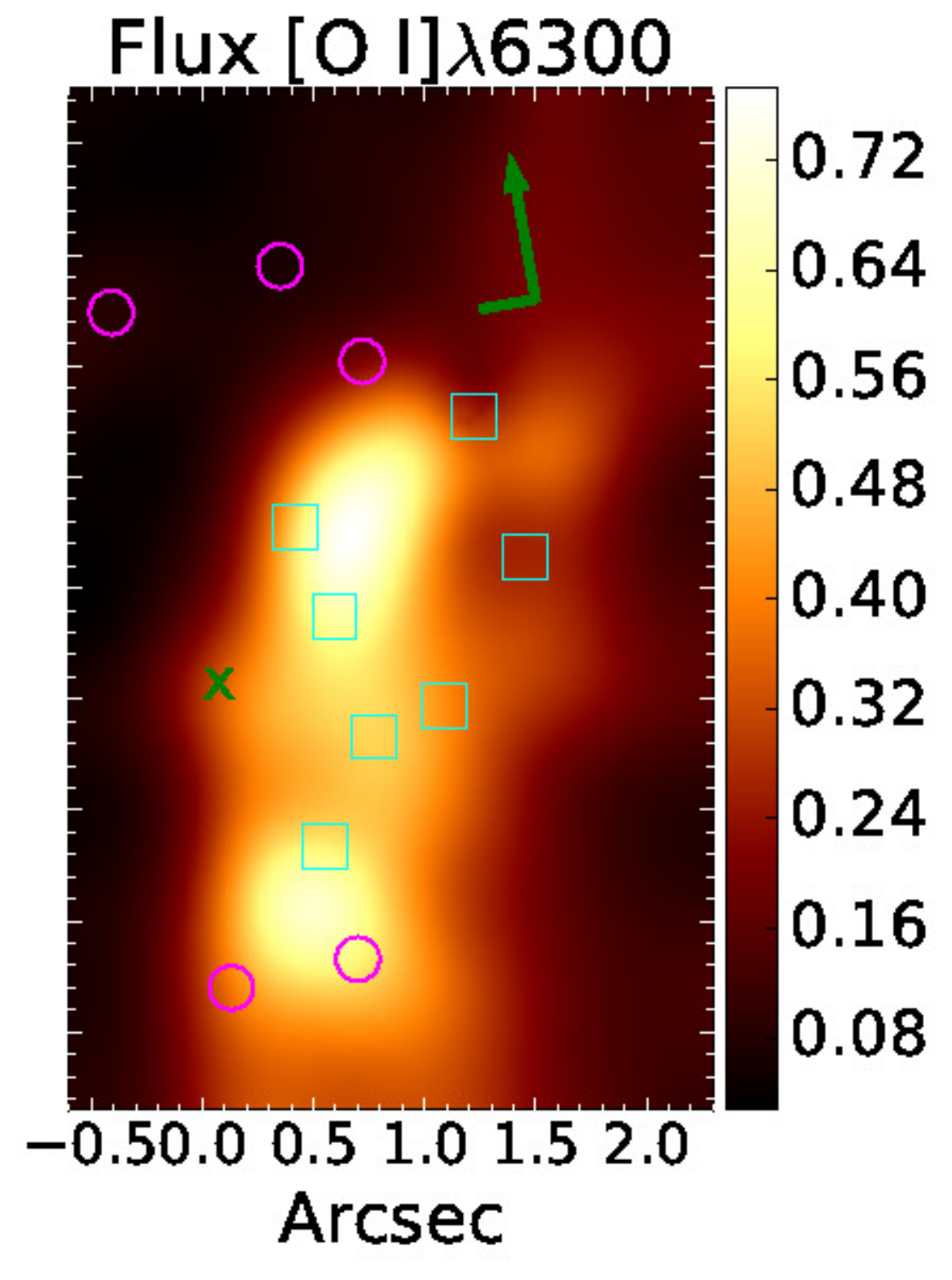}
\includegraphics[scale=0.37]{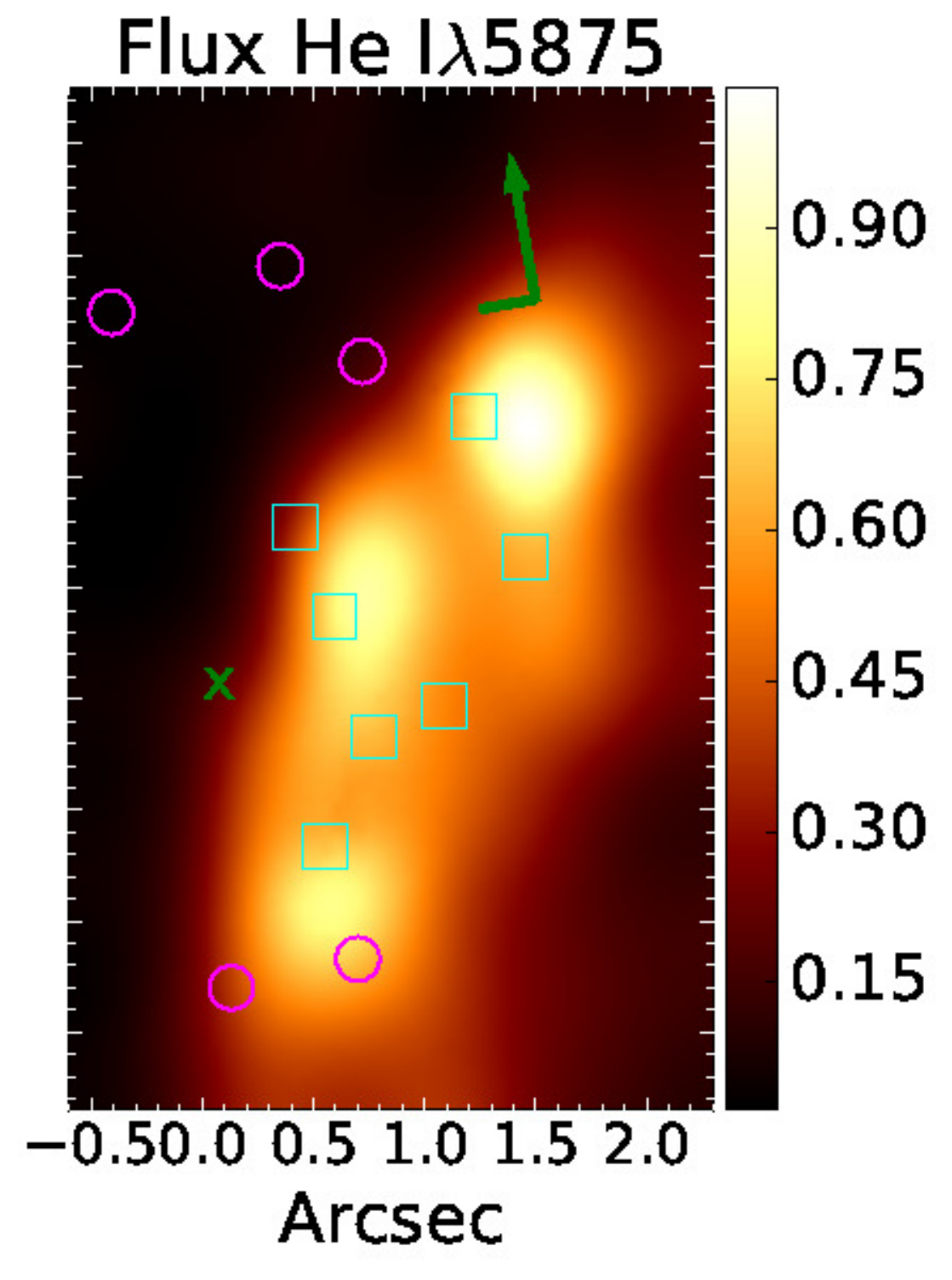}
\includegraphics[scale=0.37]{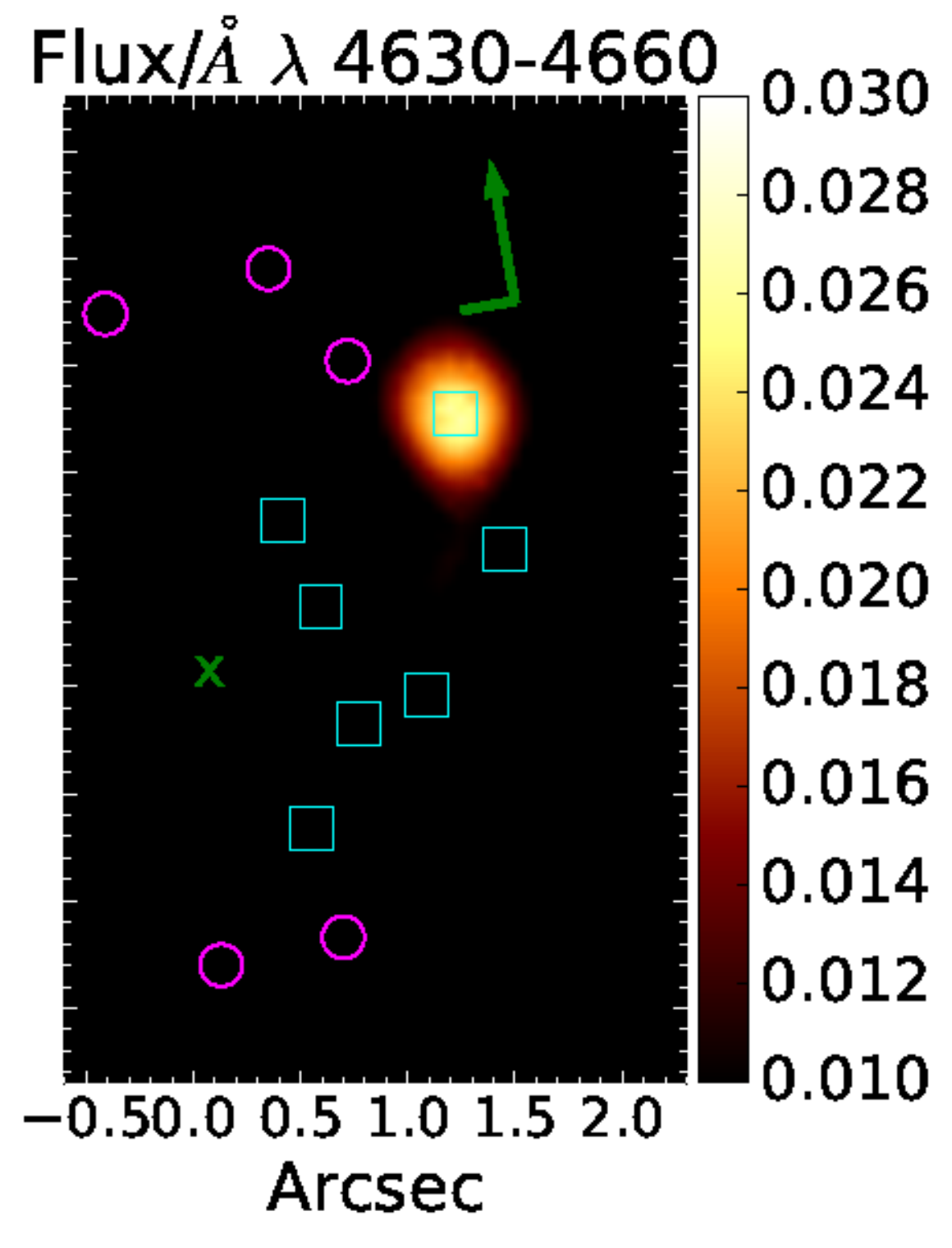}

\caption{Flux maps of the most prominent emission lines detected in the central region of NGC 7582 in units of 10$^{-17}$ erg s$^{-1}$ cm$^{-2}$, except for the average image taken from the 4630--4660\AA\ spectral range, which is in units of 10$^{-17}$ erg s$^{-1}$ cm$^{-2}$ \AA$^{-1}$ (flux density). Only the regions with A/N $>$ 10 are shown in the flux maps. The green contours shown in the H$\alpha$ image are related to the emission of the broad H$\alpha$ component detected at this position. The white contours shown in the [O III]$\lambda$5007 image delineate the regions with the same flux. The cyan squares and magenta circles are those identified in Fig. \ref{HST_image}, with the same symbol scheme also applied for the remainder of this paper. We identified five nebulae in the He II$\lambda$4686 image, named as N1, N2, N3, N4 and N5. The structure seen in the 4630--4660\AA\ image is related to a WR cluster. We used this emission to match the FOV of our GMOS$-$IFU observations with the \textit{HST} image shown in Fig. \ref{HST_image}. We propose that the WR cluster is related to the V1 structure. With such a matching, the broad H$\alpha$ emission is located at the same position of the compact structure identified as the AGN in the \textit{HST} image. \label{fig_flux_el} }
\end{figure*}

The flux maps for H$\alpha$, [N II]$\lambda$6583, [O III]$\lambda$5007, [O I]$\lambda$6300, He I$\lambda$5875, He II$\lambda$4686 and [N I]$\lambda$5198 are shown in Fig. \ref{fig_flux_el} in units of 10$^{-17}$ erg s$^{-1}$ cm$^{-2}$. The amplitude-to-noise (A/N) ratios range from 15 to 194 for H$\alpha$, from 10 to 88 for [N II]$\lambda$6583, from 8 to 153 for [O III]$\lambda$5007, from 0 to 25 for He II$\lambda$4686, from 3 to 26 for He I$\lambda$5875, from 2 to 23 for [O I]$\lambda$6300 and from 0 to 10 for [N I]$\lambda$5198. Only the regions with A/N $>$ 10 are shown in the flux maps in Fig. \ref{fig_flux_el}.

We identified five nebulae with strong He II$\lambda$4686 emission. They are indicated in the flux maps of this emission line in Fig. \ref{fig_flux_el}. We named them N1, N2, N3, N4 and N5. The He II$\lambda$4686 line is not produced by typical H II regions or starbursts due to its high ionization potential energy of 54.4 eV. It may be noticed that the diffuse emission of He II is quite similar to the diffuse emission of the [O III] line. 

The flux maps of H$\alpha$, [N II]$\lambda$6583 and He I$\lambda$5875 are very similar to each other. Two structures located north-west from the nucleus are seen in all three maps, close to N1 and N2. A third structure situated to the south of the nucleus, in the N4 region, is more clear in the [N II] and He I maps. A diffuse emission surrounding these compact regions is also detected. 

Strong [O I]$\lambda$6300 and [N I]$\lambda$5198 emissions are observed in the N4 region. These lines are produced in the transition regions between fully ionized gas and fully neutral gas. These partially ionized zones are thin in H II regions but are quite thick when high ionization photons are present \citep{1983ApJ...269L..37H}. In the N2 region, we observed only an emission from the [O I]$\lambda$6300 line.

\begin{figure}

\includegraphics[scale=0.53]{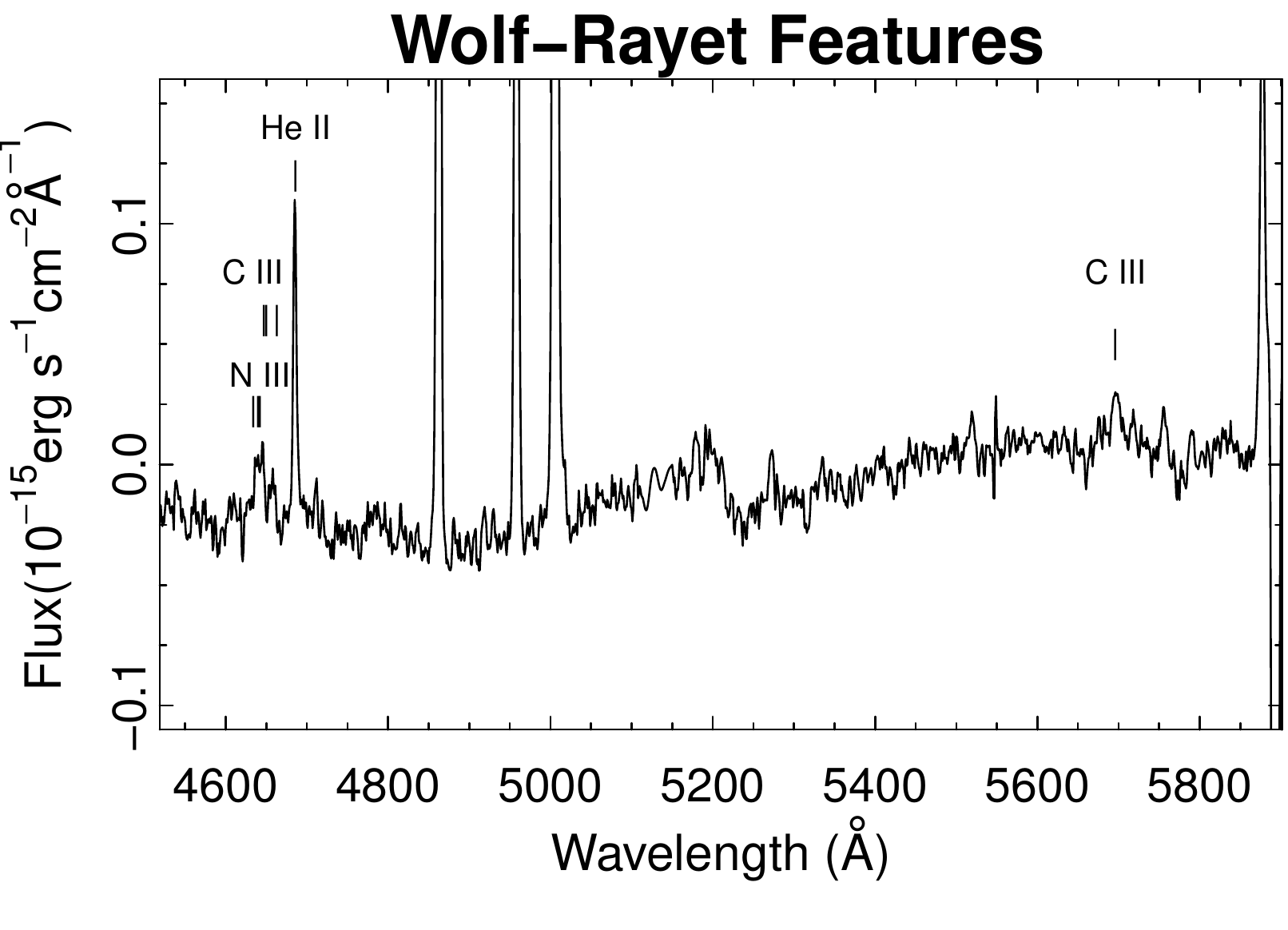}

\caption{Spectrum extracted from the position of the WR cluster. The negative flux values are caused by small differences between the observed spectrum and the model spectrum found with the spectral synthesis. The N III/C III complex at 4634-66\AA\ is clearly seen close to the He II$\lambda$4686 line. We also identify the C III line at 5696\AA. \label{wr_related_spectra}} 
\end{figure}

We extracted an image from the gas cube between 4630 and 4660 \AA, shown in Fig. \ref{fig_flux_el}. In this spectral range, emission lines that are typical of Wolf$-$Rayet (WR) stars may be present. The image shows a compact structure close to N1 and associated with V1, the brightest cluster seen in the optical \textit{HST} image (Fig. \ref{HST_image}). This image can be used to match the FOV between the GMOS and the \textit{HST}. To confirm the emission from a WR cluster, we extracted a spectrum from this region, shown in Fig. \ref{wr_related_spectra}. It reveals the the N III/C III complex at 4634$-$4666\AA, as well as the C III line at 5696\AA. Both features are normally seen in WR stars \citep{2007ARA&A..45..177C,2012A&A...540A.144S}.

We also made an image of the red wing of the broad H$\alpha$ emission. An unresolved source is seen, represented as the green contours in the H$\alpha$ flux map (Fig. \ref{fig_flux_el}). Note that this emission is located at the position of the nucleus in the \textit{HST} image. Besides being an evidence that the broad H$\alpha$ is nuclear, it also reinforces that V1 is very likely to be a WR cluster in the central region of NGC 7582.

\subsubsection{NIR data} \label{sec:el_flux_nirdata}

For the NIR data, we decided not to fit the emission lines that are present in this spectral region because the resulting images are highly contaminated by spurious emission, specially in the regions with low signal-to-noise ratios in the SINFONI 250 data set. Therefore, we extracted flux density maps of [Fe II]$\lambda$1.644$\mu$m, H$_2\lambda$2.122$\mu$m and Br$\gamma$ from the NIR gas cube by calculating the average image within a small spectral range that contains the wavelengths of these emission lines. 

\begin{figure*}
\includegraphics[scale=0.33]{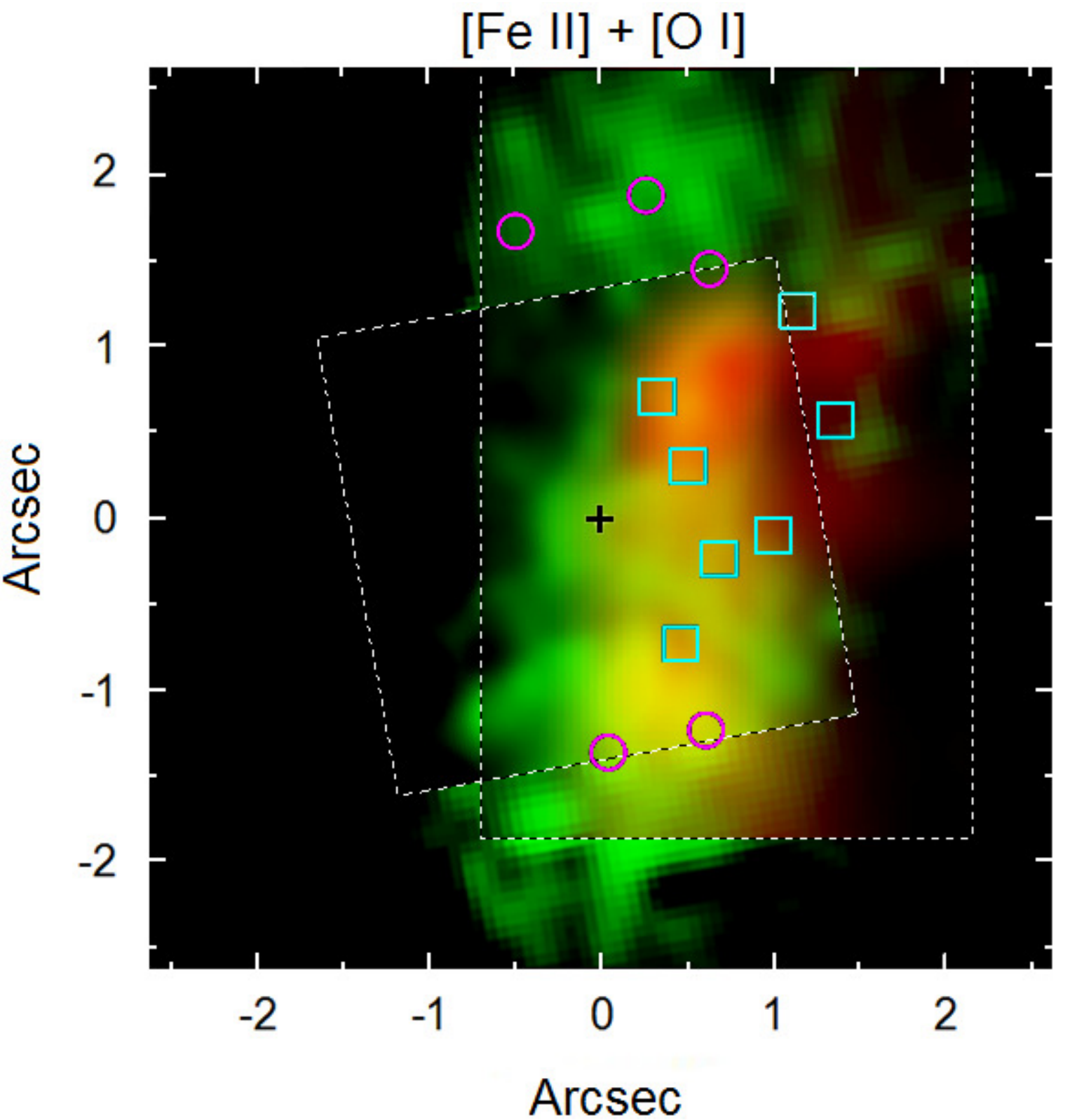}
\includegraphics[scale=0.33]{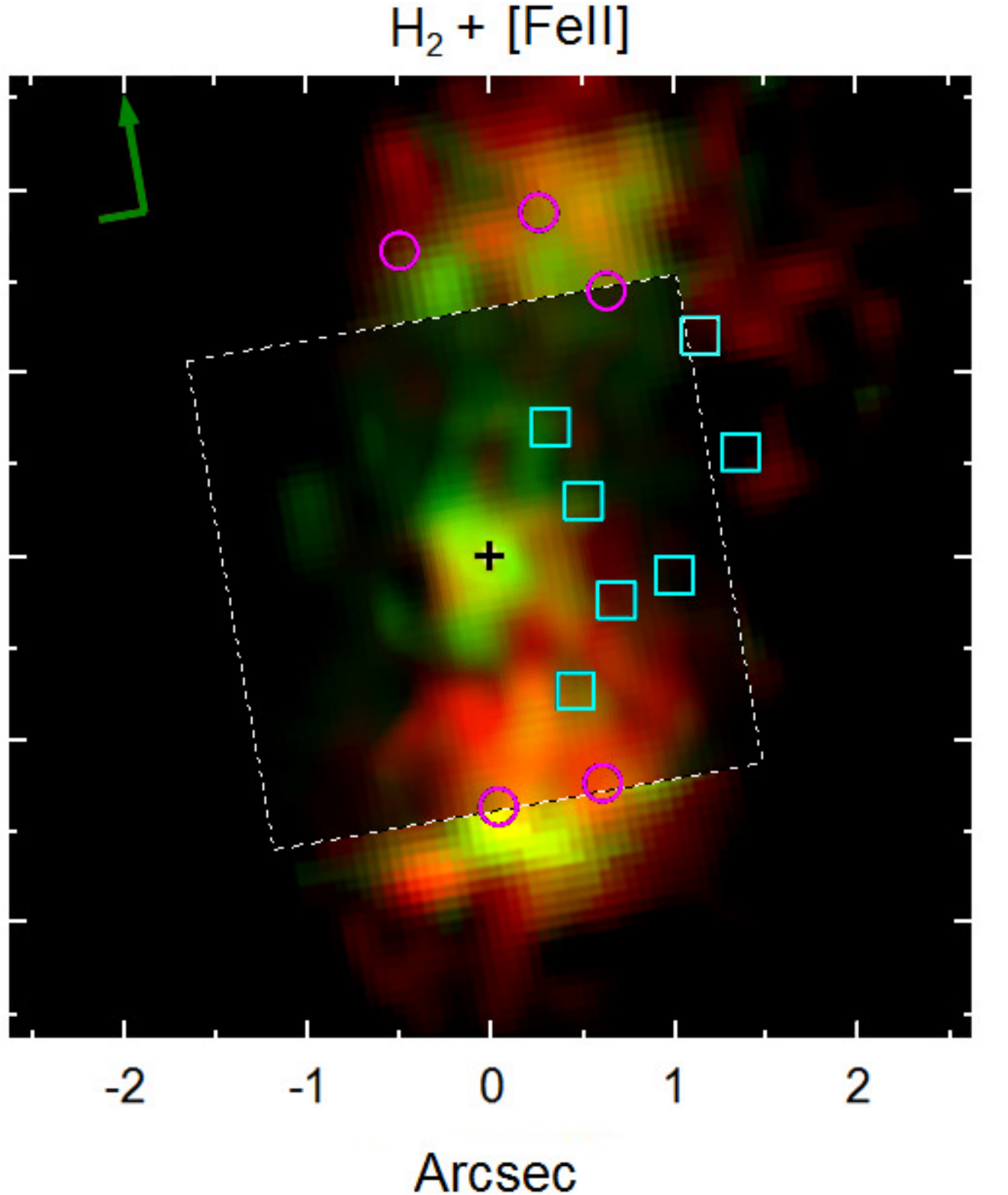}
\includegraphics[scale=0.33]{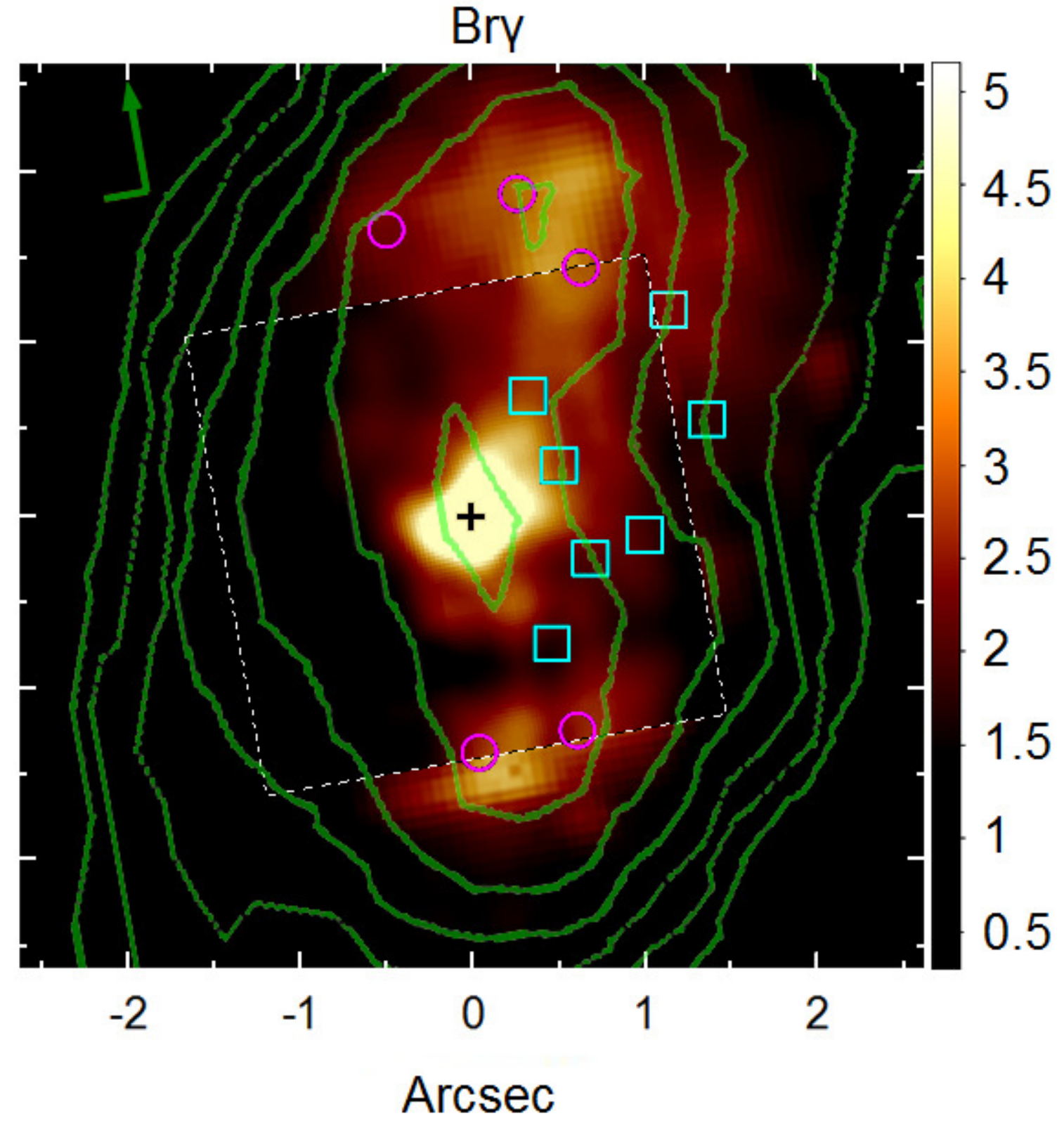}

\caption{RG images combining the [Fe II] emission with the [O I]$\lambda$6300 (left) and the H$_2$ emissions (centre). In the left-hand panel, the [Fe II] emission is shown in green, while in the central panel, the [Fe II] emission is shown in red. In the right-hand panel, we show the Br$\gamma$ map in units of 10$^{-17}$ erg s$^{-1}$ cm$^{-2}$ \AA$^{-1}$. The green contours in the Br$\gamma$ map correspond to a VLA 6 cm radio observation \citep{1999A&AS..137..457M}, centred on the bulge. The white dashed lines indicate the FOV of the data cube with a pixel scale of 100 mas (SINFONI 100), which has higher signal-to-noise ratio in the central part of NGC 7582 for these three lines. The black cross indicates the NIR position of the AGN. Note the presence of one blob north from the nucleus (NB) and another one south from the nucleus (SB), clearly seen in the Br$\gamma$ map. \label{fig:RG_jets}}
\end{figure*}

The [Fe\,{\sc ii}] and H$_2\lambda$2.122$\mu$m maps are very similar to the ones presented by \citet{2001ApJS..136...61S}. Thus, we refer the reader to their Fig. 3 to check the flux distribution of these lines individually. Here, we decided to combine the [Fe\,{\sc ii}] flux distribution with the molecular gas emission and with the [O I]$\lambda$6300 flux map in an RG image. The results are shown in Fig. \ref{fig:RG_jets}. Both the [Fe\,{\sc ii}] and [O I] emissions come from partially ionized zones \citep{1983ApJ...269L..37H,1993ApJ...406...52M,2006agna.book.....O}. These lines seem to be more intense close to M1 and M2. Also, the [Fe\,{\sc ii}] is seen in a region where the extinction is larger, as revealed by the structure map of the NICMOS image shown in Fig. \ref{HST_image}. The $H_2$ and the [Fe\,{\sc ii}] emission are also complementary to each other \citep{2013MNRAS.430.2002R}. A nuclear emission peak and two elongated structures are detected in this combined image, one to the north-west direction, close to the positions of the MIR emitters M3, M4 and M5, and the other one south from the nucleus, close to M1 and M2.

		
We also show in the right-panel of Fig. \ref{fig:RG_jets} the Br$\gamma$ emission, together with the radio emission in 6 cm based on Very Large Array (VLA) observations presented by \citet{1999A&AS..137..457M}. Again, two extended structures are identified at the same positions of those seen in the molecular emission map. Hereinafter, we will refer to such structures as northern blob (NB) and  southern blob (SB). A flux enhancement is seen at the position of both structures in the Br$\gamma$ flux map presented by \citet{2009MNRAS.393..783R}, although it is not possible to distinguish the NB and the SB from an extended emission in the image presented by these authors. Both the NB and the SB seem to be connected to the central emission peak by a fainter emission. It is hard to define a precise distance from the centre and a position angle (PA) for these clouds. Considering their emission peaks, our best estimates are 1.7 arcsec ($\sim$190 pc), with a PA = $-$27$^o$ for NB, and 1.3 arcsec ($\sim$150 pc), with a PA = 180$^o$ for SB. The apparent lack of emission between the nucleus and the SB is also seen in this map. We also see a nuclear emission that coincides with the continuum emission peak in 2.2$\mu$m, shown in Fig. \ref{HST_image}. When we compare the position of the nucleus with the location of the SB and NB, we note that not all of these objects are colinear. There seems to be an orthogonal displacement of 0.3 arcsec to the west between the position of the AGN and the imaginary line that connects the NB and the SB. Although the radio emission has a resolution that is $\sim$ 14 times lower (1.29 arcsec $\times$~0.78 arcsec) than the NIR data, one can see that this emission clearly follows the overall shape of the Br$\gamma$ emission.

\subsection{Gas kinematic maps} \label{sec:kin_ion_gas}

\begin{figure*}
\includegraphics[scale=0.35]{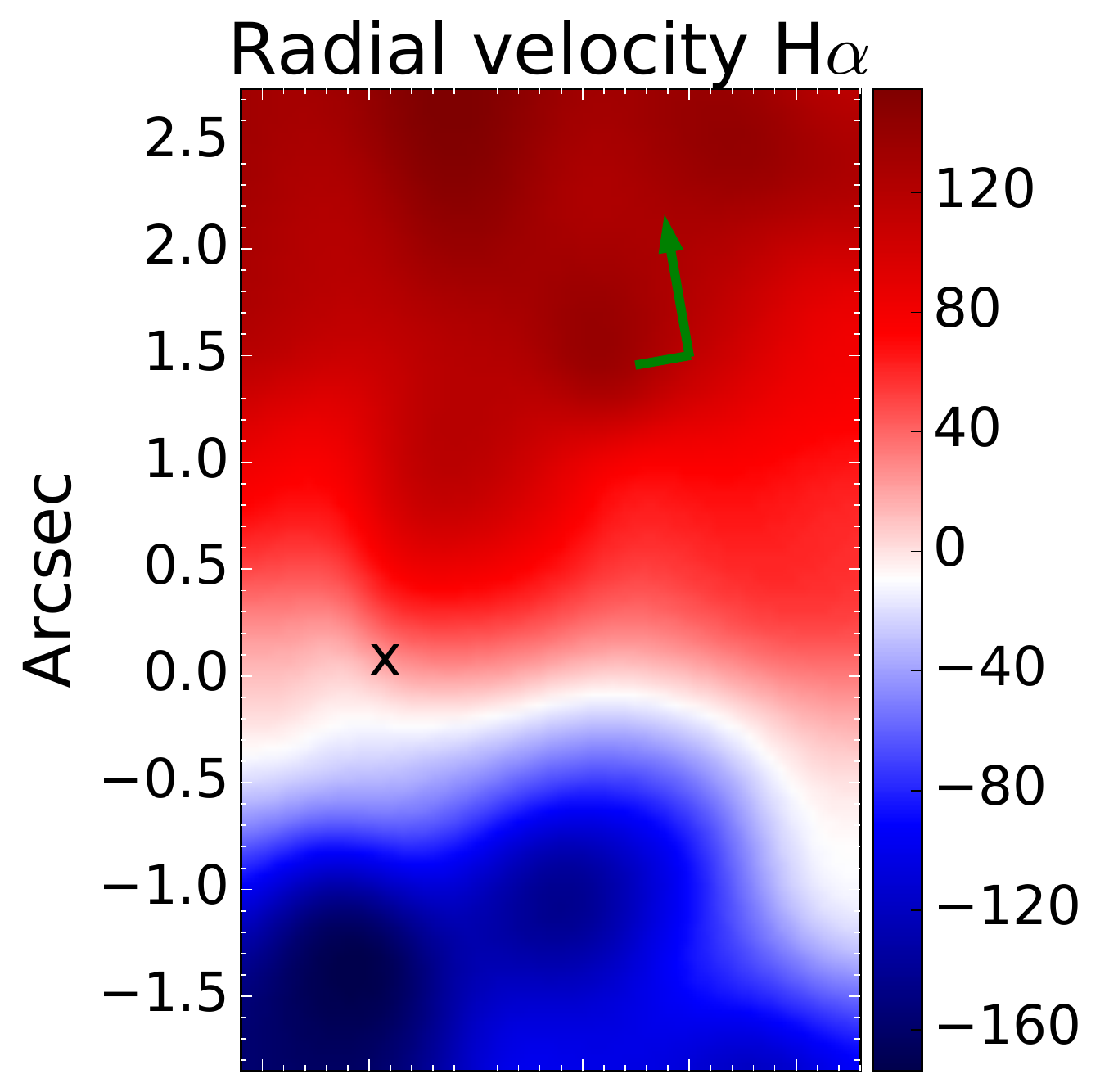}
\includegraphics[scale=0.35]{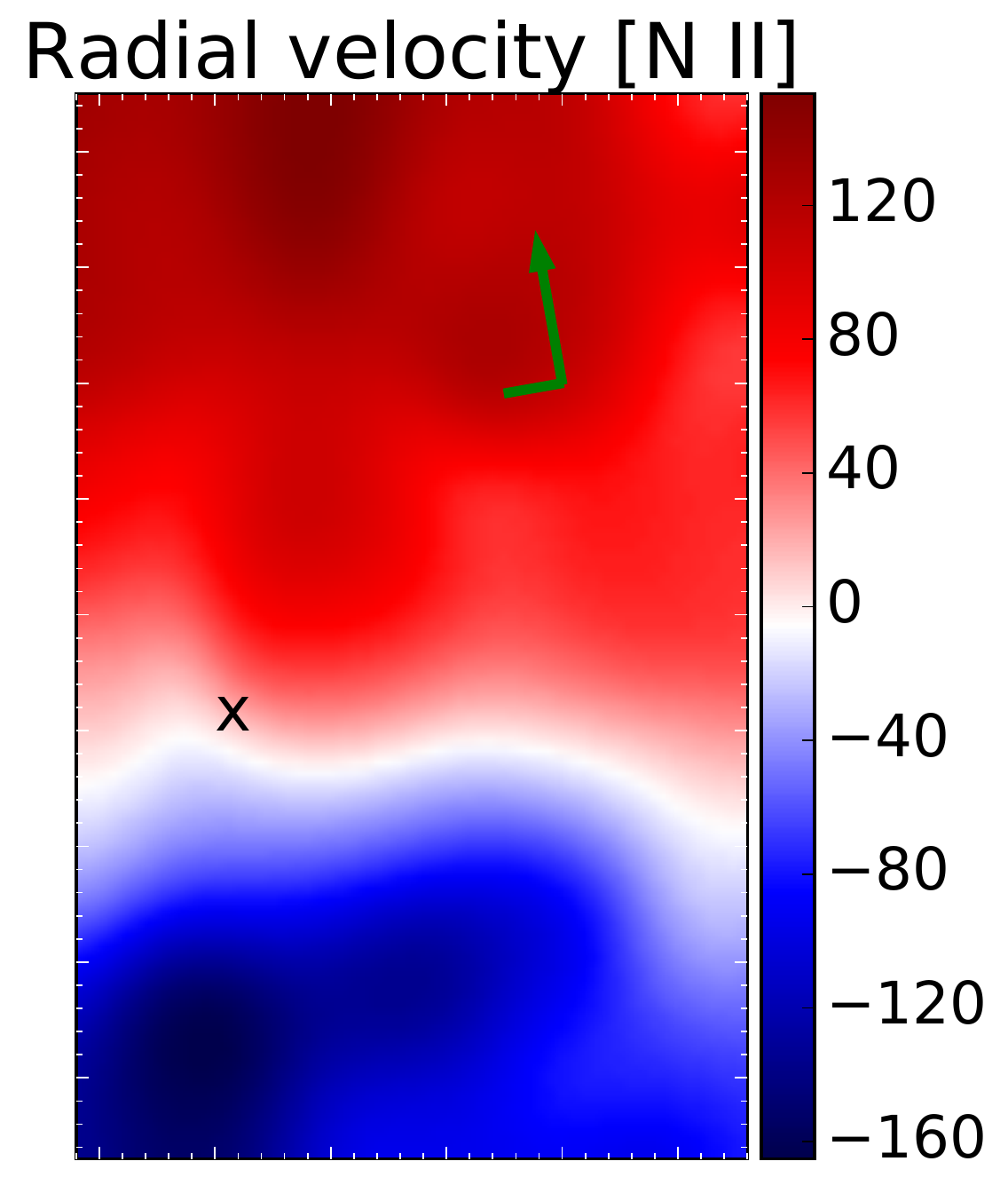}
\includegraphics[scale=0.35]{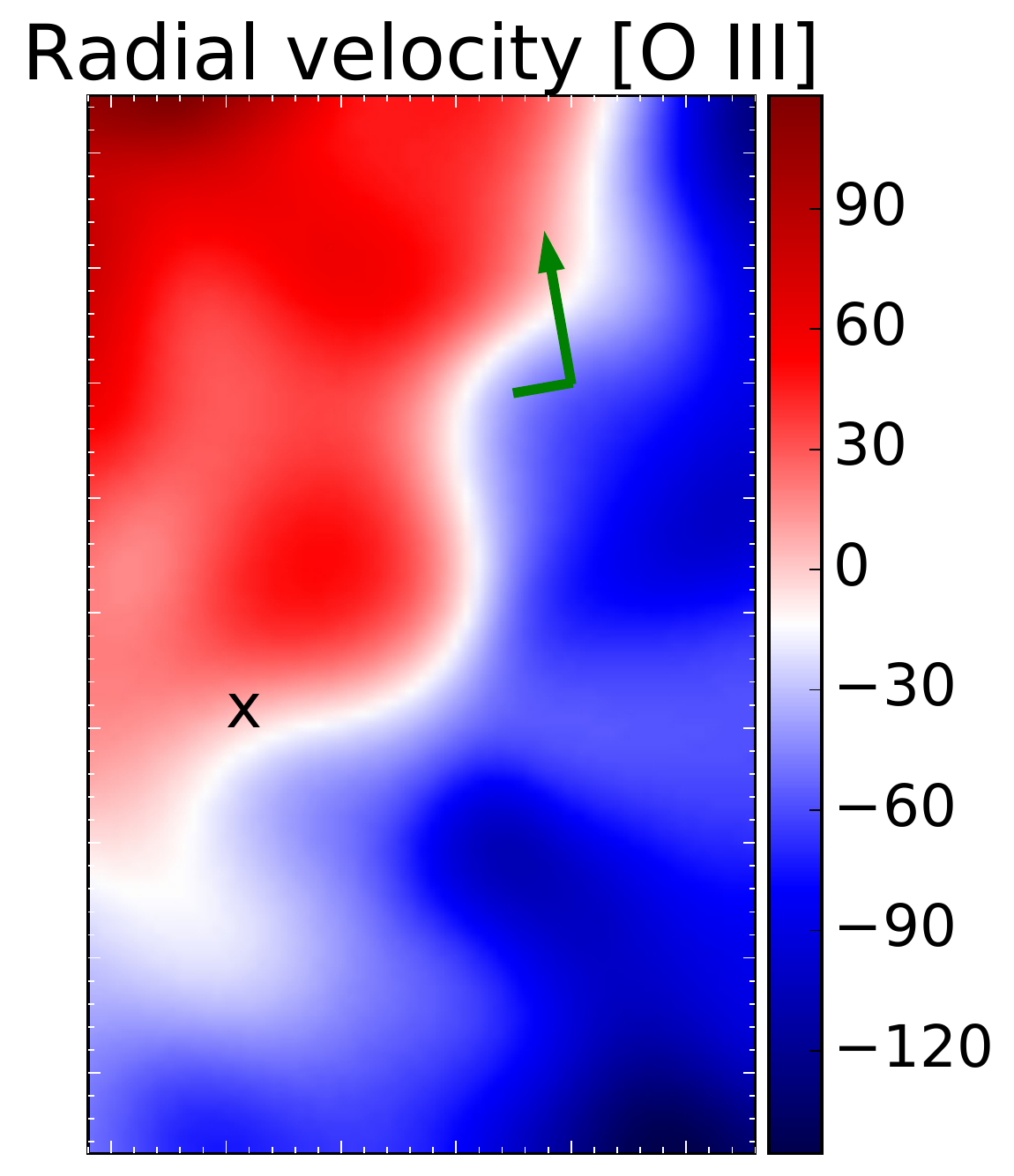}

\includegraphics[scale=0.35]{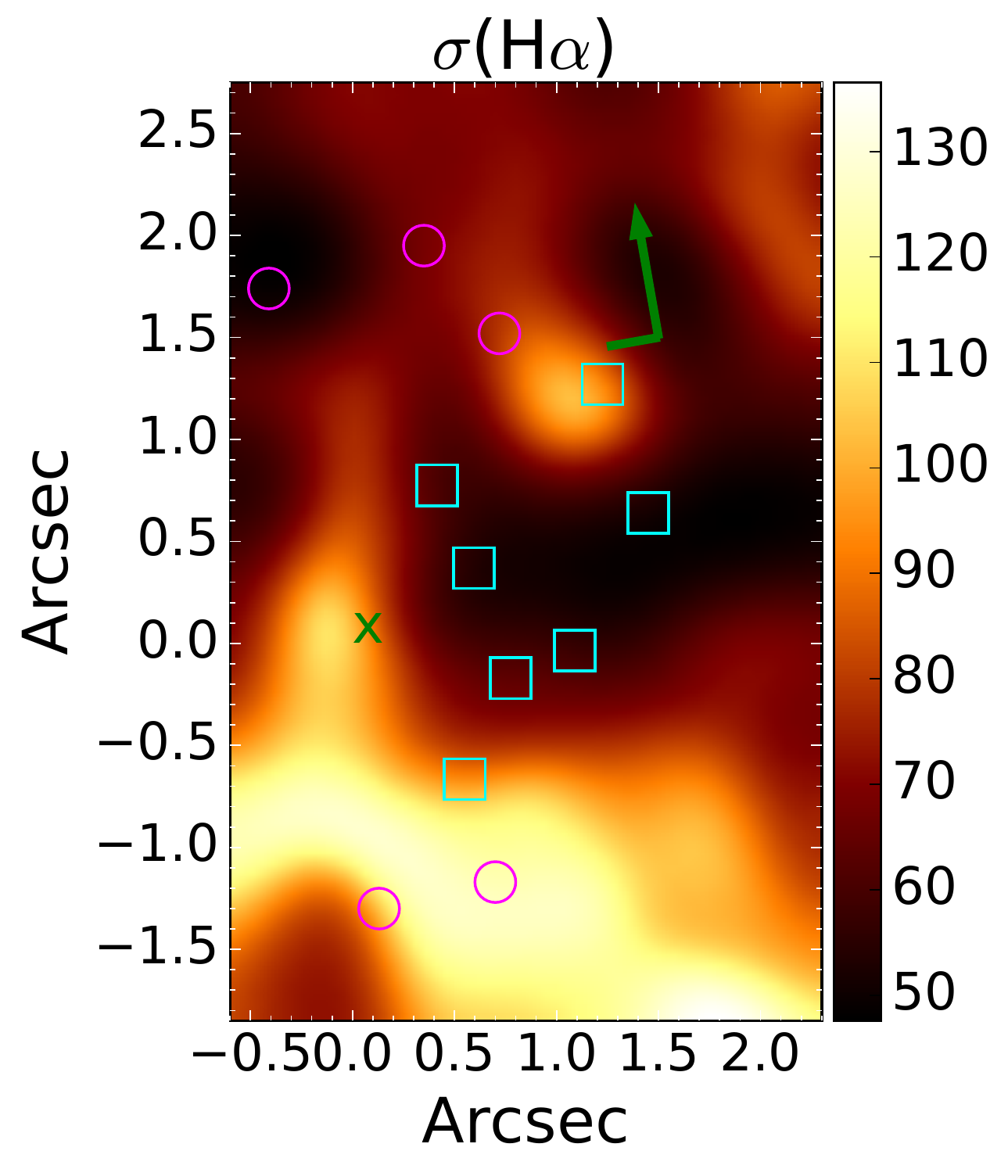}
\hspace{0.1cm}
\includegraphics[scale=0.35]{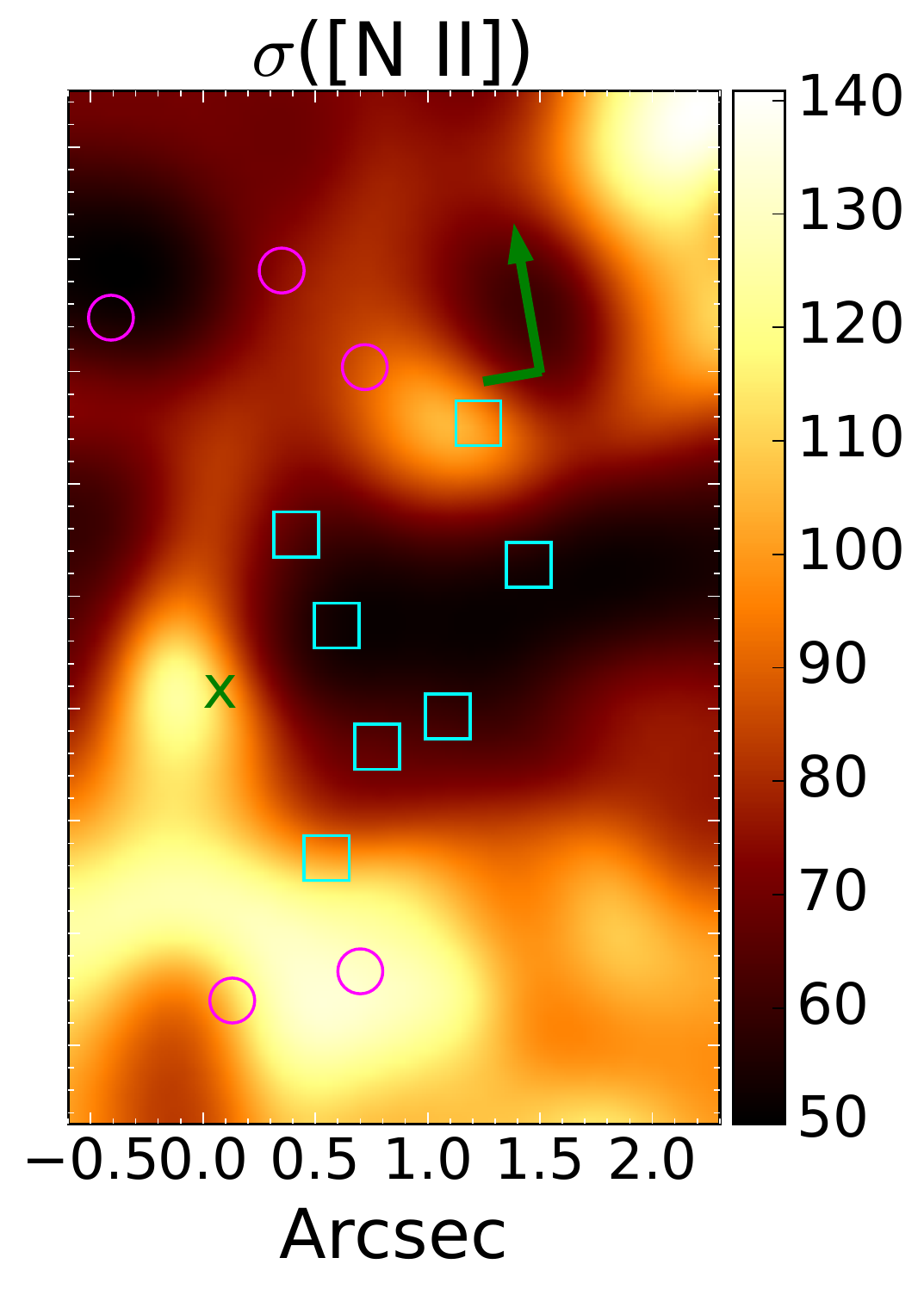}
\hspace{0.2cm}
\includegraphics[scale=0.35]{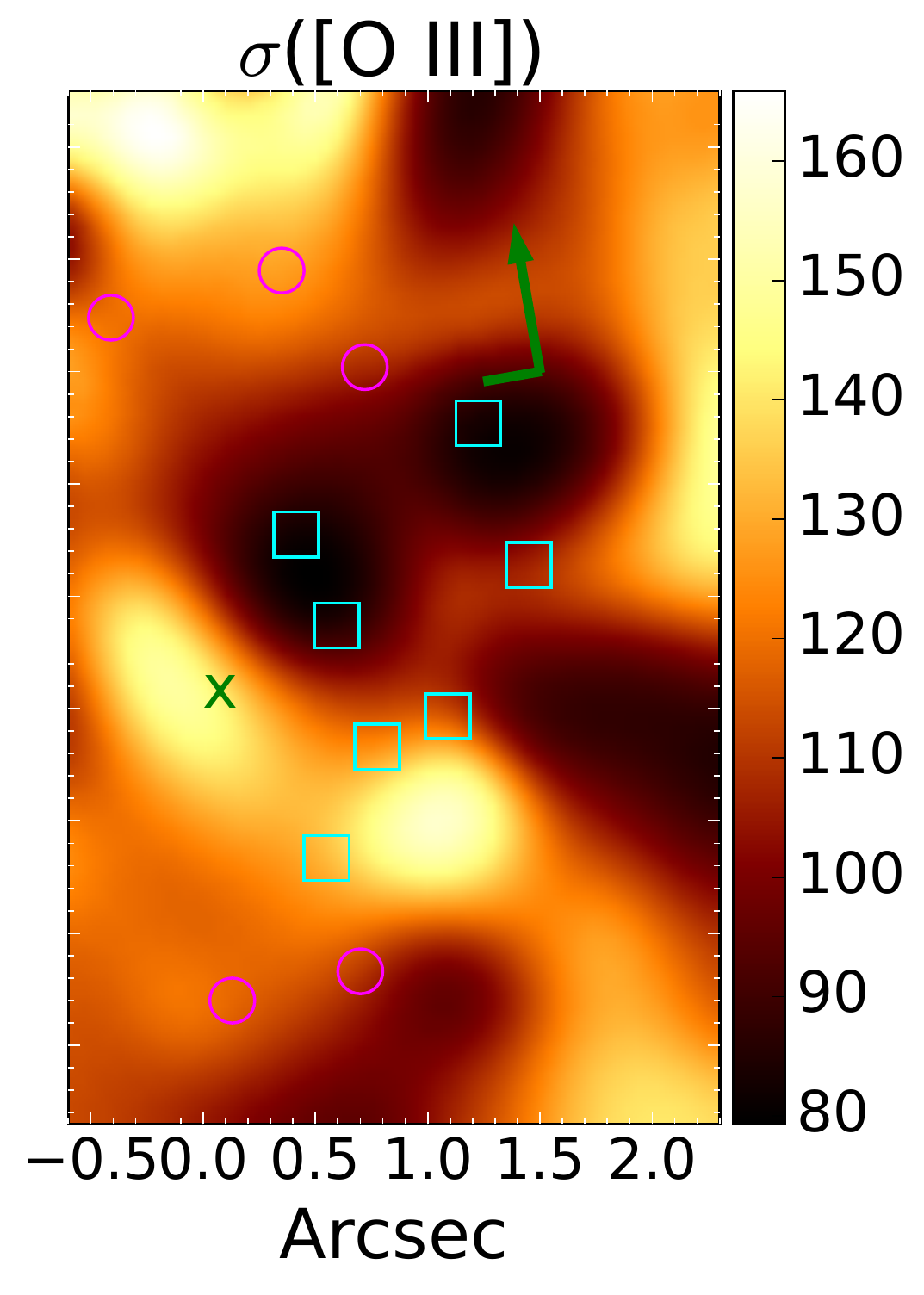}

\caption{Kinematics extracted from the H$\alpha$, [N II]$\lambda\lambda$6548, 6583 and the [O III]$\lambda\lambda$4959, 5007 lines, in units of km s$^{-1}$. The radial velocities are shown with respect to the heliocentric radial velocity of NGC 7582 of 1575 km s$^{-1}$. The H$\alpha$ and [N II]$\lambda\lambda$6548, 6583 lines have similar kinematics and are probably related to a gas disc. The structure seen in the mid-ionized gas is located in a position that is probably related to the beginning of the ionization cone seem in large scales. From the velocity dispersion maps of the H$\alpha$ and [N II] lines, we note a $\sigma$ peak at the positions of M1, M2 and N4, while the highest values for $\sigma$ in the [O III] map are seen at the position of N3. \label{fig_Vrad_el} }
\end{figure*}

As previously mentioned, we extracted the kinematics of the gas using the H$\alpha$, [N II]$\lambda\lambda$6548, 6583 and the [O III]$\lambda\lambda$4959, 5007 lines. The results are shown in Fig. \ref{fig_Vrad_el}. The radial velocities results are presented with respect to the heliocentric velocity of NGC 7582 of 1575 km s$^{-1}$, as taken from the NASA Extragalactic Database (NED). We also corrected the velocity dispersion ($\sigma$) values for the instrumental broadening effect by assuming a spectral resolution of 1.8 \AA\ (FWHM) for all wavelengths. In terms of velocity, the instrumental broadening values are $\sigma_{inst}(H\alpha) \sim$ 35 km s$^{-1}$ and  $\sigma_{inst}([O III]) \sim$ 46 km s$^{-1}$. The average uncertainties of the radial velocity maps are $\sim$ 2 km s$^{-1}$ for H$\alpha$ and [O III]$\lambda$5007 and $\sim$ 3 km s$^{-1}$ for [N II]$\lambda$6583. For the velocity dispersion maps, the average uncertainties are $\sim$ 2 km s$^{-1}$ for H$\alpha$ and $\sim$ 3 km s$^{-1}$ for [N II] and [O III].

It is clear that the gas kinematics as traced by the H$\alpha$ and the [N II]$\lambda$6583 lines are nearly the same. This means that they are mapping the same gaseous component. The radial velocity maps of these lines seem to be related to the gaseous disc in the central region of NGC 7582 that was already proposed before by \citet{1985MNRAS.216..193M} and \citet{2009MNRAS.393..783R}. This gas rotation was also shown in \citet{2009MNRAS.396..788S} using the [S II] lines detected with this same data cube. We measured the PA of this structure using the determination of the global kinematic PA proposed by \citet{2006MNRAS.366..787K} assuming that the centroid is at the position of the nucleus. We obtained PA = 0$\pm$5$^o$. It is worth mentioning that the kinematics regarding the NIR emission lines also suggest a gas rotation in a way very similar to what was detected by \citet{2009MNRAS.393..783R} and are not shown in this paper.

The kinematics of the [O III]$\lambda\lambda$4959, 5007 is quite different from the kinematics of the H$\alpha$ and the [N II]$\lambda$6583 lines, though. The structure seen in the radial velocity map of this mid-ionization line has a PA = 47$\pm$5$^o$, assuming that the centroid is at the position of the nucleus. It seems that the blueshifted structure is associated with the inner part of the ionization cone previously reported by \citet{1985MNRAS.216..193M}, \citet{1991MNRAS.250..138S} and \citet{2016ApJ...824...50D}, all using the [O III]$\lambda$5007 line, and by \citet{2007MNRAS.374..697B}, using soft X-rays image in a spectral range that is dominated by high-ionization emission lines.

All three emission lines show large values of velocity dispersion in the nuclear region of NGC 7582 ($\sim$ 130 km s$^{-1}$). A structure is also seen close to the WR cluster in the H$\alpha$ and the [N II]$\lambda$6583 lines. However, the highest values for the velocity dispersion of these lines are seen in the N4 region. In the mid-ionized gas, a $\sigma$-peak is detected in the N3 and M2 regions. The northern structure relative to the nucleus in the mid-ionization gas is probably caused by low A/N ratio of the [O III]$\lambda$5007 line in this region. The same may apply for the structure seen north-west from the nucleus in the gas traced by the [N II]$\lambda$6583 line. For both situations, A/N $\sim$ 10.

\subsection{Line ratio maps} \label{sec:EL_ratios}

\begin{figure*}

\includegraphics[scale=0.35]{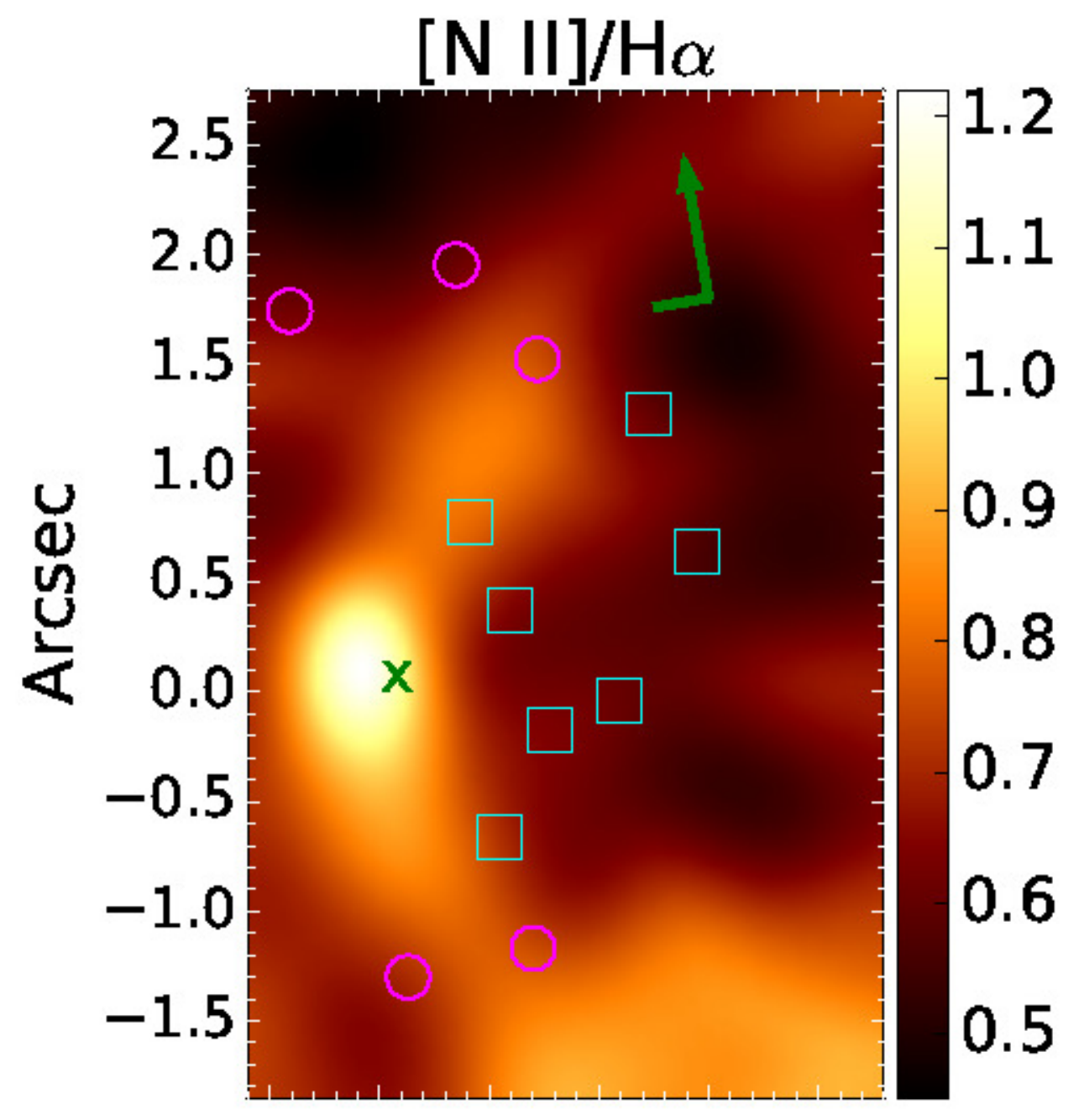}
\includegraphics[scale=0.35]{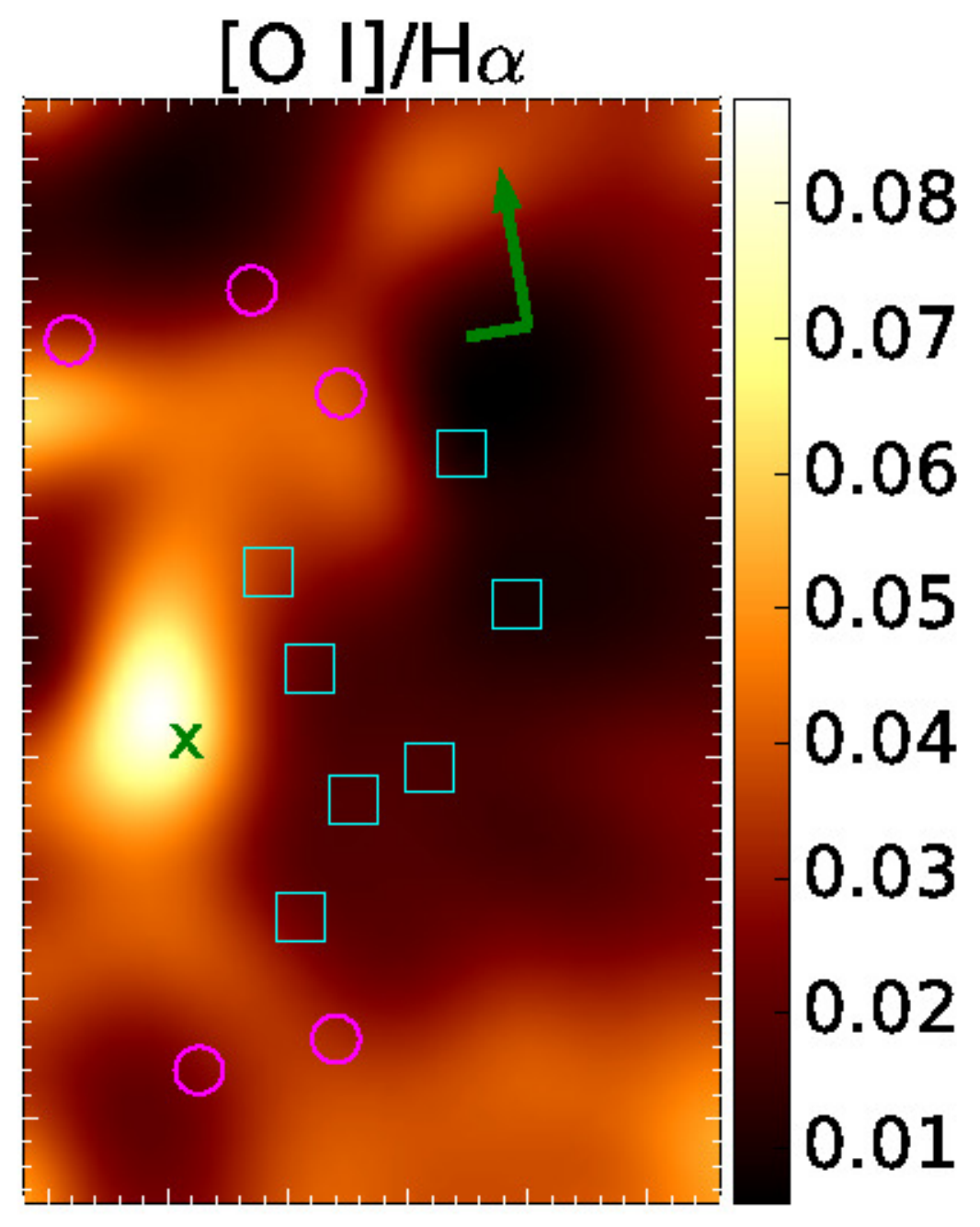}
\includegraphics[scale=0.35]{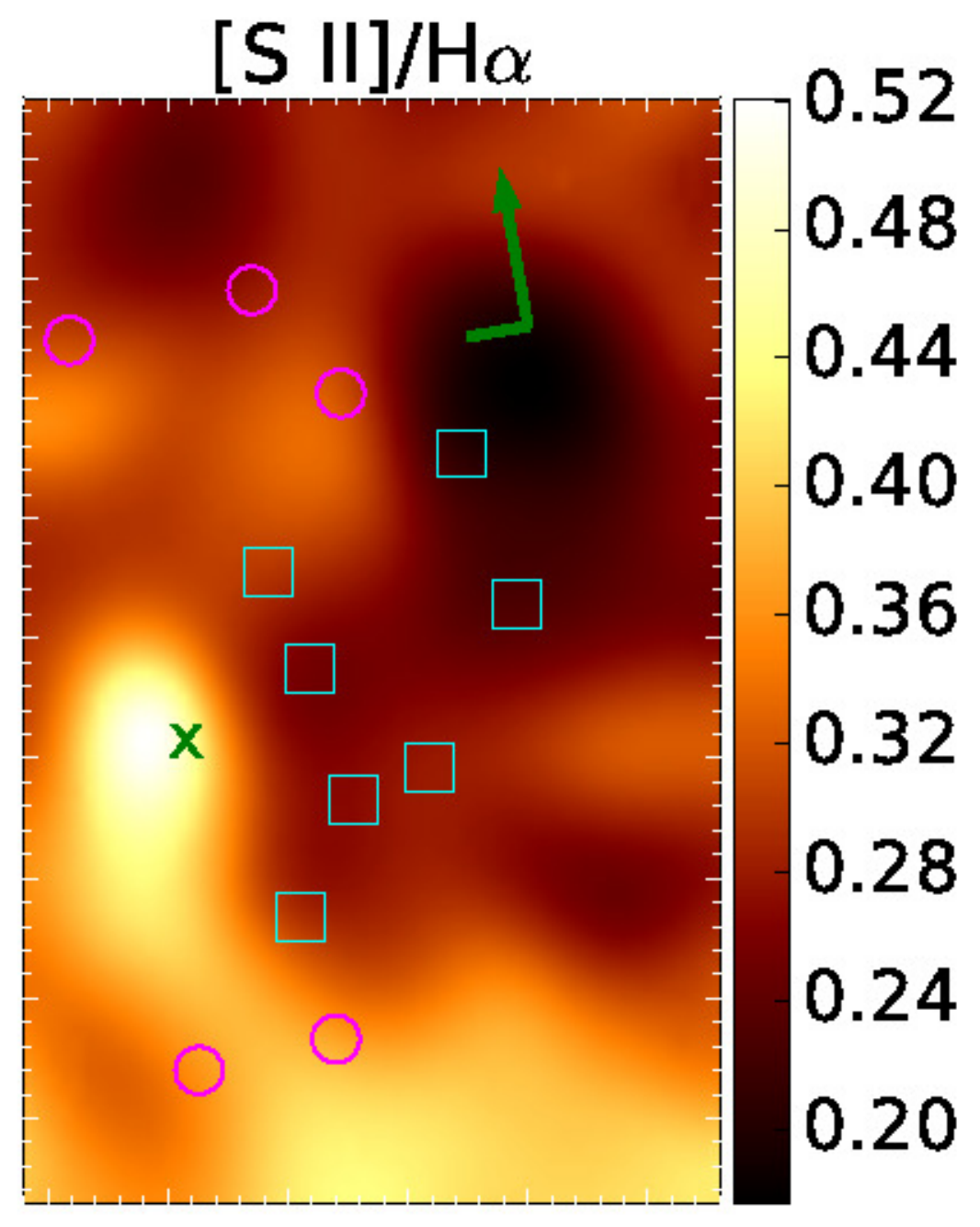}

\includegraphics[scale=0.35]{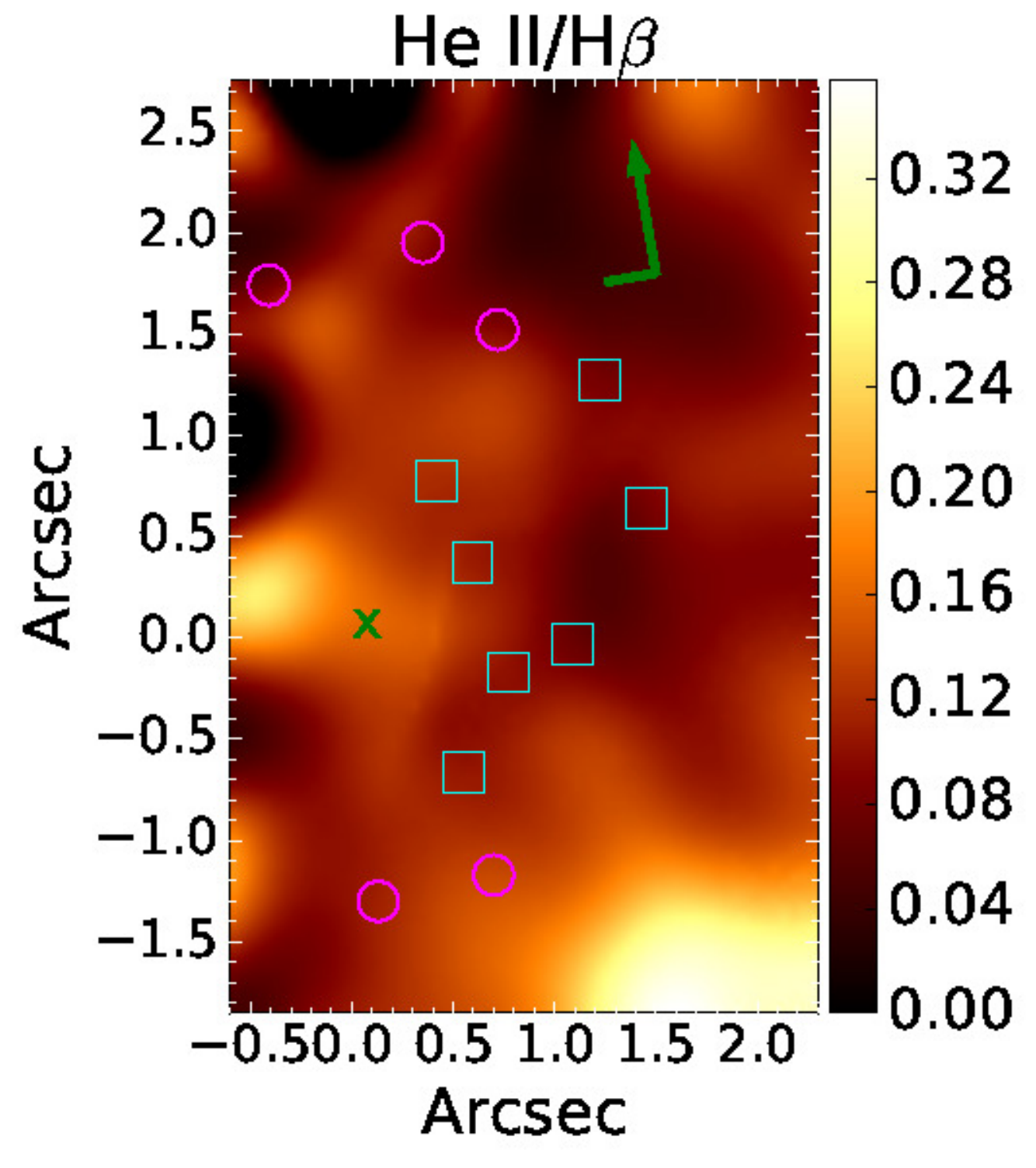}
\includegraphics[scale=0.35]{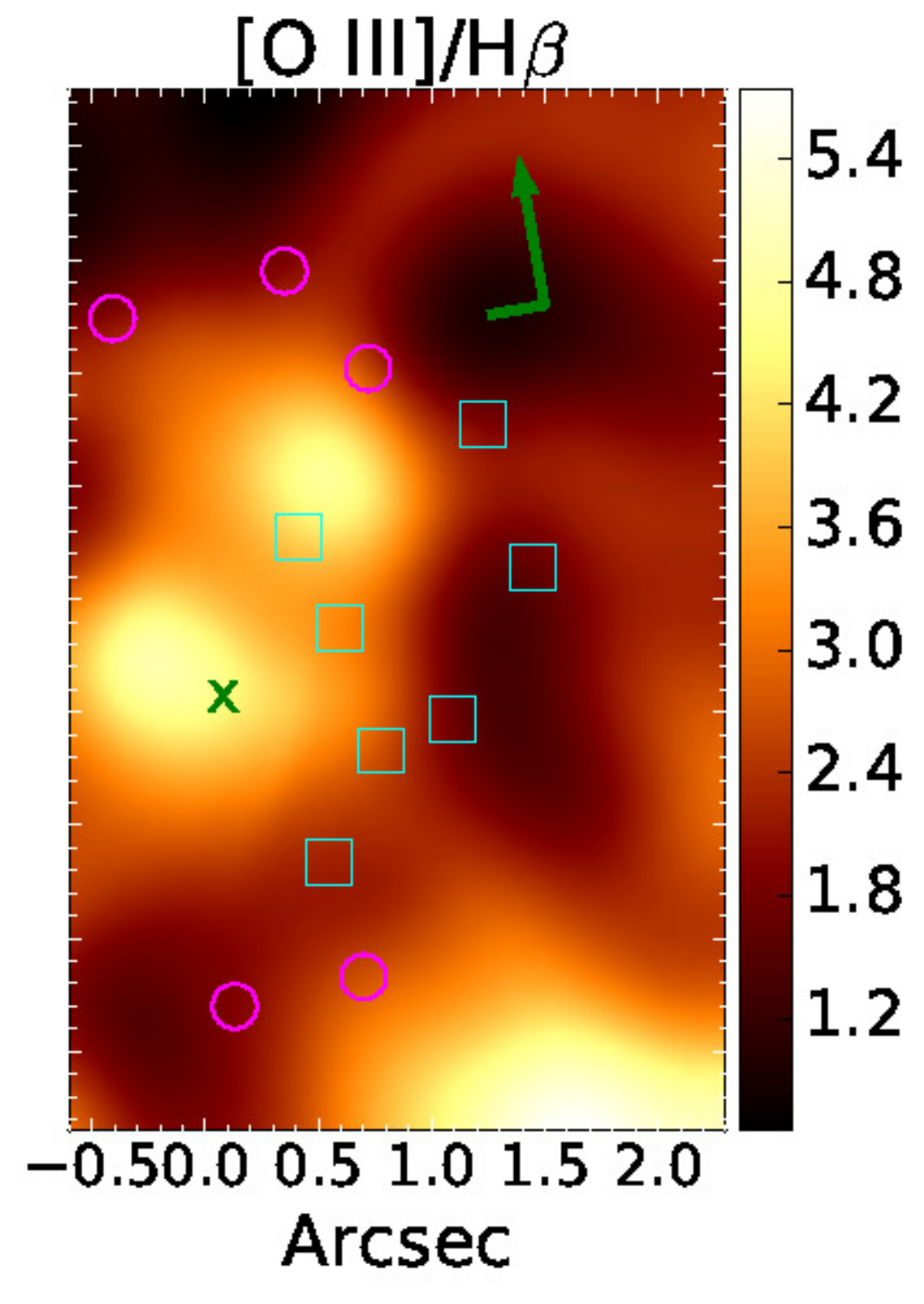}
\includegraphics[scale=0.35]{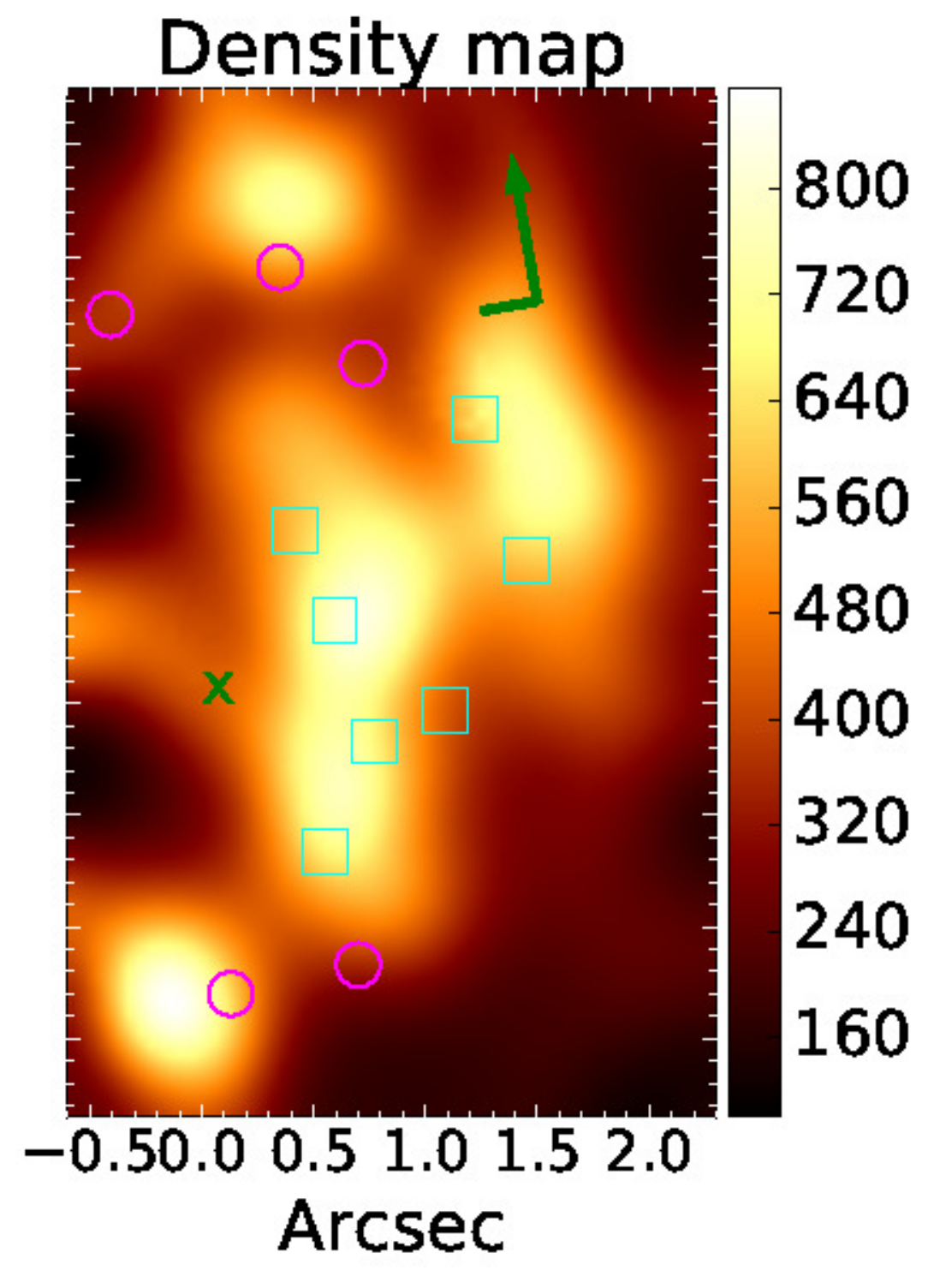}

\caption{Emission-lines ratio maps and density map in units of cm$^{-3}$. Note that the region of the AGN has strong [O I]/H$\alpha$, [N II]/H$\alpha$ and [S II]/H$\alpha$, as expected. We also see that the region of N5 has strong He II/H$\beta$ and [O III]/H$\beta$ ratios and lower density when compared to other regions in the FOV. The low density together with a region directly illuminated by the ionization cone produced a cloud with a high-ionization parameter.  \label{fig_ratio_el} }
\end{figure*}

We used the optical flux maps shown in Fig. \ref{fig_flux_el} to build maps of typical line ratios that are used in diagnostic diagrams. The maps of [N II]/H$\alpha$, [S II]/H$\alpha$, [O I]/H$\alpha$, [O III]/H$\beta$ and He II/H$\beta$ are shown in Fig. \ref{fig_ratio_el}. We also show the electron density $n_e$ map, which was calculated using the [S II]$\lambda$6716/[S II]$\lambda$6730 ratio and the \citet{2014A&A...561A..10P} relation between this line ratio and $n_e$.

\begin{figure}

\includegraphics[scale=0.35]{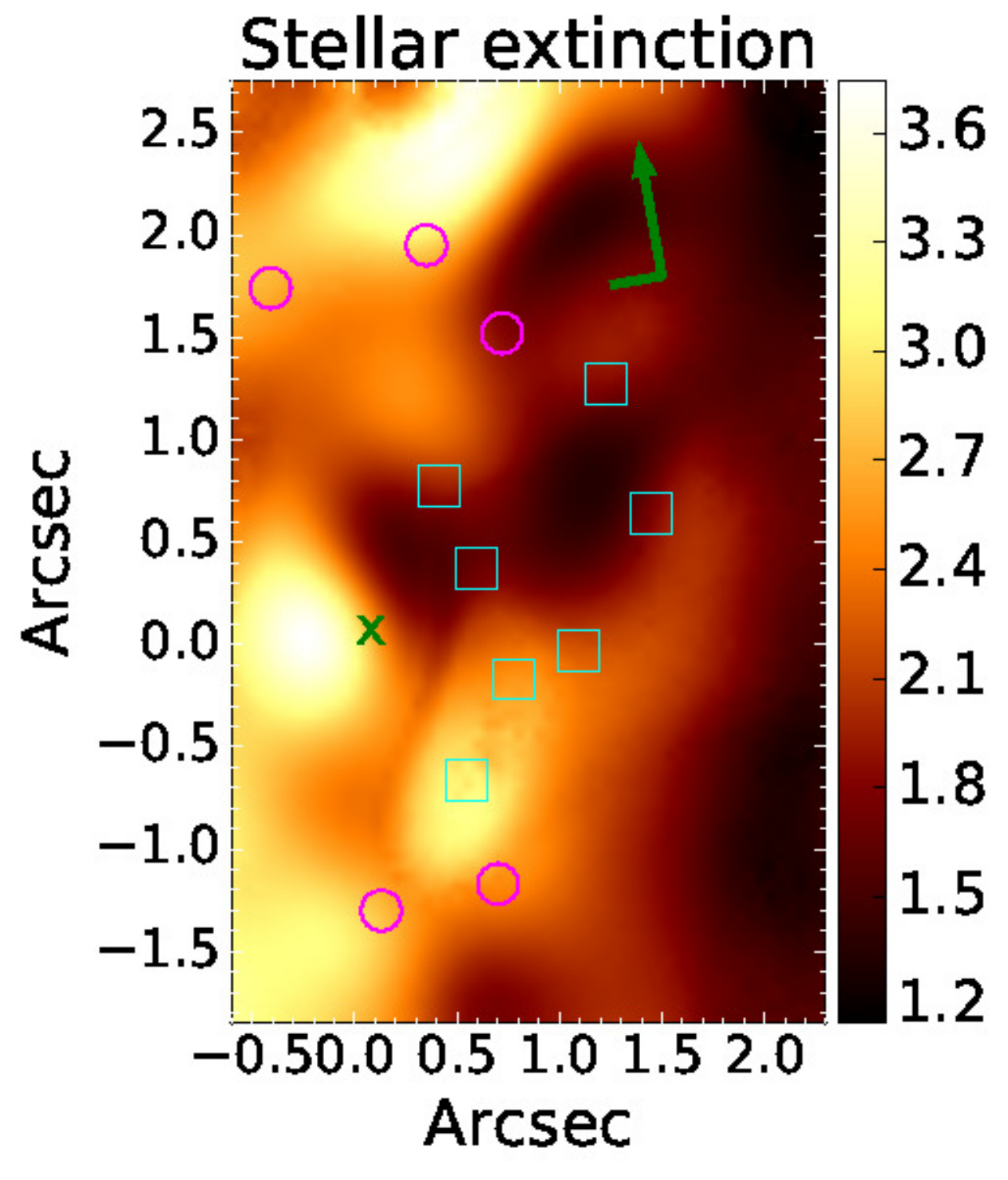}
\includegraphics[scale=0.35]{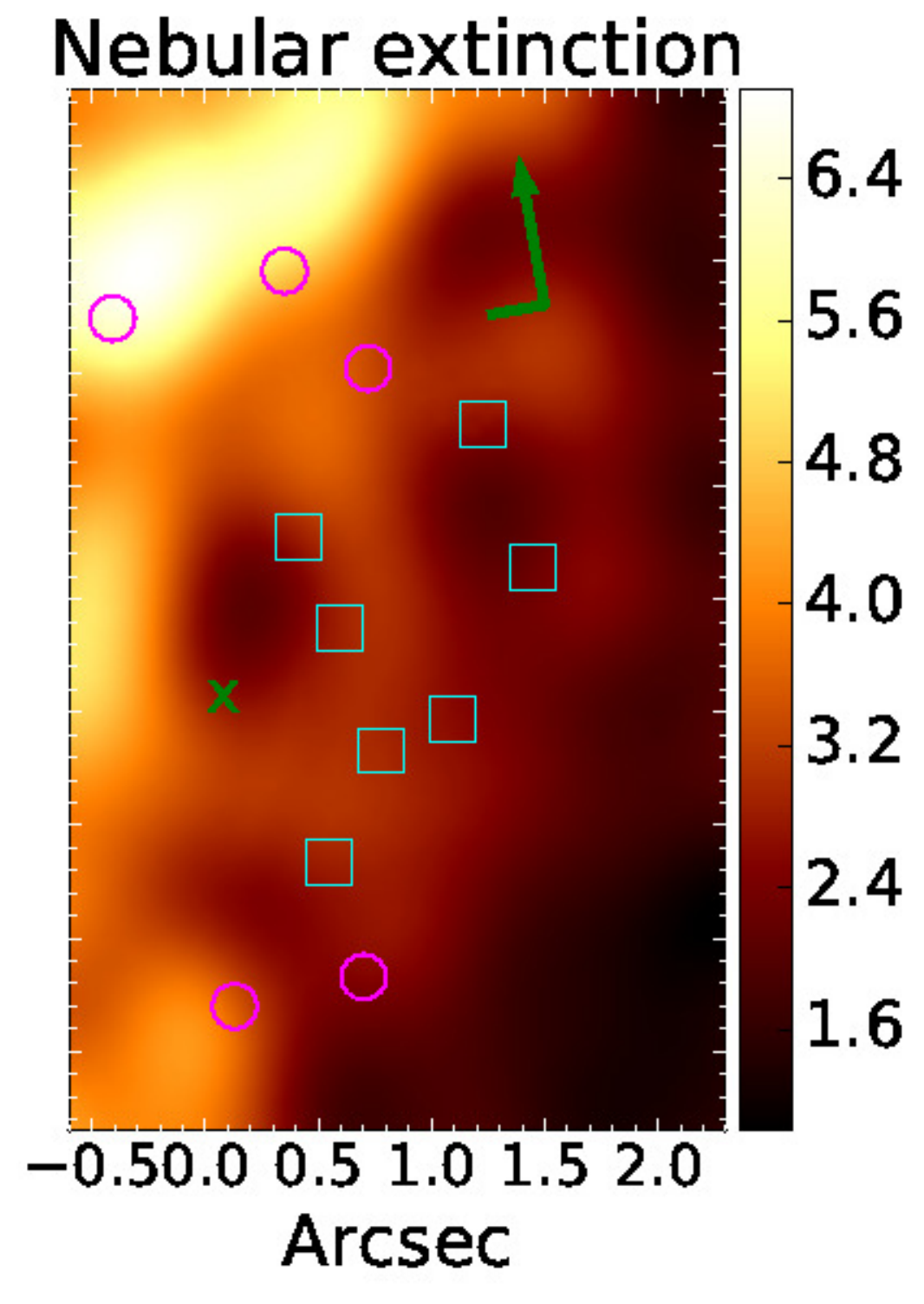}

\caption{Stellar and nebular extinction (A$_V$) within the FOV of NGC 7582. The stellar extinction was measured with the spectral synthesis applied to each spaxel of the data cube and the nebular extinction was measured with the H$\alpha$/H$\beta$ ratio assuming an intrinsic H$\alpha$/H$\beta$ = 2.87 and using the \citet{1989ApJ...345..245C} curve. It is worth mentioning that this intrinsic H$\alpha$/H$\beta$ ratio is valid for gaseous nebulae ionized by young stars. For AGNs, this intrinsic ratio is higher, so the nebular extinction at the position of the nucleus is overestimated. \label{fig_av} }
\end{figure}

In Fig. \ref{fig_av}, we present the nebular extinction map, calculated using the H$\alpha$/H$\beta$ ratio and the \citet{1989ApJ...345..245C} curve with $R$ = 3.1. We assumed an intrinsic H$\alpha$/H$\beta$ = 2.87, which holds for H II regions \citep{2006agna.book.....O}. It is worth mentioning that typical H$\alpha$/H$\beta$ = 3.1 for AGNs \citep{2006agna.book.....O}, so that our nebular extinctions
may be underestimated by 0.2 mag for the nuclear region. We also built a stellar extinction map using the spectral synthesis results for this parameter. It is worth noticing that we detect little emission from all optical lines in the region located north-east and south-east from the nucleus, where the stellar extinction is higher. On the other hand, the emission from the NIR lines are seen in such regions.

The nuclear region has high [N II]/H$\alpha$, [S II]/H$\alpha$, [O I]/H$\alpha$ and [O III]/H$\beta$ ratios. This is expected for Seyfert-like AGNs \citep{2006MNRAS.372..961K,2006agna.book.....O,2008ARA&A..46..475H}. A horizontal structure is seen, especially in the [S II]/H$\alpha$ and [O I]/H$\alpha$ maps, close to the positions of M3, M4 and M5. Since this region has both high nebular and stellar extinction, no structure is expected to be seen in the optical flux maps. An arc structure, visible across the FOV is also revealed in the [N II]/H$\alpha$ and [O I]/H$\alpha$ maps. One may also notice large He II/H$\beta$ and [O III]/H$\beta$ line ratios in the region of N5. High- and mid-ionization line ratios are related to regions with high-ionization parameter \citep{2006agna.book.....O}, defined as the ratio between the ionizing photons density and $n_e$. This may be the case for N5, since it is in the direction of the ionization cone and it has low $n_e$, as revealed by the density map.

\section{Spectral analysis} \label{sec:spectral_analysis}

\begin{figure*}

\includegraphics[scale=0.53]{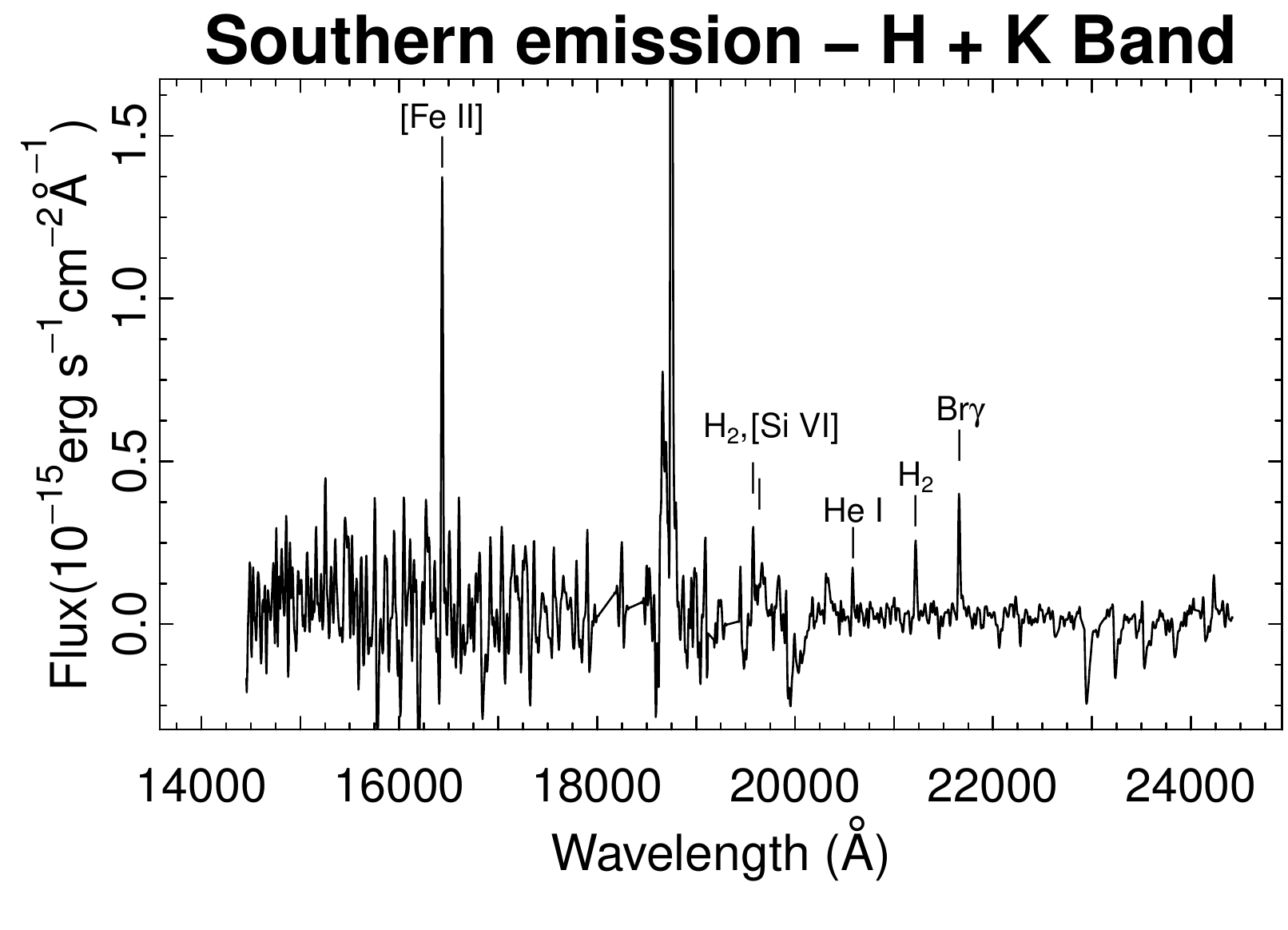}
\includegraphics[scale=0.53]{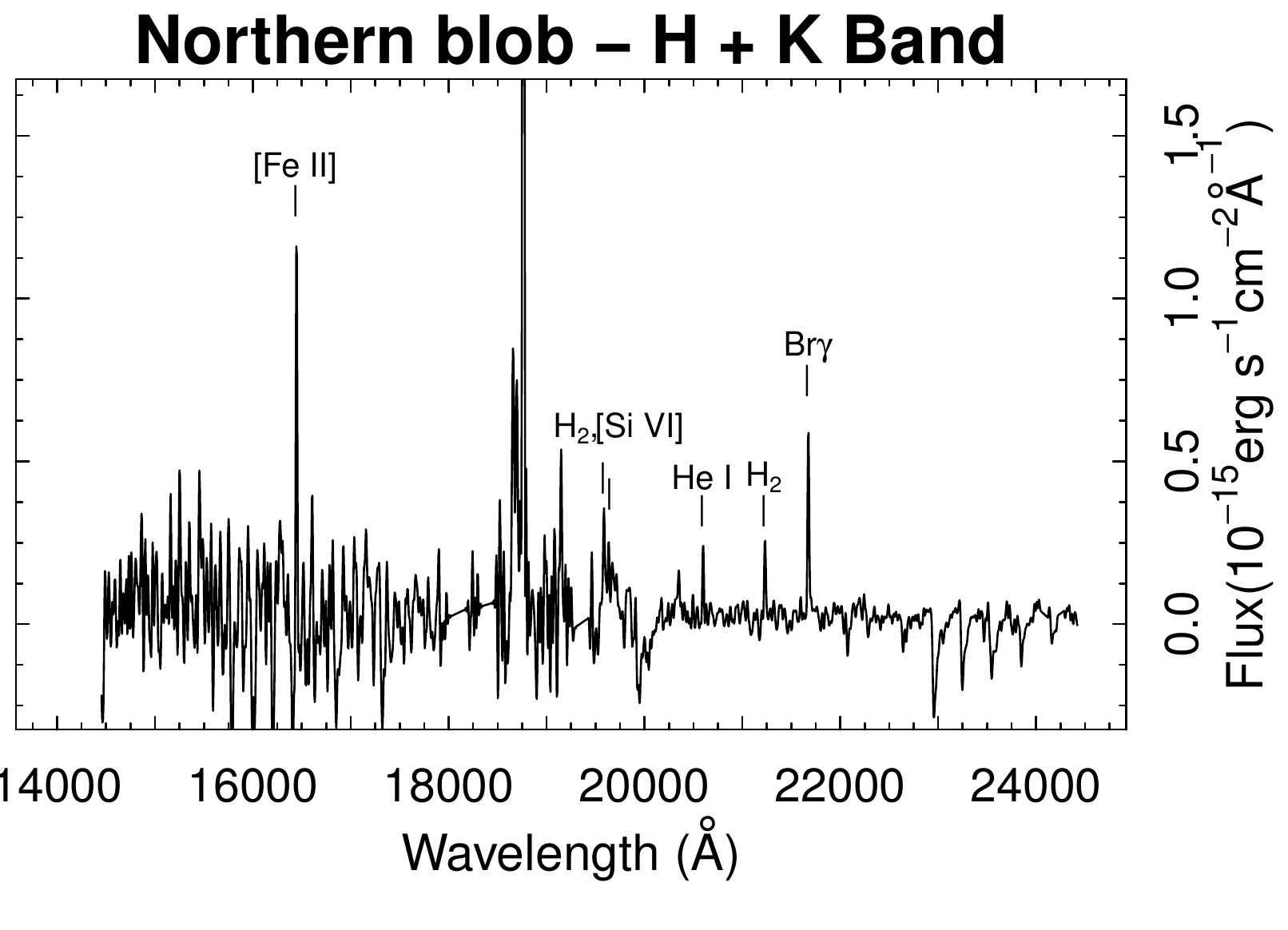}

\caption{NIR spectra extracted from the regions of the SB (left) and the NB (right).  \label{nir_emission_spectra}}
\end{figure*}

We extracted the spectra of the NB and the SB from the NIR data cubes; they are shown in Fig. \ref{nir_emission_spectra}. Both present strong Hydrogen recombination lines (Pa$\alpha$ and Br$\gamma$). In addition, [Fe II]$\lambda$1.644$\mu$m is quite strong, with [Fe II]/Br$\gamma$ $\sim$ 2 in the NB and $\sim$ 4 in the SB. He I$\lambda$2.07$\mu$m is also quite conspicuously seen in the blobs, more than in the nucleus. 

We also extracted the spectra of the nuclear region from both the optical and NIR data cubes. Spectra of the N1, N2, N3, N4 and N5 nebulae, were also taken but only from the GMOS observations. The results are analysed below.



\subsection{The nuclear spectrum} \label{sec:nuclear_spectrum}

\begin{figure*}

\includegraphics[scale=0.53]{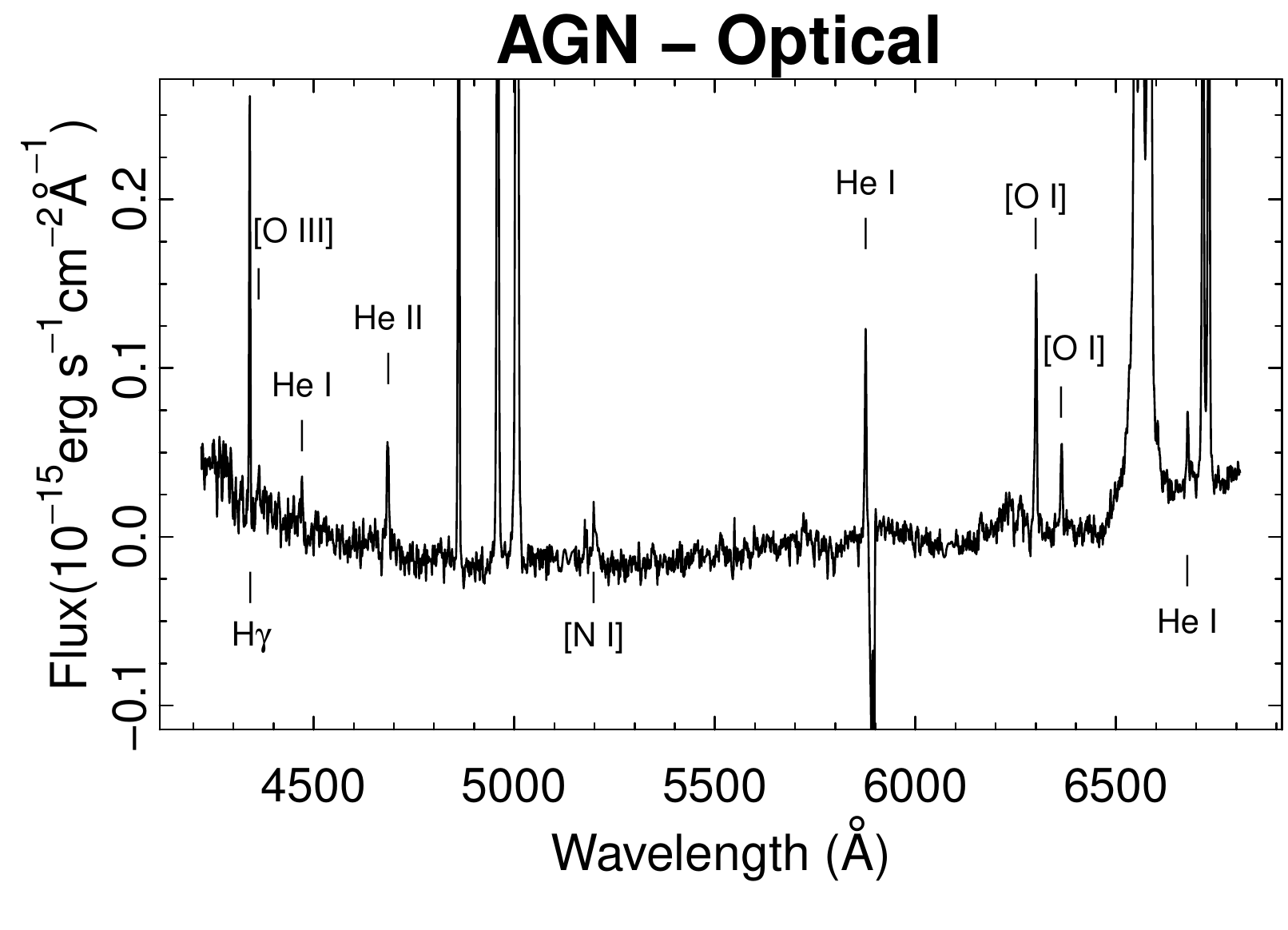}
\includegraphics[scale=0.53]{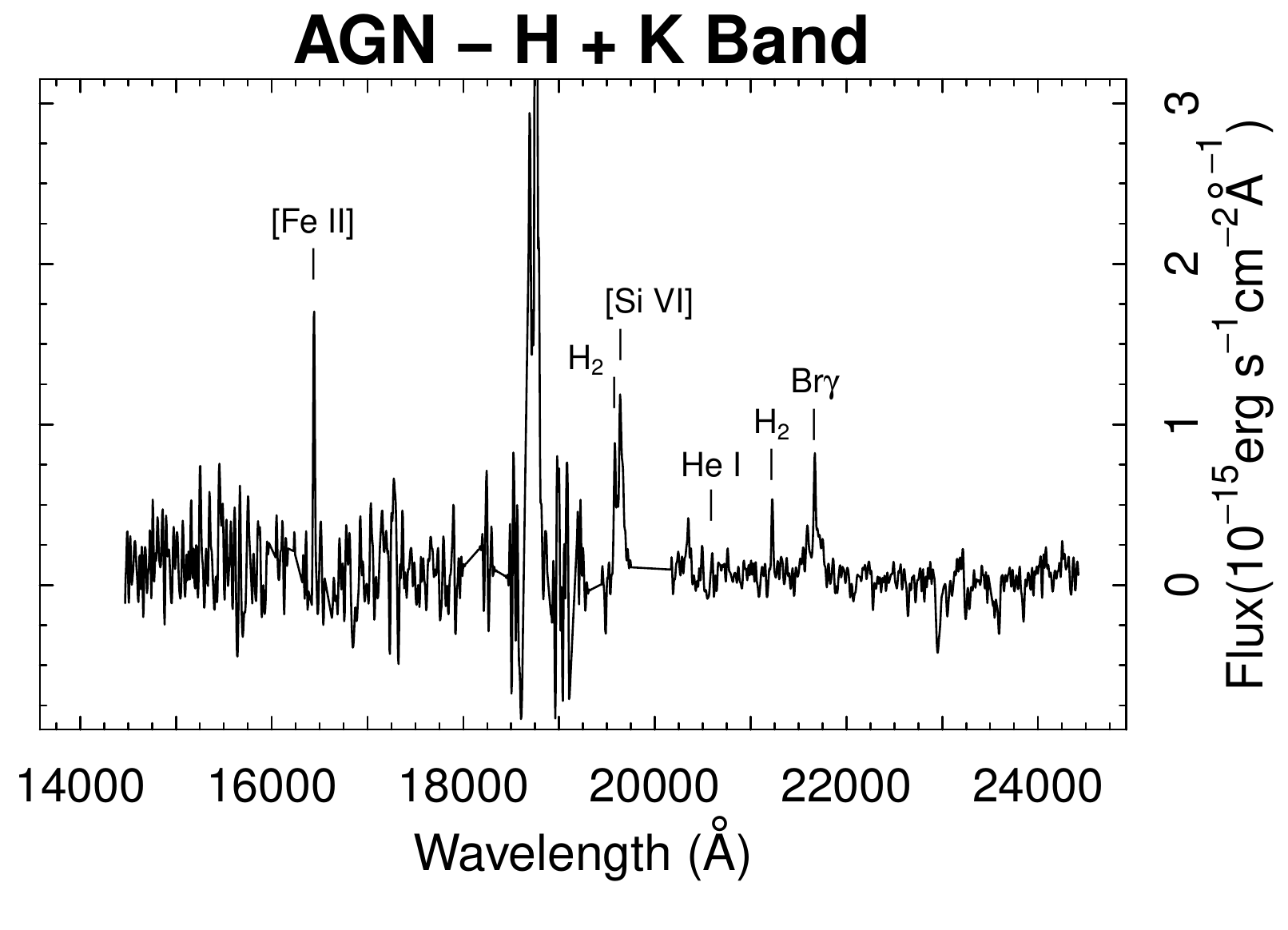}

\caption{Spectra extracted from the position of the AGN using the GMOS (left) and the SINFONI (right) data cubes. The $y$-axis was set to show the weak emission lines that are present in both spectra. A few weak lines are identified in both figures.  \label{agn_related_spectra}}
\end{figure*}

The nuclear spectrum corresponds to the spaxel taken from the peak position of the emission corresponding to the red wing of the broad component of H$\alpha$. Then, we applied an aperture correction to obtain the total flux contained within the PSF of the data cube. With this procedure, we minimize the circumnuclear emission of NGC 7582. In Fig. \ref{agn_related_spectra}, we show this spectrum focused on a few weak lines that are detected in this region.

We also fitted Gaussian functions to the most prominent emission lines of the nuclear spectrum. It is worth mentioning that the procedure here is different from what was done in Section \ref{sec:line_fittings}. First,  we began by fitting the [S II]$\lambda\lambda$6714, 6730 lines. The reason for doing this is that the [N II] + H$\alpha$ spectral region is affected by the broad component of H$\alpha$ in the nuclear spectrum. Since [S II], [N II] and narrow H$\alpha$ are produced in the same location of the NLR, we may assume that all these emission-line components have the same profile. The same procedure was adopted by \citet{1997ApJS..112..391H} to search for a broad component of H$\alpha$ in the galaxies of the Palomar survey \citep{1997ApJS..112..315H}. We assumed that the profile of each [S II] line is given by the sum of two Gaussian functions. Although two Gaussian functions are usually necessary to fit line profiles that are common in NLRs (see e.g. \citealt{1997ApJS..112..391H,2014MNRAS.440.2442R}), we will show below that only one Gaussian function is related to the NLR, while the other Gaussian function is associated with a contamination from H II regions along the line of sight of the nucleus of NGC 7582. The amplitude, FWHM and the peak of each Gaussian function were set as free parameters. After this, we fitted the [N II]$\lambda\lambda$6548, 6583 + H$\alpha$ spectral lines together. Note that a broad component for the H$\alpha$ line is needed in the nuclear spectrum. Thus, the free parameters for this spectral region are the amplitude of the Gaussian functions for the narrow components of each line plus the amplitude, peak position and FWHM of the Gaussian function related to the broad component. For all other emission lines ([O I]$\lambda\lambda$6300, 6363; [O III]$\lambda\lambda$4957, 5007; H$\beta$; [N I]$\lambda\lambda$5198, 5200; He II$\lambda$4686; [N II]$\lambda$5755 and [O III]$\lambda$4363), only the amplitude of the Gaussian functions were set as free parameters, since we also assumed that all these lines have the same profile as the [S II]$\lambda\lambda$6714, 6730 lines. 

\begin{figure*}

\includegraphics[scale=0.4]{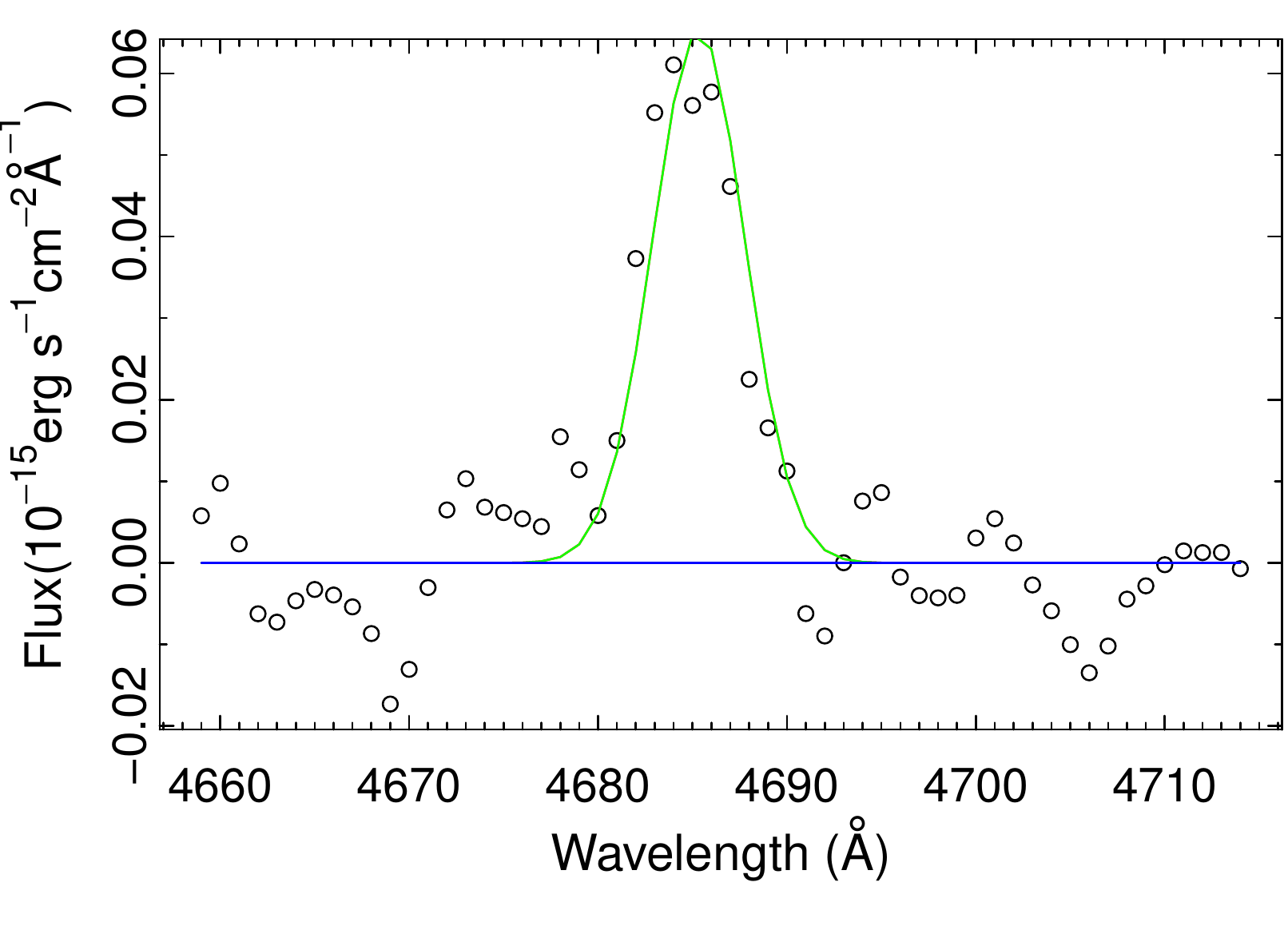}
\includegraphics[scale=0.4]{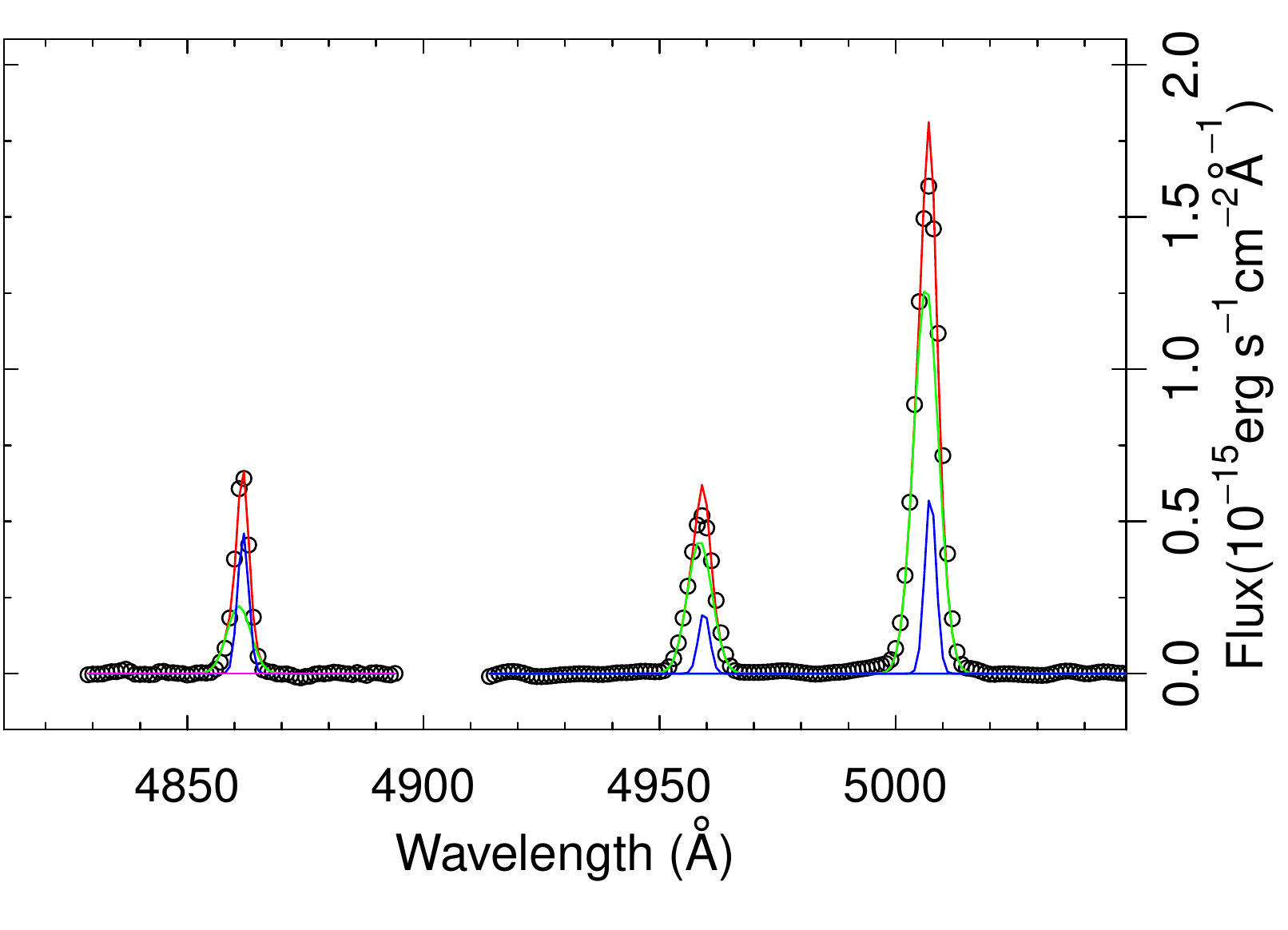}
\includegraphics[scale=0.4]{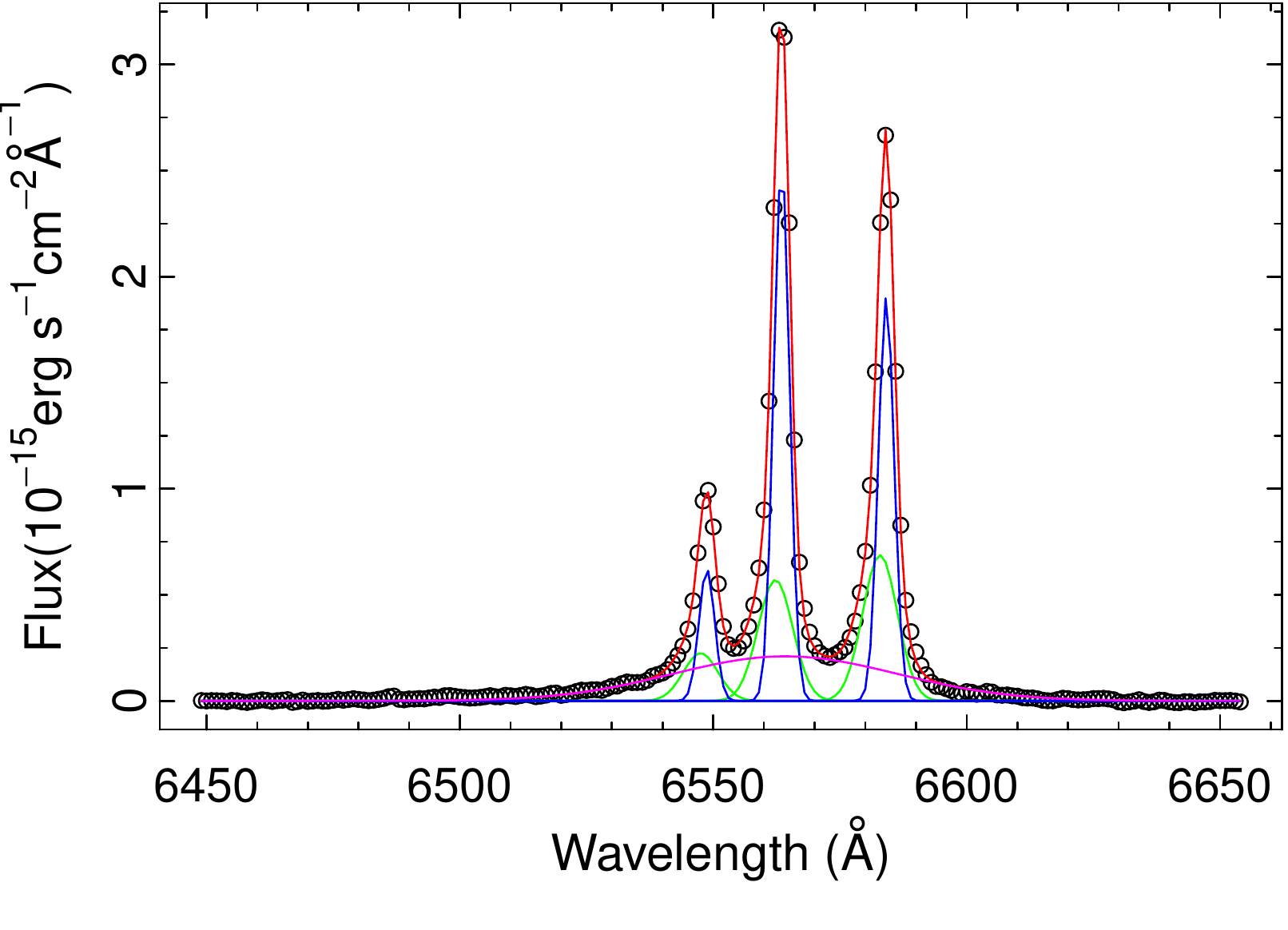}
\includegraphics[scale=0.4]{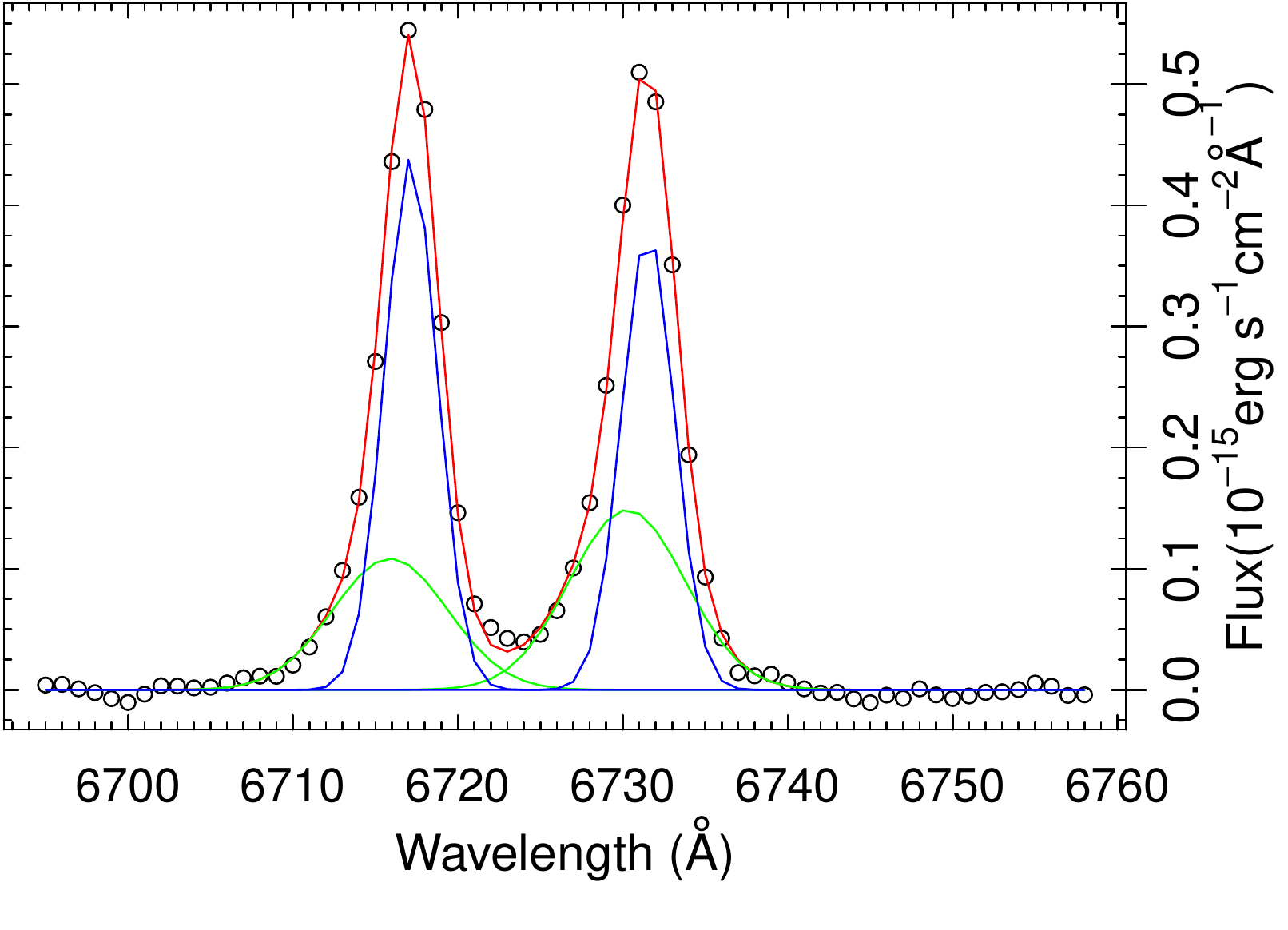}

\caption{Nuclear spectrum extracted from the position of the AGN assuming a Gaussian PSF with FWHM = 0.6 arcsec. Note that a broad H$\alpha$ component (magenta) is needed to fit the H$\alpha$+[N II] emission lines. Two Gaussians (green and blue) are needed to explain the narrower component of the H$\alpha$, [N II], [O III] and H$\beta$ lines and only one Gaussian (green) is needed to fit the He II$\lambda$4686 line. In fact, if we assume that the green Gaussian function is related to the AGN while the blue Gaussian function is associated with H II regions, it is reasonable that He II emission is well fitted only with the green Gaussian function. All emission lines were assumed to have the same kinematics. \label{nuc_spectra}}
\end{figure*}

Fig. \ref{nuc_spectra} shows the fitted profiles for the H$\alpha$, [N II]$\lambda\lambda$6548, 6583, [O III]$\lambda\lambda$4959, 5007, He II$\lambda$4686 and H$\beta$ lines. If one of the Gaussian functions used to fit the narrower component of the lines is related to the AGN, while the other Gaussian function is associated with H II regions, it is reasonable to assume that the Gaussian function with larger FWHM corresponds to the NLR, since the kinematics of this region is more affected by the SMBH. One should note that the profile of the He II$\lambda$4686 line is well fitted by the Gaussian function that is related to the NLR. This is expected, since young stars do not emit photons with an energy that is high enough to produce an He$^{++}$ region \citep{2006agna.book.....O}. Also, the [O I]$\lambda$6300 line profile is also described only by the Gaussian function related to the NLR, since H II regions do not have a large transition zone between fully ionized and neutral gas, where the [O I] emission is largely produced \citep{1983ApJ...269L..37H,2006agna.book.....O}. A caveat here is that both the He II and [O I] lines have lower signal-to-noise ratio than the other lines, which may also explain why only one Gaussian function is needed to fit their profiles. We present the H$\alpha$ flux, nebular extinction and line ratios for the NLR and for the nuclear H II region emission in Table \ref{tab_f_NLR}. It is worth mentioning that the nebular extinction was measured assuming an intrinsic H$\alpha$/H$\beta$ = 3.1 for the AGN and H$\alpha$/H$\beta$ = 2.87 for the H II region \citep{2006agna.book.....O}. The kinematic parameters related to both the NLR and the nuclear H II region are shown in Table \ref{tab_cin_objects}.

\begin{table*}
 \scriptsize
 \begin{center}
 \caption{Flux measurement for the narrow components of the AGN and the five nebulae detected in the He II$\lambda$4686 image. The $H\alpha$ flux is in units of 10$^{-15}$ erg s$^{-1}$ cm$^{-2}$.  \label{tab_f_NLR}
}
 \begin{tabular}{@{}lcccccccccc}
  \hline
  Object & Position & $f$(H$\alpha$)$_n$ & (H$\alpha$/H$\beta$)$_n$ & E(B-V) & [N II]/H$\alpha$ & [S II]/H$\alpha$ & [O I]/H$\alpha$ & [O III]/H$\beta$ & [N I]/H$\beta$ & He II/H$\beta$ \\
  \hline
  AGN & (0.0,0.0) &5$\pm$1& 3.5$\pm$0.3&0.11$\pm$0.07&1.20$\pm$0.07&0.46$\pm$0.04&0.19$\pm$0.01&5.9$\pm$0.4&0.15$\pm$0.03 & 0.28$\pm$0.03 \\
  Nuclear H II region& (0.0,0.0)  &10$\pm$1& 7.3$\pm$0.3&0.91$\pm$0.04&0.75$\pm$0.02&0.33$\pm$0.01&0.00$\pm$0.01&1.3$\pm$0.1&0.00$\pm$0.01 & 0.00$\pm$0.01\\
  N1& (1.5,1.0)  &107$\pm$1&5.8$\pm$0.2&0.61$\pm$0.03&0.57$\pm$0.01&0.23$\pm$0.01&0.01$\pm$0.01&1.6$\pm$0.1&0.02$\pm$0.01 &0.07$\pm$0.01\\
  N2& (0.6,0.3)  & 74$\pm$1&6.5$\pm$0.2&0.71$\pm$0.03&0.67$\pm$0.01&0.27$\pm$0.01&0.02$\pm$0.01&2.8$\pm$0.1&0.04$\pm$0.01&0.10$\pm$0.01\\
  N3& (1.0,$-$0.6)  & 44$\pm$1&6.3$\pm$0.1&0.68$\pm$0.02&0.60$\pm$0.01&0.28$\pm$0.01&0.02$\pm$0.01&1.9$\pm$0.1&0.04$\pm$0.01&0.10$\pm$0.01\\
  N4& (0.6,$-$1.0)  & 42$\pm$1&6.8$\pm$0.1&0.76$\pm$0.02&0.74$\pm$0.01&0.34$\pm$0.01&0.03$\pm$0.01&2.3$\pm$0.1&0.07$\pm$0.01&0.12$\pm$0.01\\
  N5& (1.4,$-$1.8)  & 11$\pm$1&5.1$\pm$0.1&0.48$\pm$0.02&0.90$\pm$0.01&0.42$\pm$0.01&0.04$\pm$0.01&4.8$\pm$0.1&0.07$\pm$0.01&0.26$\pm$0.01\\

  \hline
 \end{tabular}
 
 \end{center}
\end{table*}

\begin{table*}
 \scriptsize
 \begin{center}
 \caption{Results of the kinematics of each Gaussian set used to fit the emission lines of the spectra of the AGN, the nuclear H II region and the five nebulae detected in the He II$\lambda$4686 image. Two sets of Gaussians were used to fit the high- and mid-ionization lines (HI: [O III]$\lambda\lambda$4959, 5007 and He II$\lambda$4686) and other two sets of Gaussians were used to fit the low ionization lines (LI: H$\alpha$; H$\beta$; [N II]$\lambda\lambda$6548, 6583; [S II]$\lambda\lambda$6713, 6731; [O I]$\lambda\lambda$6300, 6363 and [N I]$\lambda\lambda$5198, 5200). For the nuclear spectrum (AGN and H II region), all narrow components have the same kinematic parameters, independent of the degree of ionization of the lines. All FWHM values were corrected for the instrumental broadening effect using the same strategy applied to the kinematic maps shown in Fig. \ref{fig_Vrad_el} and discussed in Section \ref{sec:kin_ion_gas}. All measurements are in units of km s$^{-1}$.  \label{tab_cin_objects}}
 \begin{tabular}{@{}lcccccccc}
  \hline
  Object &  FWHM$_1$(LI) & FWHM$_2$(LI) & FWHM$_1$(HI) & FWHM$_2$(HI) & V$_{r1}$(LI) & V$_{r2}$(LI) & V$_{r1}$(HI) & V$_{r2}$(HI) \\
  \hline
  AGN & $-$ &359$\pm$13 & $-$ & 359$\pm$13 & $-$&-24$\pm$8&$-$&-24$\pm$8 \\
  Nuclear H II region & 147$\pm$4 &$-$ & 147$\pm$4 & $-$ & 32$\pm$1&$-$&32$\pm$1&$-$ \\
  N1 &110$\pm$1&301$\pm$3&153$\pm$3&341$\pm$8&-58$\pm$2&24$\pm$2&-53$\pm$1&-77$\pm$2\\
  N2 &129$\pm$1&294$\pm$7&158$\pm$2&337$\pm$5&64$\pm$1&40$\pm$2&40$\pm$1&-45$\pm$5\\
  N3 &138$\pm$2&301$\pm$3&136$\pm$2&379$\pm$2&-37$\pm$1&-118$\pm$4&-16$\pm$1&-157$\pm$2\\
  N4 &61$\pm$8&276$\pm$2&117$\pm$2&343$\pm$2&21$\pm$3&-123$\pm$1&9$\pm$1&-111$\pm$2\\
  N5 &94$\pm$3&298$\pm$3&81$\pm$2&312$\pm$2&-43$\pm$1&-151$\pm$3&-40$\pm$1&-139$\pm$1\\
 \hline
 \end{tabular}
 
 \end{center}

\end{table*}

In Table \ref{tab_cin_NLR_NFWHM}, we list the luminosity of the H$\alpha$ line, corrected for the effects of the nebular extinction and assuming a distance of 22.5$\pm$2.2 Mpc \citep{2013AJ....146...86T}. One should be aware that the H$\alpha$ luminosity of the AGN that is presented in Table \ref{tab_cin_NLR_NFWHM} is the sum of the luminosities of the narrow and the broad components. Our nuclear H$\alpha$ luminosity value is 10 times lower than the \citet{1980MNRAS.193..563W} measurement. However, these authors used a slit 1.5 arcsec wide, thus their spectrum is probably contaminated by the H II regions close to the nucleus, in particular the one detected in the N1 region (see Section \ref{sec:spec_five_nebulae} and Table \ref{tab_cin_NLR_NFWHM}). The electron density was calculated in the same way as mentioned in Section \ref{sec:EL_ratios}. The mass of ionized gas was calculated using the relation \citep{2006agna.book.....O}:

\begin{equation}
    M_{ion}=\frac{L(H\alpha)m_H}{\epsilon n_e}=\frac{2.2\times10^7L_{40}(H\alpha)}{n_e} M_\odot,
    \label{eqionizedmass}
\end{equation}
where $L_{40}(H\alpha)$ is the H$\alpha$ luminosity in units of 10$^{40}$ erg s$^{-1}$ and $\epsilon$ = 3.84$\times$10$^{-25}$ erg s$^{-1}$ cm$^3$ is the H$\alpha$ line emissivity, assuming case B. Finally, the electron temperatures were calculated with the task {\sc temden} of the {\sc stsdas nebular} package under the {\sc iraf} environment \citep{1995PASP..107..896S}, using both the [O III]$\lambda$4959 + 5007/[O III]$\lambda$4363 and the [N II]$\lambda$6548 + 6583/[N II]$\lambda$5755 line ratios (see Table \ref{tab_cin_NLR_NFWHM}). Since we did not detect the [N II]$\lambda$5755 line for the nuclear H II region, only the temperature related to the [O III] lines is shown in this case.

\begin{table*}
\scriptsize
 \begin{center}
 \caption{Parameters extracted using the emission lines for the five nebulae, the AGN and the nuclear H II region. Column 1: object. Column 2: luminosity of the H$\alpha$ line in units of erg s$^{-1}$. This luminosity was corrected for the nebular extinction using the E(B-V) parameter shown in Table \ref{tab_f_NLR}. For the AGN, this luminosity is the sum of the narrow components with the broad component of H$\alpha$. Column 3: electron density in units of cm$^{-3}$. Column 4: mass of ionized gas in units of solar mass. Column 5: electron temperature measured with the [N II]$\lambda\lambda$6548, 6583 and the [N II]$\lambda$5755 lines in units of Kelvin. Column 6: electron temperature measured with the [O III]$\lambda\lambda$4959, 5007 and the [O III]$\lambda$4363 lines in units of Kelvin.\label{tab_cin_NLR_NFWHM}}
 \begin{tabular}{@{}lccccc}
  \hline
  Object & log $L$(H$\alpha$) & $n_e$ & $M_{ion}$ & $T_e$([N II]) & $T_e$([O III]) \\
  (1) & (2) & (3) & (4) & (5) & (6) \\
  \hline
  AGN & 39.12$\pm$0.13&1326$\pm$700&(2.2$\pm$1.2)$\times$10$^3$ & 1.0$\times$10$^4$ & 1.2$\times$10$^4$ \\
  Nuclear H II region & 39.74$\pm$0.22&223$\pm$80&(5.4$\pm$2.3)$\times$10$^4$ & $-$ & 1.3$\times$10$^4$ \\
  N1 &40.46$\pm$0.20&466$\pm$40&(1.4$\pm$0.3)$\times$10$^5$ & 6.4$\times$10$^3$ & 1.2$\times$10$^4$\\
  N2 &40.41$\pm$0.21&593$\pm$70&(9.5$\pm$2.3)$\times$10$^4$ & 7.2$\times$10$^3$ & 1.1$\times$10$^4$\\
  N3 &40.15$\pm$0.20&425$\pm$50&(7.3$\pm$1.7)$\times$10$^4$ & 6.4$\times$10$^3$ & 1.5$\times$10$^4$\\
  N4 &40.21$\pm$0.20&411$\pm$130&(8.7$\pm$3.3)$\times$10$^4$& 6.4$\times$10$^3$ & 1.4$\times$10$^4$\\
  N5 &39.34$\pm$0.20&187$\pm$145&(2.6$\pm$2.1)$\times$10$^4$& 7.5$\times$10$^3$ & 1.4$\times$10$^4$\\
 \hline
 \end{tabular}
 
 \end{center}

\end{table*}

We also calculated a few parameters for the broad component of H$\alpha$, such as the line flux, luminosity, FWHM and the radial velocity. The results are listed in Table \ref{tab_broad_component}. It is worth mentioning that the H$\alpha$ luminosity of the broad component was corrected for the extinction using the E(B-V) parameter found for the narrow components. Thus, one should bear in mind that this luminosity may be underestimated, since it is known that the nuclear region of this galaxy is highly obscured by a clumpy torus around the BLR \citep{2015ApJ...815...55R}.

\begin{table}
\scriptsize
 \begin{center}
 \caption{Parameters for the nuclear broad component. \label{tab_broad_component}}
 \begin{tabular}{@{}lcc}
  \hline
  Parameter & H$\alpha$ & Br$\gamma$  \\
  \hline
  Flux (10$^{-15}$ erg s$^{-1}$ cm$^{-2}$) & 12$\pm$1 & 0.43$\pm$0.06 \\
  log luminosity (erg s$^{-1}$) & 39.41$\pm$0.05 & 37.4$\pm$0.2\\
  FWHM (km s$^{-1}$) & 2382$\pm$44 & 2410$\pm$227  \\
  V$_r$ (km s$^{-1}$)& 64$\pm$18 & 112$\pm$97  \\
 
    \hline
 \end{tabular}
 
 \end{center}

\end{table}

\begin{figure}

\includegraphics[scale=0.5]{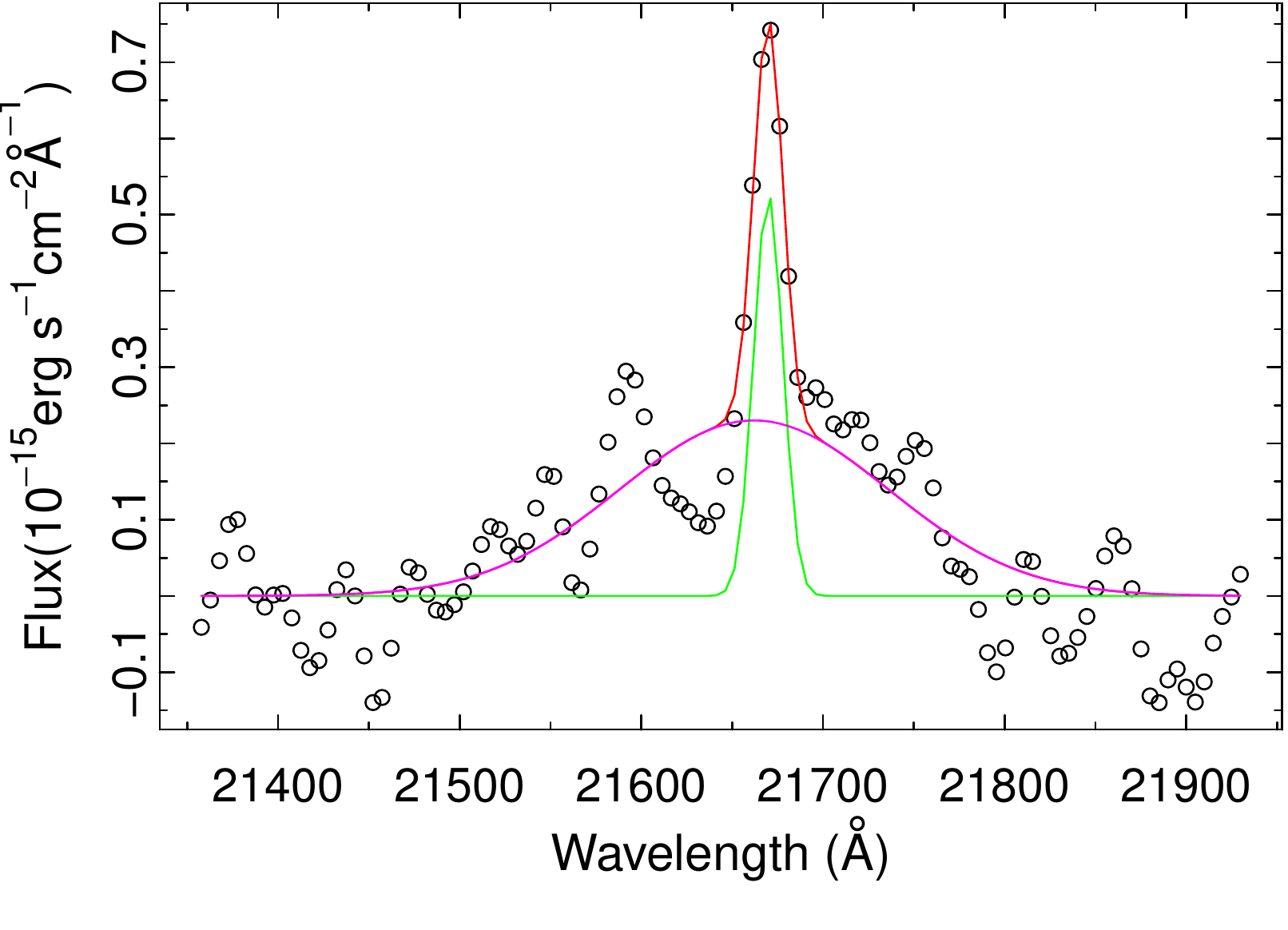}
\centering
\caption{Line profile of the nuclear Br$\gamma$ line. We fitted one Gaussian function for the narrower profile (green) and another Gaussian function for the broad component (magenta). We estimate an FWHM = 2410$\pm$227 km s$^{-1}$ for the BLR. Such a result agrees with the FWHM = 2382$\pm$44 km s$^{-1}$ of the BLR measured from the H$\alpha$ line. \label{fig:nuc_brg_spectrum}}
\end{figure}

In the NIR spectrum, also shown in Fig. \ref{agn_related_spectra}, the nucleus shows strong Hydrogen recombination lines: Pa$\alpha$ and Br$\gamma$. The Br$\gamma$ profile clearly presents the broad component (Fig. \ref{fig:nuc_brg_spectrum}) already reported in \citet{2001ApJS..136...61S}. The flux, luminosity, FWHM and the radial velocity of such component are shown in Table \ref{tab_broad_component}. This component is not seen in Pa$\alpha$ because of strong telluric absorption. The FWHM(Br$\gamma$) = 2410$\pm$227 km s$^{-1}$ agrees with the FWHM(H$\alpha$) = 2382$\pm$44 km s$^{-1}$. We also estimate the Br$\gamma$/H$\alpha$ $\sim$ 0.04. Assuming that H$\alpha$/H$\beta$ = 3.1 (this value may be higher for the BLR due to the collisional enhancement of H$\alpha$), thus the theoretical Br$\gamma$/H$\alpha$ = 0.009 \citep{2006agna.book.....O}. If this discrepancy is caused by absorption, then E(B-V)$_{BLR}$ $\sim$ 0.7. However, one should be aware that the NIR and the optical observations were taken with a difference of three years and the covering factor of the BLR possibly changed along this time. Also, the absolute flux calibration of both NIR and optical data cubes are very uncertain, with an error of $\sim$ 50 \% for both observations.

H$_2$ molecular lines are also seen in the NIR nuclear spectrum, as well as strong [Fe II]$\lambda$1.644$\mu$m, both typical in Seyfert galaxies. Luminous Seyfert 1 galaxies frequently show coronal lines in the $K$ band. In the present case, only [Si VI]$\lambda$1.965$\mu$m is seen with an intensity comparable to Br$\gamma$. [Ca VIII]$\lambda$2.321$\mu$m, frequently seen in luminous Seyfert galaxies, is not detected.

\subsection{Spectral characteristics of the optical nebular regions} \label{sec:spec_five_nebulae}

We extracted representative spectra of the N1, N2, N3, N4 and N5 regions. We fitted the most prominent emission lines of the five nebulae in the same way as is presented in Section \ref{sec:line_fittings}, the only difference being that we fitted two Gaussian functions to describe the line profiles in each spectrum. That means that we have, for each nebula, a set of kinematic parameters related to the mid- and high-ionization lines ([O III]$\lambda\lambda$4959, 5007, [O III]$\lambda$4363 and He II$\lambda$4686) and another set of kinematic parameters associated with the low-ionization lines (H$\alpha$, H$\beta$, [N II]$\lambda\lambda$6548, 6583, [N II]$\lambda$5755, [S II]$\lambda\lambda$6716, 6731, [O I]$\lambda\lambda$6300, 6363 and [N I]$\lambda\lambda$5198, 5200). For these nebulae, both Gaussian functions that were used to fit the line profiles do not have a specific interpretation, since it is hard to disentangle the kinematics of the gas that is mainly ionized by the AGN from the kinematics of the gas that is mainly ionized by young stars. The H$\alpha$ flux, nebular extinction and line ratios are shown in Table \ref{tab_f_NLR}. The kinematic results are shown in Table \ref{tab_cin_objects}. Finally, the H$\alpha$ luminosity, $n_e$, mass of ionized gas and the electron temperatures for both the low- and high-ionization gas are given in Table \ref{tab_cin_NLR_NFWHM}.

\subsection{Diagnostic diagrams} \label{sec:diagnostic_diagrams}

\begin{figure*}

\includegraphics[scale=0.4]{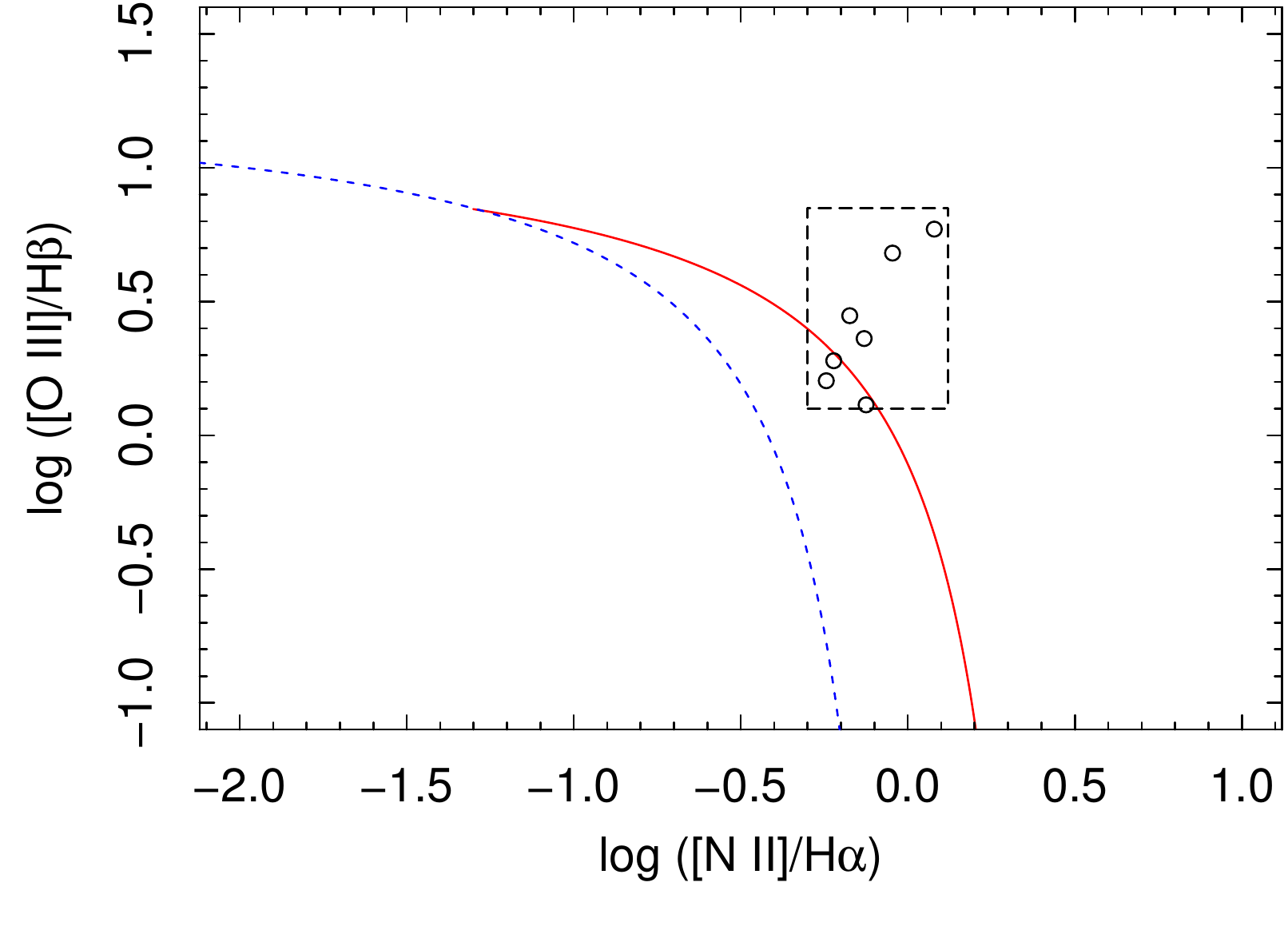}
\includegraphics[scale=0.4]{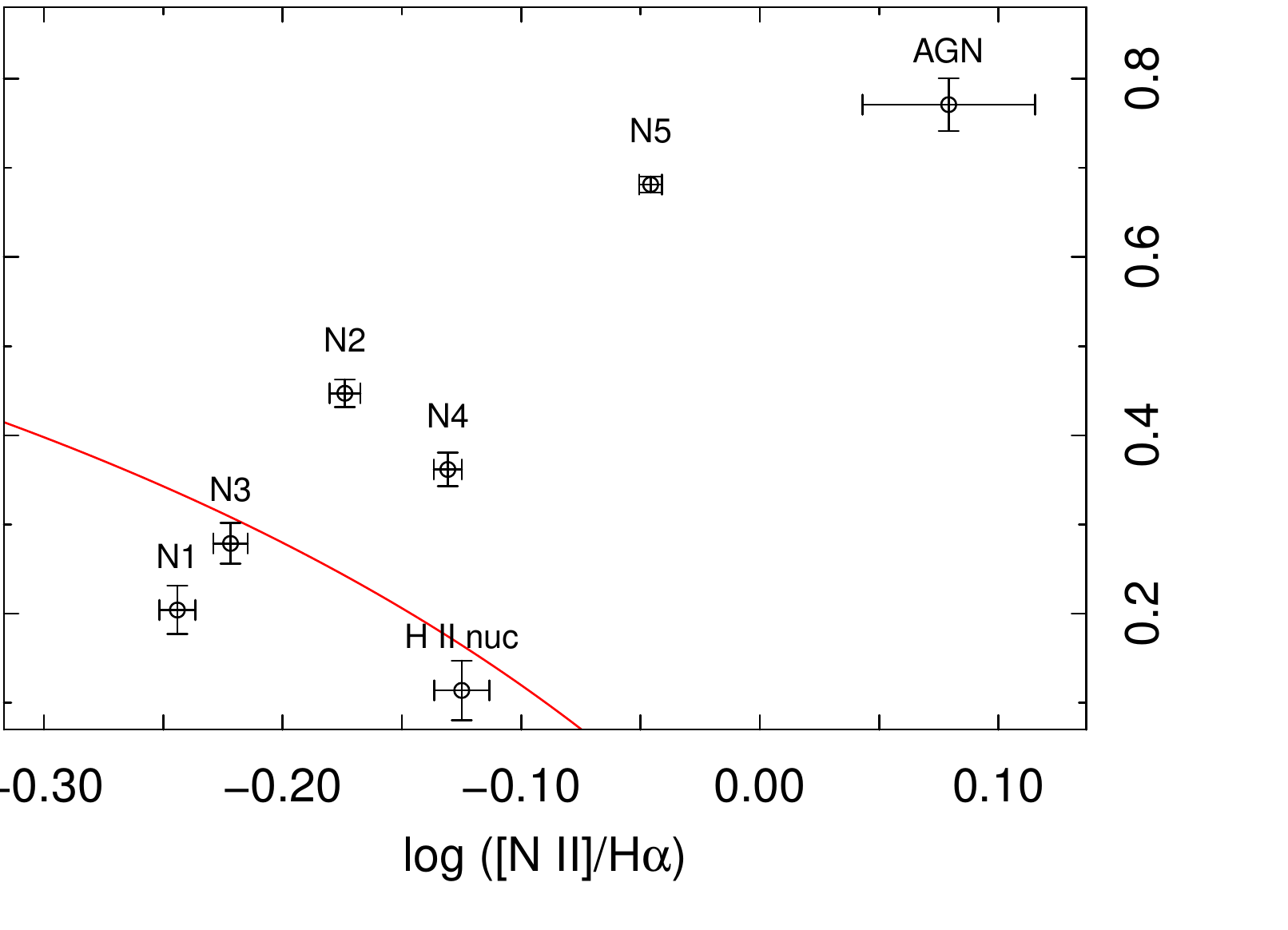}

\includegraphics[scale=0.4]{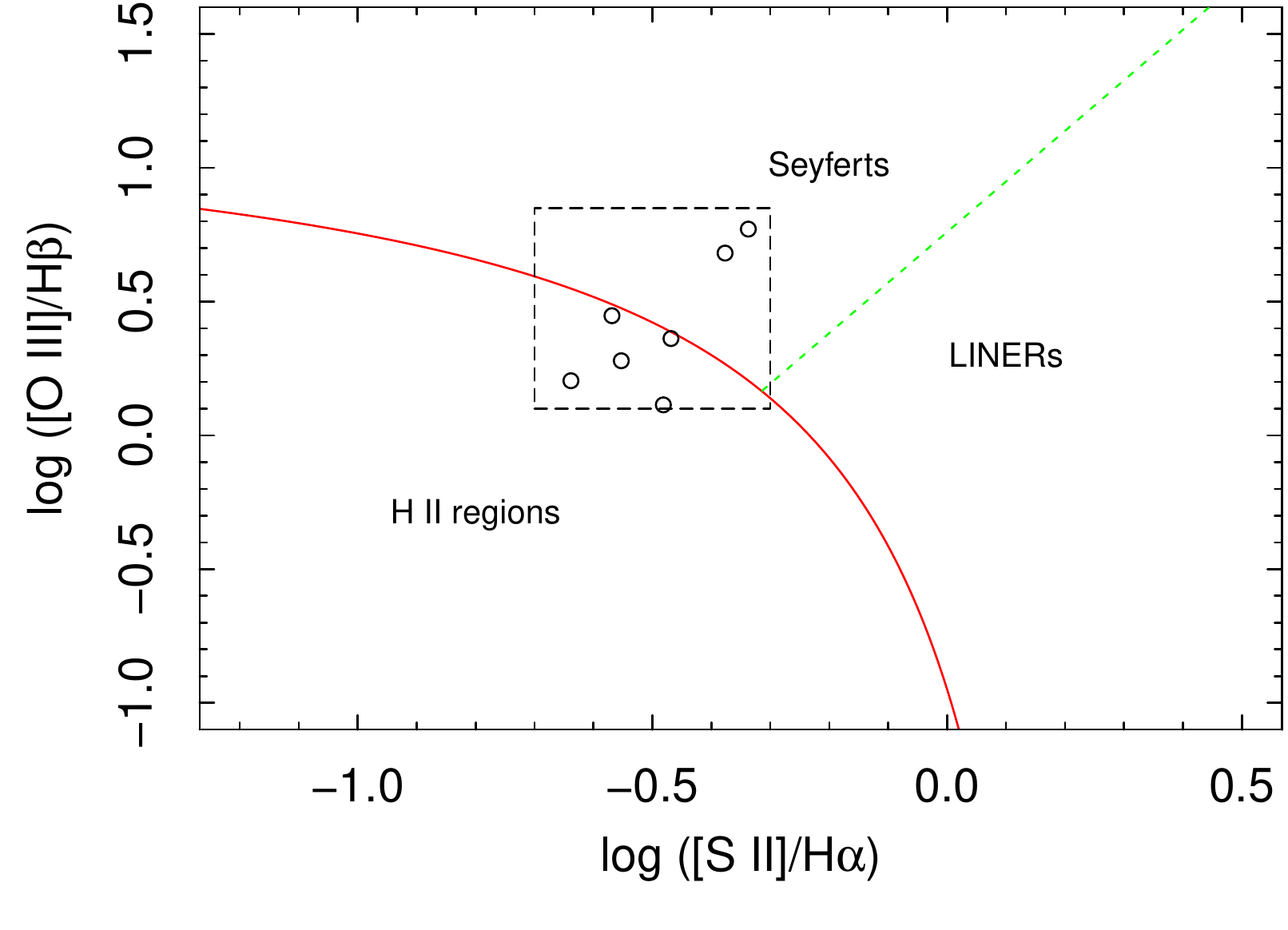}
\includegraphics[scale=0.4]{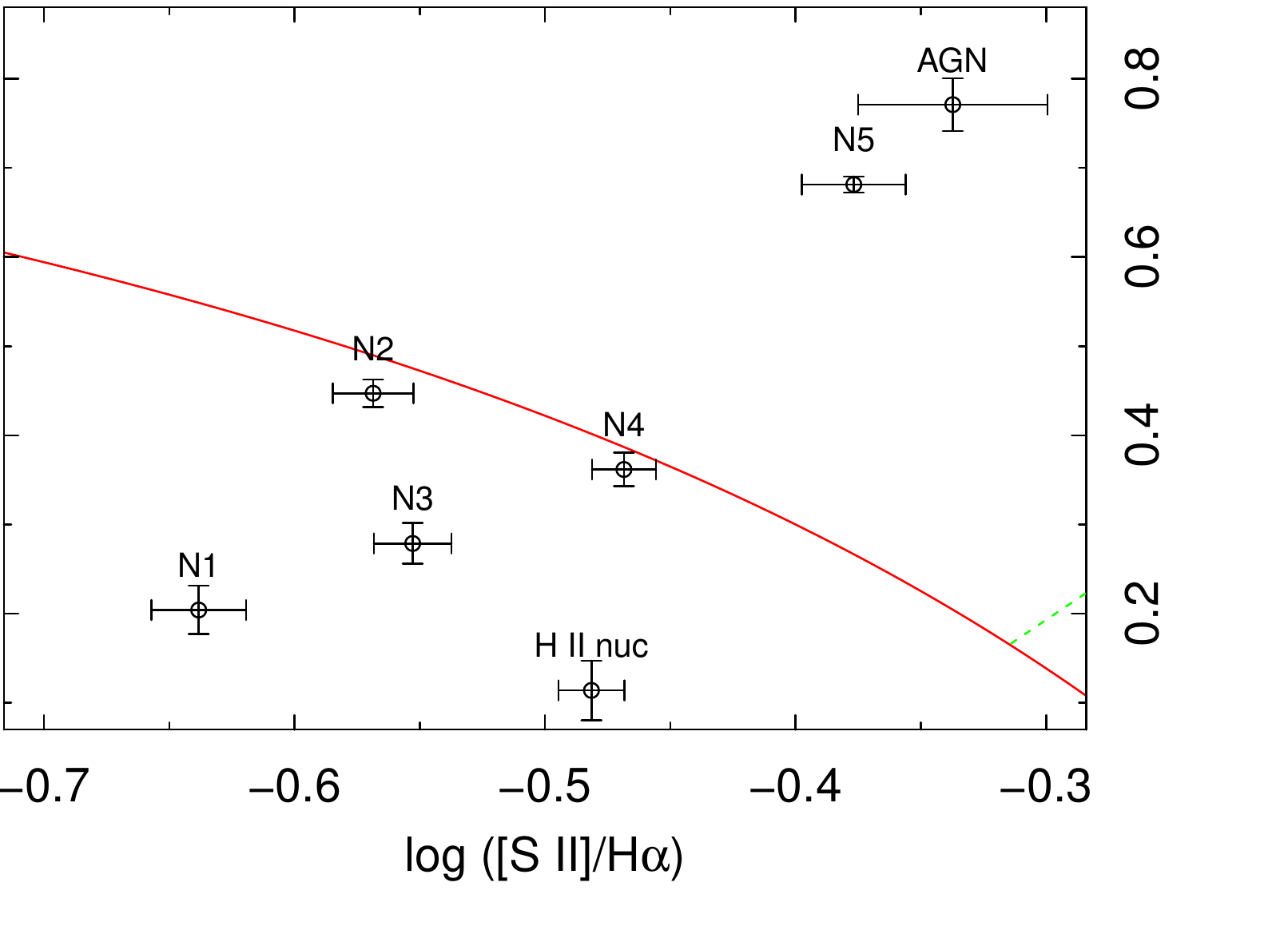}

\caption{BPT diagrams using different line ratios (from top to bottom) and with different ratio ranges (from left to right). The five nebulae, the nuclear H II region, and the AGN are identified in the figures. The continuous red line is the maximum starburst line proposed by \citet{2001ApJ...556..121K}, the dashed blue line is the empirical division between H II regions and AGNs \citep{2003MNRAS.346.1055K} and the dash green line is the LINER$-$Seyfert division suggested by \citet{2006MNRAS.372..961K}. The dashed boxes indicate the ratio ranges used in the diagrams located on the right-hand side. \label{bpt}}
\end{figure*}

In order to analyse the ionization source and structure of the five nebulae, the nuclear H II region and also of the AGN, we built BPT diagnostic diagrams. These diagrams, originally proposed by \citet{1981PASP...93....5B}, compare typical emission line ratios and are very useful to identify the contribution of starbursts and AGNs in the ionization balance of gaseous nebulae. They are shown in Fig. \ref{bpt}. The maximum starburst line proposed by \citet{2001ApJ...556..121K}, the empirical division between H II regions and AGNs of \citet{2003MNRAS.346.1055K} and the Seyfert$-$LINER division suggested by \citet{2006MNRAS.372..961K} were also inserted in the BPT diagrams. 

We note that each Gaussian function used to fit the narrower components of the nuclear emission lines lie in different positions of the BPT diagram. The Gaussian function related to the NLR has line ratios typical of a Seyfert nucleus. Given the presence of a broad H$\alpha$ line, we suggest that, at least at the time of our observations, NGC 7582 has a Seyfert 1.9 nucleus. The line ratios associated with the other Gaussian function are located in the H II region area of the BPT diagram, thus confirming the nature of this nuclear component. N5 also lies inside the Seyfert region. N1 and N3 may be classified as transition objects, i.e. clouds that are photoionized by both starbursts and AGNs (see e.g. \citealt{1997ApJS..112..315H,2006MNRAS.372..961K,2008ARA&A..46..475H}). N2 and N4 have [N II]/H$\alpha$ typical of photoionization by AGNs, but their [S II]/H$\alpha$ ratio would classify these clouds as transition objects.

\section{Discussion} \label{sec:discussion}

\subsection{The ionization cone} \label{sec:ion_cone}

NGC 7582 has a central extended emission that fits in the FOV of the GMOS$-$IFU. It is clear that the low-ionization gas reveals a different geometry when compared to the high-ionization gas. \citet{1985MNRAS.216..193M} have shown that their observations could be described by a model of gaseous disc seen in H$\alpha$ plus a conical outflow seen in [O III]$\lambda$5007. This outflowing gas seen in the mid- and high-ionization species is in the same direction as the ionization cone revealed by \citet{1991MNRAS.250..138S} and \citet{2016ApJ...824...50D}. We suggest that the five nebulae detected in the He II map are clouds located inside the ionization cone. It is worth mentioning that the He II$\lambda$4686 emission should trace the gas that is photoionized by an AGN, since young stars do not emit photons with energies $>$ 54.4 eV. That means that the clouds of the NLR of NGC 7582 are resolved in our images. 

\citet{2009MNRAS.393..783R}, using a distinct NIR data cube observed with the Gemini Near-Infrared Spectrograph, also revealed an outflow that is cospatial with our high-ionization emission. These authors found this outflow after subtracting the stellar kinematics from the gas kinematics obtained with the Br$\gamma$ line.

\subsection{The compact nebulae regions} \label{sec:compact_regions}

\subsubsection{Optical nebulae} \label{sec:compact_regions_optical}

The spectra extracted from the regions of the five nebulae detected in the He II image suggest that the low- and the high-ionization emission may come from different locations along the line of sight of these positions. Recalling that each line profile was fitted with two Gaussian functions, we can see that the kinematics of the broader component of the low-ionization lines are detached from the kinematics of the broader component of the high-ionization lines. Also, the electron temperatures are very different for the low ($T_e$ $\sim$ 7000 K) and the high ($T_e$ $\sim$ 14000 K) ionized gas (see Table \ref{tab_cin_NLR_NFWHM}). We proposed in Section \ref{sec:ion_cone} that the high-ionization He II emissions are originated in clouds inside the ionization cone. But what about the low-ionization emission? For this case, the five regions may have different characteristics, as discussed below:

\begin{enumerate}
\item \textit{N1}: this region has high H$\alpha$, [N II]$\lambda$6583, and He I$\lambda$5875 emission. It is also located very close to the WR emission and to V1. By the way, the high-velocity dispersion seen in this region may be related to winds from the WR stars. Line ratio diagnostic reveals that this region may be classified as a transition object (see Fig. \ref{bpt}). Thus, the low-ionization emission is probably originated by star-forming regions that are dominant in this location.

\item \textit{N2}: high H$\alpha$, [N II]$\lambda$6583 and He I$\lambda$5875 emission and a stellar cluster are located right in the centre of this region (V2). However, a strong [O I]$\lambda$6300 emission suggests a large partially ionized zone that is probably associated with the ionization cone. The low [N I]$\lambda$5198 emission may be an indication that this partially ionized zone may have densities between 2$\times$10$^3$ and 2$\times$10$^6$ cm$^{-3}$ [critical densities of the [N I]$\lambda$5198 and [O I]$\lambda$6300 lines, respectively \citep{1997iagn.book.....P}]. The line ratio diagnostic suggests that this region has Seyfert like emission.

\item \textit{N3}: there is a significant low-ionization emission, but not as high as in N1 and N2. The H$\alpha$, [N II]$\lambda$6583 and He I$\lambda$5875 emission seems to be related to the V4 and V5 clusters. The [O I]$\lambda$6300 emission also suggests the presence of high-energy photons from the AGN. This region may also be classified as a transition object in the diagnostic diagram.

\item \textit{N4}: part of the H$\alpha$, [N II]$\lambda$6583 and He I$\lambda$5875 emission may be related to V3. The line ratio diagnostic suggests that this region has typical Seyfert emission. The strong [Ne II]$\lambda$12.8$\mu$m line \citep{2006MNRAS.369L..47W} and a significant H$_2$ molecular line emission \citep{2001ApJS..136...61S} suggest a molecular cloud photoionized by the AGN. We will revisit this region in Section \ref{sec:ir_emissions}.

\item \textit{N5}: the low-ionization emission of this region is weak. The high-ionization emission associated with the low density results in a region with high-ionization parameter. This is in accordance with the line ratio diagnostic for this region that suggests a Seyfert-type emission. No stellar cluster is located close to this region.

\end{enumerate}

It seems that the bulk of the low ionization emission of, at least, two regions (N1 and N3) is associated with clouds that are mainly photoionized by young stars. The only exception would be N5. Obviously, a fraction of the low-ionization emission is originated in the same clouds that emit the He II line inside the ionization cone; part of the [O III]$\lambda$5007 emission is related to the nebulae photoionized by the starbursts. 

\subsubsection{NIR nebulae} \label{sec:compact_regions_nir}

The [Fe II] and Br$\gamma$ characteristics are quite distinct in the NB and SB; this is noticeable not only in the spectra (Fig. \ref{nir_emission_spectra}) but also in the images shown in Fig. \ref{fig:RG_jets}. The [Fe II], usually strong in shocked regions \citep{1988ApJ...328L..41K,1993ApJ...416..150F}, seems to indicate that the SB must be strongly shocked. This conclusion is supported by the fact that this region is also a strong emitter of [O I] and [N I], as clearly seen in Fig. \ref{fig_flux_el}. 

The NB has stronger Br$\gamma$ emission than the SB and seems to be more photoionized. It is not clear why such an asymmetry is introduced. Perhaps noticeable is the fact that, in the Br$\gamma$ emission, there is also a link between the nucleus and the NB. Three MIR spots seem to be associated with the NB while the other two are associated with the SB.

\subsection{The IR emission: evidence for jets} \label{sec:ir_emissions}

\begin{figure}
\begin{center}

\includegraphics[scale=0.15]{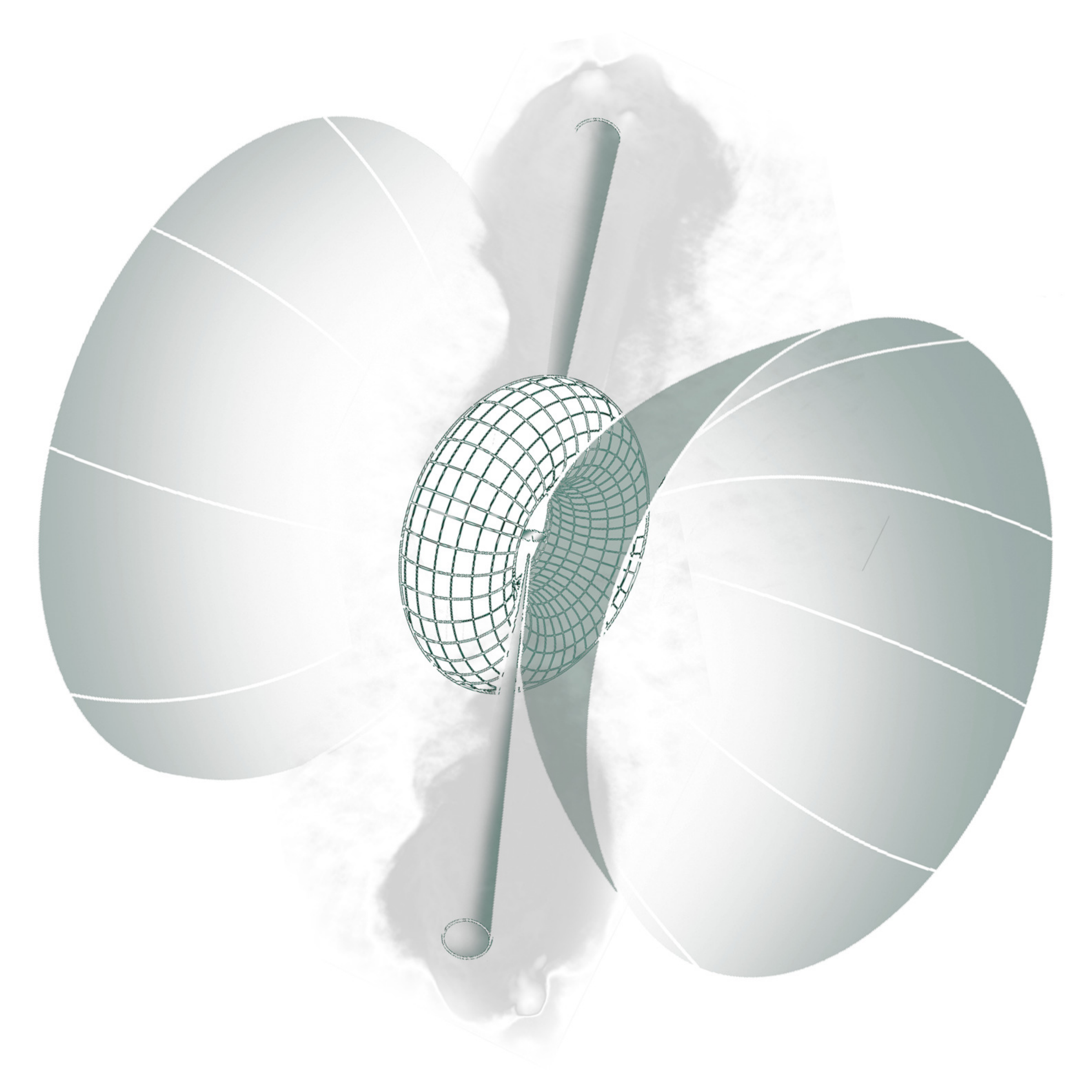}
\caption{Jet+ionization cone model for the central region of NGC 7582. This is the representation of the central region of NGC 7582 projected on the sky, as seen by the observer. In this scheme, the accretion disc, that sets the direction of the jets, is twisted with respect to the molecular torus, which defines the direction of the ionization cone. The high-energy photons across the ionization cone are responsible for the He II emission, while shocks caused by the jet explain both the northern and southern [Fe II] emissions. \label{fig:RG_razaoNIIHa_HeII}}
\end{center}
\end{figure}

The NIR images of [Fe II] and Br$\gamma$ (see Fig. \ref{fig:RG_jets}) show the NB and SB, in addition to the nucleus. It is clear, from the NIR images, that the bulk of the [Fe II] and Br$\gamma$ emission are not associated with the ionization cone, but with the MIR emitters M1, M2, M3, M4 and M5. The nearly aligned structure, with a PA $\sim$ 20$^o$, is almost orthogonal to the axis of the ionization cone. We interpret this as being associated with jets of, perhaps, supra-thermal wind that is ejected from the nucleus. Reconciling this with the orientation of the ionization cone demands a hypothesis that the central disc is quite tilted with respect to the molecular torus, responsible for defining the orientation of the ionization cone. This idea is depicted in Fig. \ref{fig:RG_razaoNIIHa_HeII}, where the jet is nearly perpendicular to the cone. Although this is somewhat unexpected, there are numerous cases in the literature in which the disc/jet orientation does not match that of the molecular/dusty torus (see, for instance, \citealt{2016MNRAS.457..949M,2017MNRAS.469..994M}). 

Another aspect that should be noticed is that the locations of the two blobs are not exactly symmetric with respect to the nucleus. This may be partially explained by dust, which affects more the eastern side than the western one, as is demonstrated by the structure map of the NICMOS image (Fig. \ref{HST_image}). However, the radio contours also show such an asymmetry and this indicates that the effect is real. Perhaps the case here is similar to the one observed in NGC 1068 where the radio structure as well as the ionized gas displays a structure that has its apex on a molecular cloud exposed to the nucleus, but not in the nucleus itself (see \citealt{2017MNRAS.469..994M} for a recent discussion).

\citet{2006MNRAS.369L..47W} interpreted the six MIR compact structures they detected as young massive clusters deeply embedded in dust to justify the absence of stellar clusters in optical images. However, the M2 region shows strong He II$\lambda$4686 emission (the N4 nebula), as well as strong [O I]$\lambda$6300 and [N I]$\lambda$5198 emission. As mentioned before, these two lines are produced in partially ionized gas; these zones are very small in H II regions but large in the presence of high-energy photons that are emitted by an AGN. Moreover, this region is in the Seyfert position of the BPT diagram. All these results point to the fact that the [Ne II]$\lambda$12.8$\mu$m emission of M2 is related to photoionization caused by the AGN. The strong [O I]$\lambda$6300, [N I]$\lambda$5198 and [Fe II]$\lambda$1.644 $\mu$m emissions together with a high molecular emission indicate that this cloud is not fully ionized. In addition, M2 seems to be located at the boundary of the ionization cone. The high-velocity dispersion of the gas in this region may be related to the fact that M2 (or N4) is exposed to the wind from the AGN and it is itself a source of a secondary wind. Thus, we propose that M2 is a molecular cloud that is exposed to high-energy photons from the AGN and that is hit by the jet that emerges from the nucleus, instead of being photoionized by a young massive cluster.

We do not see any optical line emission at the positions of M1, M3, M4 and M5; the nebular extinction is quite high in these regions. M3 stands out in the [O I]/H$\alpha$, [S II]/H$\alpha$ and [N II]/H$\alpha$ ratio maps, which indicates a thicker partially ionized zone. It also has strong [Fe II]$\lambda$1.644 $\mu$m emission. The molecular H$_2$ image reveals some emission in this area. M1 and M4 have high gas density, which is typical of star-forming regions. M5 has the highest nebular extinction along the FOV, which does not allow a precise optical analysis. The important thing to be pointed out here is that these five compact MIR emitters seem to be related to the NIR emission and, thus, to the Seyfert jet. However, some of the structures might also be associated with young stellar clusters, as do some optical nebulae.

\subsection{The Seyfert 1 AGN of NGC 7582} \label{sec:Sy1AGN}

The presence of a broad emission in H$\alpha$ at the position of the AGN is not surprising at all. There are evidences that the torus around the BLR of this galaxy is clumpy \citep{2015ApJ...815...55R}. In fact, a BLR is necessary to explain the nuclear X-ray absorption events of NGC 7582 \citep{2009ApJ...695..781B,2015ApJ...815...55R}. A broad component was already observed before in NIR spectra of this galaxy \citep{2005ApJ...633..105D,2009MNRAS.393..783R}, as well as in the optical \citep{1999ApJ...519L.123A}. These authors discussed the idea of a reddening change caused by the torus, although they proposed that it would be more likely that the broadening of these lines was caused by an SN explosion.

We propose that the broad H$\alpha$ and Br$\gamma$ emissions presented in this work do not come from an SN explosion, but from the AGN instead. We see an unresolved emission from the red wing of the broad H$\alpha$ , right at the position of the AGN, as set by the \textit{HST} image. We also noticed that the kinematics of the BLR extracted from the optical and from the NIR lines are nearly the same. It is worth mentioning that these emissions may come partially from a direct view of the BLR and/or partially reflected by the inner part of the ionization cone. With these results, we may propose that there was a clear view to the BLR in the dates of our observations (2004 July and 2007 August) and, thus, NGC 7582 may be classified as a Seyfert 1.9 galaxy.

\section{Conclusions} \label{sec:conclusions}

We studied the optical and NIR properties of the nuclear region of the galaxy NGC 7582 using archival GMOS$-$IFU and SINFONI observations. The FOV of the data cubes covered the central part of the galaxy on a spatial scale of hundreds of parsecs, with a superb spatial resolution of 67 pc (0.61 arcsec) in the optical and of 30 pc (0.26 arcsec) in the NIR. We focused our analysis mainly on the emission lines of the nuclear and circumnuclear region. Our main conclusions are summarized as follows:

\begin{itemize}

 \item In the optical data we resolve the NLR of the inner ionization cone. On the basis of the He II$\lambda$4686 emission we identified five nebular regions, all exposed to the emission of the central AGN.

\item On the basis of the location in the diagnostic diagrams, the nebular region N1 (and, possibly, N3) is partially ionized by stellar emission. N1 is near the brightest optical stellar cluster, which is associated with Wolf-Rayet features.

\item We detect broad optical H$\alpha$ emission at the position of the AGN. This may be partially due to direct view through a clumpy torus or reflected in the inner ionization cone. Broad emission is also seen in Br$\gamma$. The nucleus can also be identified in the optical by enhanced line ratios of [N II]/H$\alpha$, [O I]/H$\alpha$, [S II]/H$\alpha$ and [O III]/H$\beta$. Enhancements in forbidden line widths such as [N II] and [O III] at the nucleus are also noticed.

\item The NIR view is quite distinct from that of the optical. In addition to the strong emission from the nucleus, two major emitting regions are identified: the NB at a separation of 1.5$-$2.0 arcsec from the nucleus and the SB, at 1.0$-$1.5 arcsec from the nucleus. Hydrogen recombination lines are stronger in the NB than in the SB while [Fe II] is stronger in the SB when compared to the NB. 

\item Previously identified MIR spots coincide with the NB and SB. They have been proposed as star-forming regions. However, the two strongest ones are not noticed in the unobscured regions associated with the SB in \textit{HST} images. We propose that these MIR spots are related to the NB and SB, both excited by the AGN through a jet/wind mechanism and exposed to the central source.

\item We propose that the optical and NIR observations are explained by a model in which an accretion disc is quite tilted relative to the torus. Perpendicular to the disc, jets/wind of suprathermal electrons hit interstellar molecular clouds, forming the NB and SB. The NB is highly obscured by dust from the galactic plane. The SB is partially visible in the optical, identified as N4. This cloud also presents strong and distinct emission of [N I]$\lambda$5198 and [O I]$\lambda$6300 and shows evidence of strong turbulence, a probable consequence of the interaction of the jet with an interstellar molecular cloud.

\end{itemize}

\section*{Acknowledgements}

This paper is based on observations obtained at the Gemini Observatory, which is operated by the Association of Universities for Research in Astronomy, Inc., under a cooperative agreement with the NSF on behalf of the Gemini partnership: the National Science Foundation (United States), the National Research Council (Canada), CONICYT (Chile), the Australian Research Council (Australia), Minist\'{e}rio da Ci\^{e}ncia, Tecnologia e Inova\c{c}\~{a}o (Brazil) and Ministerio de Ciencia, Tecnolog\'{i}a e Innovaci\'{o}n Productiva (Argentina) and also on observations collected at the European Organisation for Astronomical Research in the Southern Hemisphere under ESO programmes 079.C-0328(A) and 079.C-0328(B). This research has also made use of the NED, which is operated by the Jet Propulsion Laboratory, California Institute of Technology, under contract with the National Aeronautics and Space Administration. Some of the data presented in this paper were based on observations made with the NASA/ESA Hubble Space Telescope, obtained from the data archive at the Space Telescope Institute. STScI is operated by the association of Universities for Research in Astronomy, Inc. under the NASA contract NAS 5-26555. This publication also makes use of data products from the Two Micron All Sky Survey, which is a joint project of the University of Massachusetts and the Infrared Processing and Analysis Center/California Institute of Technology, funded by the National Aeronautics and Space Administration and the National Science Foundation. 

TVR acknowledges CNPq for the financial support under the grant 304321/2016-8. JES acknowledges FAPESP for the financial support under the grant 2011/51680-6.  DM acknowledges CNPq for the financial support under the grant 312987/2016-1. The optical data were obtained under grant support from Instituto do Mil\^enio - MEGALIT (AGR). RBM acknowledges CAPES (Coordena\c{c}\~ao de Aperfei\c{c}oamento de Pessoal de N\'ivel Superior).

Finally, we thank Roberto Cid Fernandes for carefully reading the manuscript, which improved a lot the quality of this work, and also Juliana Cristina Motter, for English revision. We also thank the anonymous referee for valuable insights that also enriched this paper. 

\bibliographystyle{mn2e}
\bibliography{bibliografia}



\end{document}